\documentclass{ws-ijmpd}

\begin{document}
\markboth{Pankaj S. Joshi and Daniele Malafarina}
{Gravitational Collapse}
%
\catchline{}{}{}{}{}
%

\title{\bf RECENT DEVELOPMENTS IN GRAVITATIONAL COLLAPSE
AND SPACETIME SINGULARITIES}
\author{Pankaj S. Joshi\footnote{email: psj@tifr.res.in}
 and Daniele Malafarina\footnote{daniele.malafarina@polimi.it}}
\address{Tata Institute of Fundamental Research\\
Homi Bhabha Road, Mumbai 400005\\
India}

\maketitle
\begin{history}
\received{Day Month Year}
\revised{Day Month Year}
\comby{Managing Editor}
\end{history}

\begin{abstract}
It is now known that when a massive star collapses under
the force of its own gravity, the final fate of such a continual
gravitational collapse will be either a black hole or a naked
singularity under a wide variety of physically reasonable
circumstances within the framework of general theory of relativity.
The research of recent years has provided considerable
clarity and insight on stellar collapse, black holes and the
nature and structure of spacetime singularities. We
discuss several of these developments here. There are also
important fundamental questions that remain unanswered
on the final fate of collapse of a massive matter cloud in
gravitation theory, especially on naked singularities which are
hypothetical astrophysical objects and on the nature of cosmic
censorship hypothesis. These issues have key implications for
our understanding on black hole physics today, its astrophysical
applications, and for certain basic questions in cosmology
and possible quantum theories of gravity.
We consider these issues here and summarize recent results
and current progress in these directions.
The emerging astrophysical and observational perspectives
and implications are dicussed, with particular reference
to the properties of accretion discs around black holes and
naked singularities, which may provide characteristic
signatures and could help distinguish these objects.

\end{abstract}
\makeatletter
\newcommand\@pnumwidth{1.55em}
\newcommand\@tocrmarg{2.55em}
\newcommand\@dotsep{4.5}
\setcounter{tocdepth}{3}
\newcommand\l@section{\@dottedtocline{2}{0em}{1.5em}}
\newcommand\l@subsection{\@dottedtocline{2}{1.5em}{2.3em}}
\newcommand\l@subsubsection{\@dottedtocline{3}{3.8em}{3.2em}}
\newcommand\tableofcontents{%
\section*{Contents}%
\@starttoc{toc}%
}
\makeatother
\tableofcontents

\section{Introduction}
What is the final fate of a massive star that has exhausted
its internal nuclear fuel and which then undergoes a catastrophic
gravitational collapse under the force of its own gravity?
What are the implication of such a phenomenon on our basic
understanding of gravitation theory, and what will be the
observational implications as far as very high energy astrophysical
phenomena are concerned?

This is the arena where we in fact come face to face
with the regime of extreme and ultra-strong gravity fields,
and the answer must be sought within the framework of a
gravitation theory such as the Einstein's theory of gravity.
This is where the new and unfamiliar universe encompassing
the ultra-strong gravity effects actually reveals itself
and we must encounter exotic astrophysical objects
such as the spacetime singularities, black holes, and
other ultra-compact entities in nature.

>From such a perspective, considerable work has been done in
past years to understand the dynamical gravitational collapse
in general relativity, and many new insights have been obtained.
At the same time, several basic issues also remain unclear
or unanswered, and there is a general consensus that the nature
of cosmic censorship, black holes and naked singularities remain
one of the most important unresolved issues in gravitation theory
and black hole physics today. These
issues would
necessarily have far-reaching implications and applications
for our understanding
on fundamental aspects of gravity theories and applications
to high energy astrophysics.

Our purpose here is to discuss several key
problems
and questions on gravitational collapse, formation of black holes
and naked singularities as final state for a continual collapse,
and the related
issues regarding the
nature and structure of spacetime
singularities. These problems are closely related to the nature
of cosmic censorship hypothesis and our current understanding
on black hole physics.
The understanding of such questions is
also basic to the theoretical
developments as well as the modern astrophysical applications
of black holes which are being vigorously pursued today.

Further, we discuss and consider here certain interesting
astrophysical implications emerging from the current
work on gravitational collapse within the framework of
general relativity. An important question would be, if naked
singularities, which are hypothetical astrophysical objects,
actually formed in gravitational collapse of massive stars,
how these would look observationally different from
black holes. A lead that is emerging from the
recent work is that the accretion discs around these
ultra-high gravity objects, namely black holes and
naked singularities, would be significantly
different from each other providing characteristic signatures.
Therefore there exists
a possibility to distinguish these two different outcomes
of collapse observationally. We discuss some of these
current developments here.

We thus consider and take up below a series of outstanding
questions on these topics, where we attempt to clarify
what is already known, while separating the issues which
remain unanswered as yet, and on which more work still remains
to be done. It is our hope that such a treatment will clarify
where the research frontiers are moving on these problems and
what remain the major outstanding questions where more work is
needed. We intend to review here in this manner some of the
major challenges in black hole physics today, and the current
progress on the same. It is emphasized that to secure a
concrete foundation for the basic theory of black hole
physics as well as to understand the high-energy astrophysical
phenomena, it is essential to gain a suitable insight
into these questions. This will be of course from a perspective
of what we think are the important problems, mainly
within an analytical treatment of the Einstein's theory
of gravity in the framework of the general relativity
field equations, and no claim to completeness is made.

\section{Stellar collapse}
The stars have a life cycle wherein they are born in
gigantic clouds of dust and galactic material, they then
evolve and shine for millions of years, and eventually
enter the phase of dissolution and extinction. Stars
shine by burning their nuclear
fuel within, which is mainly hydrogen, fusing it into helium and later
into
other heavier elements. Eventually, when all matter is
converted to iron, no more nuclear processes
capable of producing energy
are possible and no
new internal energy is produced within the star. The all-pervasive
force of gravity then takes over to determine the final fate
and evolution
of such a star. Earlier there was a balance between the force
of gravity that pulled matter of the star towards its center and
the outwards pressures generated by internal fusion processes.
This balance kept the star stable and going,
while it lived its
normal life span of shining and radiating light and energy
produced within. Once the internal pressures subside,
gravity takes over, and the star begins to contract
and collapses onto itself.

A star as massive as ten or twenty times the sun would burn
much faster and live only a few million years, as compared to the
lifetime of several billion years for a smaller star such as the sun.
When the sun runs out of internal fuel, its core will contract under
its own gravity, but it will then be eventually supported by
a new force within, created by fast moving electrons, called
the electron degeneracy pressure.
Such an object is called a white dwarf. Similarly, stars up to three
to five times the mass of the sun would settle to the final state,
which is a neutron star, after an initial collapse and losing
some of their original mass. These are pure neutron objects
created in the collapse under the strong crush of gravity which
collapses atoms too. The
quantum pressure of these neutrons then support the star, which is
barely some ten to twenty kilometers in size. The final outcome
of collapse thus depends on the initial mass of the star, which
again stabilizes at a much smaller radius due to the balancing
pressures generated by either electrons or neutrons.

The more massive stars cannot, however, settle to a white dwarf
or neutron star state because these quantum pressures are then just
not sufficient to
balance gravity and
stabilize the collapsing star beyond the neutron star
mass limit. Then a continual gravitational collapse, which no known
physical forces are able to halt, becomes inevitable once the star
exhausted its internal fuel. So the life-history of a star of large
mass is essentially different from the small mass stars, and as
the astrophysicist Subrahmanyan Chandrasekhar pointed out
way back in 1934, ``...one is left speculating
on other possibilities.''
\cite{Chandra}
What will be the final fate of such a continual gravitational
collapse of a massive star? The answer must be determined by
the Einstein theory of
gravitation, as gravity now is the sole force deciding the future
evolution of the star.

Gravitation theory and relativistic astrophysics have gone
through extensive developments in past decades, further to the
discovery of quasars in 1960s, and also other very high energy
phenomena in the universe such as the gamma-ray bursts. For
compact objects such as neutron stars and for situations involving
very high energy densities and masses, strong gravity fields
governed by the general theory of relativity play an important
role and dictate the observed high energy phenomena having
intriguing physical properties.

Several models to explain the gamma-ray bursts, which emit in
a few seconds energy of the sun's entire lifetime,
have been proposed
in terms of a collapsar, invoking collapse of a single massive star
as the mechanism required to produce
the extreme burst of ultra-high energies.
The gravitational collapse of a massive star or much larger
matter clouds, lie at the heart of astrophysics of such
phenomena. It is the key physical process which is
basic to the formation of a star itself from interstellar
clouds, in formation of galaxies and galaxy clusters, and
in a variety of cosmic happenings including structure
formation in the universe.

\section{Final fate of gravitational collapse}
What is the final fate of a massive star towards the end
of its life cycle, when it exhausted its internal nuclear fuel
and started shrinking and collapsing under the pull of its
own gravity? This is one of the most important and outstanding
unresolved problems in astrophysics and cosmology today.

When the massive star runs out of its nuclear fuel, the
force of gravity takes over and a catastrophic gravitational
collapse of the star takes place. The star that lived for millions
of years and which stretched to millions of kilometers in size,
now collapses catastrophically within a matter of seconds.
According to the general theory of relativity, the outcome of
such a continual collapse will be a spacetime singularity
where all physical quantities such as densities and spacetime
curvatures diverge.

The fundamental question of the fate of a massive star,
when it collapses continually under the force of its own gravity,
was highlighted by Chandrasekhar
\cite{Chandra},
when he pointed out: `Finally, it is necessary to emphasize one
major result of the whole investigation, namely, that the life
history of a star of small mass must be essentially different from
the life-history of a star of large mass. For a star of small mass
the natural white-dwarf stage is an initial step towards complete
extinction. A star of large mass ($ > M_c$) cannot pass
into the white-dwarf stage, and one is left speculating on
other possibilities.'

It is possible to see the seeds of modern black hole
physics already present in the above inquiry made on the
final fate of massive stars. The issue of endstate of large
mass stars has, however, remained unresolved and elusive
for a long time of many decades after that. In fact,
a review of the status of the subject many decades later
notes, `Any stellar core
with a mass exceeding the upper limit that undergoes
gravitational collapse must collapse to indefinitely high
central density... to form a (spacetime) singularity'.
\cite{Report1986}
The reference above is to the prediction by general
relativity, that under reasonable physical conditions,
the gravitationally collapsing massive star must terminate
into a spacetime singularity.
\cite{HE}

While Chandra's work pointed out the stable configuration
limit for the formation of a white dwarf, the issue of final
fate of a star which is much more massive with tens of
solar masses, remains very much open even today. Such a star
cannot settle either as a white dwarf or a neutron star
final state.

The issue is clearly important both in high energy
astrophysics as well as cosmology. For example, our observations
today on the existence of dark energy in the universe and
the cosmic acceleration it produces
are intimately connected to the
observations of supernovae in the universe, which are
the product of
collapsing stars.
It is the observational evidence coming from
supernovae, that are exploding in the faraway universe,
which tells us
how the universe may be accelerating away and
the rate at which this acceleration takes place. At the
heart of such a supernova underlies the phenomenon of
catastrophic gravitational collapse of the massive star,
wherein a powerful shockwave is generated, blowing
off the outer layers of the star.

If such a star is able to throw away enough of its
matter in such an explosion, it might eventually settle
as a neutron star. But in the case otherwise, or if further
matter accreted onto the neutron star, there will be a
continual collapse again, and we shall have to then explore
and investigate the question of final fate of such a massive
collapsing star. But the stars, which are more massive
and well above the normal supernovae mass limits must
straightaway enter a continual collapse mode at the end
of their life cycles, without any intermediate neutron
star stage. The final fate of the star in this case
must be then decided by the general theory of
relativity alone.

The important point here is, more massive stars which
are tens of times the mass of the sun burn much faster and
are far more luminous. Such stars then cannot endure more
than about ten to twenty million years, which is a much
shorter span of life as compared to stars such as the sun,
which live much longer. Therefore, the question of final
fate of such short lived massive stars is of central
importance in astronomy and astrophysics.

What needs to be investigated then is what happens
in terms of the final outcome,
when such a massive star dies on exhausting its internal
nuclear fuel. As we indicate here, the general theory of relativity
then predicts that the collapsing massive star must
terminate into a spacetime singularity, where the matter
energy densities, spacetime curvatures and other physical
quantities blow up. It then becomes crucial to know whether
such super-ultra-dense regions, forming in the stellar collapse,
are visible to an external observer in the universe, or
whether they will be always hidden within a black hole
and an event horizon of gravity that could form as the
star collapses. This is one of the most important
issues in the physics of black holes today.

\subsection{A black hole is born: The Oppenheimer-Snyder-Datt model}\label{OSD}
To understand the final state of collapse for a massive star,
we need to trace the time evolution of the system or its dynamical
progression using the Einstein equations of gravity. The star
shrinks under the force of its own gravity, which comes to
dominate other basic interactions of nature such as the weak
and strong nuclear forces that typically provided the outwards
pressure to balance the pull of gravity.

This problem was considered for the first time by Oppenheimer
and Snyder, and independently by Datt, in the late 1930s.
\cite{OS,Datt}
In order
to deal with the rather complex Einstein equations, they
assumed the
density to be homogeneous and the same everywhere within the
spherical star. They also neglected gas pressure, taking it to be
zero. Their calculations then showed that an event horizon
develops as collapse progresses, such that no material particles
or photons from the region escape to faraway observers.
Once the star collapses to a radius smaller than the horizon,
it enters the black hole, finally collapsing to
a spacetime singularity with extreme densities that is
hidden inside the black hole and
invisible to any external observers. For the collapsing star
to create a black hole, an event horizon must develop prior
to the time of the final singularity formation.

Very considerable amount of research and astrophysical
applications have been
developed on
such black holes in past decades,
which occupy a major role in astrophysics and cosmology today.
To understand how an event horizon and black hole can form
when a massive star collapses, let us consider
the model for an homogeneous star
that eventually collapses to a singularity of
infinite density and spacetime curvatures.
The relativistic calculations imply that as the star collapses
the force of gravity
on its surface keeps growing and eventually a stage
is reached when no light emitted from its
surface is able to escape away to faraway observers. This is
the epoch when an event horizon formed, and the star then enters
the black hole region of spacetime. The infalling emitter does
not feel any thing special while entering the horizon, but any
faraway observer stops seeing the light from him. The strong gravity
of the star causes this one-way membrane, that is the event horizon
to form. Within the horizon
collapse continues further to crush the star into a
singularity. Such black holes can suck in more matter from
surroundings and grow bigger and bigger.

The physics that is accepted today as the backbone of the
general mechanism describing the formation of black holes as the
endstate of collapse relies on this very simple and widely studied
Oppenheimer-Snyder-Datt (OSD) dust model, which describes the
collapse of a spherical cloud of homogeneous dust.
In the OSD case, all matter falls into the spacetime singularity
at the same comoving time, while the event horizon forms
earlier than the singularity, thus covering it. A black hole
region in the spacetime results as the endstate
of collapse.

It is of course clear that the homogeneous
and pressureless dust is a rather highly idealized and
unphysical model of matter. Taking into account inhomogeneities
in the initial density profile it is possible to show that the
behaviour of the horizon can in fact change drastically,
thus leaving two different kinds of outcomes as the possible result
of generic dust collapse: the black hole, in which the horizon
forms at a time anteceding the singularity, and the naked
singularity, in which the horizon is delayed, thus allowing
the null geodesics or light rays to escape the central singularity
where the density and curvatures diverge, to reach faraway observers.
\cite{DJ,waugh1,waugh2}
It is clear that once the light rays escape, then the material
particles or the timelike geodesics will also escape from the
singularity.

\begin{figure}
\centerline{\includegraphics[width=9cm]{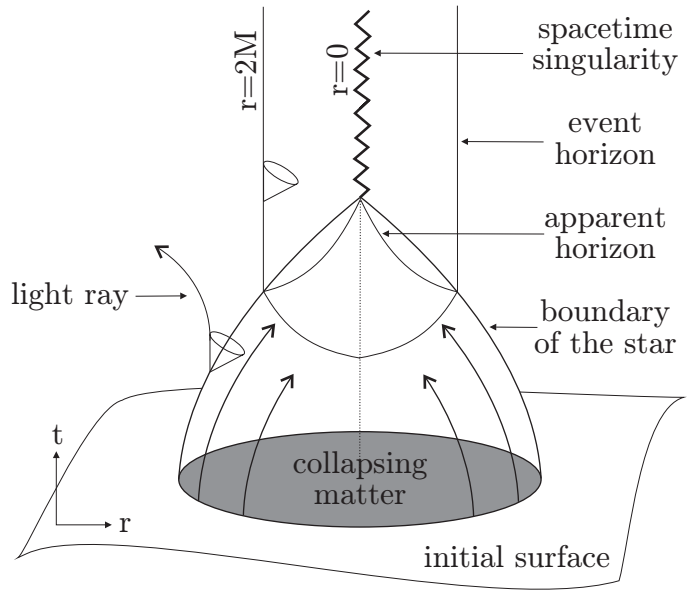}}
\caption{Dynamical evolution of a homogeneous
spherical dust cloud collapse, as described by
the Oppenheimer-Snyder-Datt solution. \label{f:one}}
\end{figure}

The issue of such a collapse has to be probed necessarily
within the framework of a suitable theory of gravity, because
the ultra-strong gravity effects will be necessarily important
in such a scenario. This purpose was achieved by the
OSD model that used the general theory of relativity to examine
the final fate of an idealized massive matter cloud, namely
a spatially homogeneous ball with no rotation or internal
pressure, and assumed to be spherically symmetric.
As said, the dynamical collapse created the spacetime singularity,
preceded by an event horizon, thus developing a black hole
in the spacetime. The singularity would be hidden inside such a
black hole, and the collapse eventually settled to a final
state which is the Schwarzschild geometry
(see Fig. \ref{f:one}).

Interestingly, there was not much attention paid to
this model at that time, and it was widely thought by
gravitation theorists and astronomers that it would be absurd
for a star to reach such a final ultra-dense state during
its evolution.  It was in fact as late as 1960s only, that
a resurgence of interest took place in such black holes,
their dynamical formation and physical properties that
they would exhibit for the surrounding regions of spacetime.
This was mostly due to some important observational
developments in astronomy and astrophysics that happened around that time,
such as the discovery of several very high energy phenomena in the
universe, like quasars, radio galaxies and such others,
where no known laws of physics were able to explain
the observations
that were related to
such extremely high energy phenomena in the cosmos.
Attention was drawn then to the dynamical gravitational
collapse and its final fate, and in fact the term
`black hole' was coined just around the same time in 1969,
by John Wheeler.

\subsection{Predictions of General Relativity}
According to Einstein's general theory of relativity, the
collapse must proceed to create a space-time singularity. This is
a region where the physical parameters such as the energy density
and the space-time curvatures blow up taking arbitrarily large
values. Thus the usual laws of physics break down near such a
singularity. This is the regime of ultra-strong gravity fields,
with other basic forces of nature playing only a secondary role.
Quantum effects must also become important in such strong fields
at ultra-small length scales near the singularity, and
eventually what is really needed would be a quantum theory
of gravity which would possibly resolve the singularity.
\cite{Ashtekar}

Specifically, the outcome of a continual collapse in a
fully general scenario of a spacetime with an evolving
matter field is
described by the singularity theorems in general relativity.
Subject to the following conditions, namely,
(i) An energy condition requiring the positivity
of energy density for matter fields, (ii) A reasonable causal
structure of the spacetime in terms of a causality condition
such as chronology or strong causality condition, and
(iii) A condition that trapped surfaces exist or develop
in the spacetime, which is a sufficient condition ensuring
that sufficient mass is packed in a small enough region,
the singularity theorems in general relativity
\cite{HE}
imply that the spacetime must contain a singularity in the
form of geodesic incompleteness.

It follows that for any general relativistic gravitational
collapse developing from regular initial data, in
a spacetime without any symmetry conditions such as spherical
symmetry necessarily holding, if the above physically reasonable
conditions are satisfied
then the collapse must create a spacetime singularity necessarily.
In all physically reasonable scenarios, the densities, curvatures
and all other physical quantities would typically blow up
in the limit of approach to such a spacetime singularity.

The OSD collapse scenario discussed above would be a special
case of such a general gravitational collapse, which terminates
into a final simultaneous spacetime singularity, which is
the final endstate of the collapsing matter cloud.

In view of the generality of the singularity theorems, it
would be expected that any physically realistic gravitational
collapse, where the massive star collapses continually towards
the end of its life cycle, must terminate into a spacetime
singularity of ultra-high densities and extreme spacetime
curvatures.

\subsection{What the singularity theorems do not predict}
It must be noted, however, that the singularity theorems
in general relativity predict only the existence of spacetime
singularities, under a set of physically reasonable conditions.
The above theoretical result on the existence of singularities
is of a rather general nature, and provides no information
at all of any kind on the nature and structure of such
singularities.

In particular, these theorems give us no information as
to whether such spacetime singularities, whenever they form
especially in a gravitational collapse, will be necessarily covered
in the event horizons of gravity and thus hidden from us, or
whether these could also be visible to external observers
in the universe.

Specifically, the possibility remains very much open
that a spacetime singularity develops in gravitational collapse,
however,
which
is no longer covered by an event horizon and may
be causally connected to faraway observers in the universe.
In such a case, the ultra-high density and curvature regions
would be
able to communicate with and send out signals to
exterior faraway observers. In this sense, gravity predicts
exciting outcomes for the final fate of a massive collapsing
star, with profound implications for fundamental physics.

Therefore, whether a strong curvature singularity that formed
in a realistic collapse would be visible or hidden from a faraway
observer in the universe remains very much an open question
in the Einstein's
theory of gravity. The key physical feature that decides
the visibility or otherwise of the singularity is the
interplay between the structure and time-curve
of the singularity and that of the
trapped surface formation in the spacetime.

\subsection{The cosmic censorship conjecture}
Unlike the idealized and rather special model that
the OSD homogeneous collapse scenario described above,
real stars have an inhomogeneous density (namely, higher
at their centers), and they also have non-zero pressures
within them as they collapse. Moreover, the stars also rotate.
Would every massive star collapsing towards the end of its
life cycle turn into a black hole necessarily, just as
the OSD case?
The cosmic censorship conjecture supposes that the answer
to this question is in the affirmative, namely, that the
singularity forming in collapse always hides within an
event horizon, never to be seen by external observers.

Theorists generally believed that in such
circumstances, a black hole will always form covering the
singularity, which will then be always hidden from external
observers. Such a black hole is a region of spacetime from
which no light or particles can escape. The assumption
that spacetime singularities of collapse would be
always covered by black holes is called the
Cosmic Censorship Conjecture (CCC).
\cite{CCC,IsraelCCC}
Thus, whatever the physical conditions and forces within
the massive stars may be, their continual collapse must
yield a black hole. Censorship assumption amounts to a
constraint on the nature of allowed dynamical
evolutions for collapsing clouds in general relativity.

As of today, we do not have any proof or any specific
mathematically rigorous formulation of the CCC available
within the framework of gravitation theory.
Therefore,
one of the key questions in black hole physics
today is,
is it possible that such singularities of collapse,
which are super-ultra-dense regions forming in spacetime,
be visible to external observers in the universe?
This is one of the most important unresolved issues in gravitation
theory currently.

If the singularities were always covered in horizons and if
CCC were true, that would provide a much needed basis for the
theory and astrophysical applications of black holes. On the
other hand, if the spacetime singularities which result from a
continual collapse of a massive star were visible to external
observers in the universe, we would then have the opportunity to
observe and investigate the super-ultra-dense regions in the
universe, which form due to gravitational collapse and where
extreme high energy physics and also the quantum gravity
effects will be at work.

The crucial physical question here is, can we really
see and observe such super ultra-dense regions which develop
when massive stars collapse, in violation to the CCC?
If CCC is to hold then the
singularity must be necessarily enveloped and
hidden in a region of spacetime from which no particles or
light can escape to faraway observers, and which cannot
communicate with external universe.
Such a region is called a black hole, and its boundary
is termed the event horizon.
The event horizon is, by definition, the boundary
of the region of spacetime that is accessible to an observer
at infinity, and is a one-way membrane that allows no matter
or light to escape away but
lets these fall in through it and
into the black hole.

In other words, this conjecture means that collapse of a massive
star must necessarily produce a black hole, which hides
within an event horizon the spacetime singularity of gravitational
collapse. But since past many decades it is only a conjecture.
At times it is
taken to be true, but the inevitability of event horizons
covering the singularities of collapsing stars has yet
to be demonstrated.

The point here is that, as discussed above, general relativity
predicts the necessary existence and formation of a spacetime
singularity from gravitational collapse, but it does not  tell
that the singularity must be within the black hole region only.
In other words, amongst the two, namely the horizon and singularity,
which one must come first as the star collapses is not answered
by the general relativity.  Therefore
the  hypothesis, which is the foundation of all of black
hole physics today, remains of crucial importance for all of
the basic theory and astrophysical applications of black holes
which are extensively pursued in recent years.

To decide on the validity or otherwise of censorship, one
must therefore study the dynamical collapse of massive stars to
determine their final fate. In recent years, detailed
analytical
research on
gravitational collapse has given insights into when black
holes form and when they would not, in fact, form. Some models
suggest that visible
ultra-dense regions, or naked singularities, may arise
naturally and generically as an outcome of collapse. If so, the
implications would be enormous and would touch on nearly
every aspect of astrophysics and fundamental physics. They
might account for extreme high-energy phenomena such as
the gamma-rays bursts that have defied explanation. They
might also offer a laboratory to explore quantum gravity effects
that are otherwise extremely difficult to observe.

The CCC came into existence in 1969,
when Penrose
suggested
and assumed that what happens in the OSD picture of
the gravitational collapse as discussed above, would be
the generic final fate of a realistic collapsing massive
star in general. In other words, it was assumed that
the collapse of a realistic massive star will terminate
into a black hole only, which hides the singularity,
and thus no visible or naked singularities must develop
in gravitational collapse.

To be precise, the key elements that are crucial to
the definition of the CCC are three:
\begin{itemize}
  \item[-] Dynamics: meaning that singularities must arise
from regular matter fields through the dynamical time evolution
that is governed by Einstein equations.
  \item[-] Physical validity: meaning that the models must
describe some physically realistic matter fields, typically those
obeying to some energy conditions.
  \item[-] Generic: meaning that the collapse outcome must
not be different, or change suddenly, whenever a slight change is
introduced in the model (in the form of slightly different matter
fields or perturbing the system's symmetries).
\end{itemize}
It is easy to see that the last two conditions are subject to
a rather broad interpretation, leaving some room for different
formulations and interpretations of the cosmic censorship
conjecture.

Further to the CCC formulation, many important developments
then took place in the
black hole physics which started in the earnest, and
several important theoretical aspects as well as
astrophysical applications of black holes started developing.
The classical as well as quantum aspects
of black holes were then explored and interesting
thermodynamic analogies for black holes were also developed.
Many astrophysical applications for the real universe
then started developing for black hole,
as for example in
models using black holes for
the description of
phenomena such as jets
emitted
from the centers of galaxies, extremely energetic gamma
rays burst, and such others.

The key issue raised by the CCC, however, still
remained very much open, namely whether a real star will
necessarily go the OSD way only as for its final state of collapse,
and whether the final singularity will be always necessarily
covered within an event horizon of gravity. This is
because, real stars are inhomogeneous, have internal pressure
forces and so on, as opposed to the idealized OSD assumptions.
Thus this remains an unanswered question, which is one
of the most important issues in gravitation physics and
black hole physics today.

\begin{figure}
\centerline{\includegraphics[width=9cm]{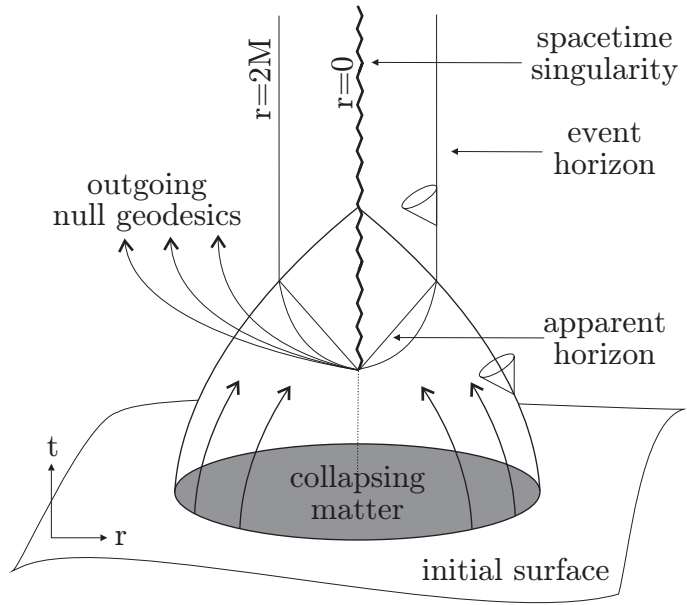}}
\caption{A spacetime singularity of gravitational collapse
which is visible to external observers in the universe,
in violation to the cosmic censorship conjecture. \label{f:two}}
\end{figure}

A spacetime singularity that is visible
to faraway observers in the universe is called a naked singularity
(see Fig. \ref{f:two}).
The point here is, while the general relativity predicts the
existence of singularity as the endstate for collapse, it gives
no information at all on the nature or structure of such
singularities, and whether they will be covered by
event horizons, or would be visible to external
observers in the universe.

As there is no proof, or even any mathematically
rigorous statement available for CCC after many decades
of serious effort, what is really needed to resolve the
issue is
a meticulous study of
gravitational collapse models for a realistic collapse
configuration, with inhomogeneities and pressures included.
These scenarios need
to be worked out and analyzed in detail within
the framework of Einstein gravity. Only such considerations
will allow us to determine the final fate of collapse in terms
of either a black hole or naked singularity final state.

It was hoped, as theorists kept developing black hole physics,
that a derivation from basic physical laws, of censorship would soon
arrive. However, despite numerous attempts over past four decades,
this is still not realized, and we do not even have any rigorous
mathematical formulation of the hypothesis. The reasons for this are
becoming clear now, as we discuss below. According to the
principles as laid out by the singularity theorems, singularities are
inevitable, but no such principles apply to the event horizon
as to when it exists or covers the singularity. In fact, the
initial singularity
in cosmology that created the observed universe is not hidden
inside a horizon, it is visible in principle. Whether this
is also true for stars has been an elusive question because
the Einstein equations are highly non-linear and complex.

The censorship hypothesis nevertheless provided a major impetus
to developments in black hole physics. Assuming, provisionally, that
it holds, physicists have investigated the properties of black holes
and created detailed laws of black hole dynamics and related aspects.
Meanwhile, they have also applied the concept of black holes to
explain various ultra-high energy processes observed in the
universe, such as quasars and X-ray-emitted binary star
systems.

\section{Recent developments on collapse}
It is clear that in view of the lack of any theoretical
progress on CCC, the only important and meaningful way to make
any progress on this problem is to make a detailed and
extensive study of gravitational collapse in general relativity.
Some recent progress in this direction is summarized below.
While we seem to have now a good understanding of the black
hole and naked singularity formations as final fate of
collapse in very many gravitational collapse models,
the key point now is to understand the genericity and stability
of these outcomes, as viewed in a suitable framework.
We discuss these issues here in some detail. Recent
developments on throwing matter into a black hole and the effect
it may have on its horizon and certain quantum aspects
will be discussed later. The issue of predictability or
its breakdown in gravitational collapse is of interest
as we shall point out later.

In particular, we discuss below the recent results
investigating the final fate of a massive star within
the framework of the Einstein gravity, and the stability
and genericity aspects of the gravitational collapse
outcomes in terms of black holes and naked singularities.
It is pointed out that some of the new results obtained in
recent years in the theory of gravitational collapse imply
interesting possibilities and understanding for the
theoretical advances in gravity as well as towards
new astrophysical applications.

Over the past years, many gravitational
collapse models have been worked out and studied in detail
at first considering simplified models such as dust or self similar
collapse (see e.g. Refs.
\refcite{Joshi-selfsim1}--\refcite{Lake-selfsim}).
and later refining to more elaborated models with the inclusion of
pressures.
The generic conclusion that has emerged out of these
studies is that, both the black holes and naked singularity
final states do develop as collapse endstates, for a
realistic gravitational collapse that incorporates inhomogeneities
as well as non-zero pressures within the interior of
the collapsing matter cloud. Subject to various regularity
and energy conditions to ensure the physical reasonableness
of the model, it is the initial data, in terms of the
initial density, pressures, and velocity profiles for
the collapsing shells, that determine the final fate of
collapse as either a naked singularity or a black hole
(for further detail and references see e.g. Ref.
\refcite{Joshi2008}).

\subsection{Collapse studies}
As we noted, while the censorship conjecture
is at the heart of the modern day black hole physics and its
astrophysical applications, not to speak of the proof, even
any rigorous mathematical formulation of the same is also
far from the sight today.
How does one resolve this profound issue at the heart
of physics and astrophysics of black holes? Because a direct
proof that applies under all conditions is so difficult,
our group and several other teams have considered a variety
of specific collapse scenarios, which provide many useful
insights on the final fate of a massive star. In essence,
we have conducted a series of computational experiments
to compile a list of conditions under which gravitational
collapse leads to black hole or a naked singularity.

The basic strategy here is to examine all possible courses
of evolution for a massive star with a given set of physically
reasonable properties and regularity conditions, which
are basic consistency requirements such as having a positive
energy, regularity of the initial data from which the collapse
evolves as the cloud collapses under self-gravity, and such others.
Mathematically, these are the allowed dynamical solutions
to Einstein equations. We supply the initial data for collapse in
terms of the initial density and pressure profiles of matter within
the star, and the velocities of the collapsing concentric shells of
gas. We then compute the causal structure within the collapsing cloud,
which reveals how the light and matter are trapped as gravity grows
stronger as collapse progresses. This decides whether the
final singularity of collapse would be visible or
covered by a horizon.

It should be noted that gravitational collapse studies
have in fact a long history beginning
with the Oppenheimer-Snyder-Datt models that we mentioned
above. In particular,  strict homogeneity is only an idealization
and we must allow for inhomogeneities of density in a matter
cloud. One such model was considered by Seifert and collaborators
\cite{seifert1,seifert2},
wherein neighboring shells of matter intersected to
create momentary singularities that were not covered by horizons.
However, these were not  taken seriously in that, an observer
would not be destroyed and crushed to a zero volume while
passing through the same, which is the true sign for a genuine
singularity. The observer would just sail through undamaged
to another region of spacetime while going through
such a weak singularity.

A realistic star has typically a higher density at its
center, slowly decreasing as one moves outwards. In 1979 Eardley
and Smarr
\cite{Eardley}
performed a numerical simulation of such a model with zero
pressure, and an exact mathematical treatment by
Christodoulou
\cite{Chris}
followed in 1984. The model revealed a naked singularity
at the center where the physical radius of the cloud becomes
zero. However, Newman soon showed
\cite{Newman}
that this singularity is again gravitationally weak. Many
researchers, including Newman and Joshi in 1988 (see e.g. Ref.
\refcite{PSJNew}),
unsuccessfully tried to formulate a rigorous theorem
that naked singularities are always gravitationally weak.
What we realized
then was that we simply did not know enough about
gravitational collapse to formulate a censorship theorem that
holds generally. We had to continue on the longer road of slowly
building up our knowledge by considering case studies
involving realistic collapse models.

Subsequently, researchers found many scenarios of
inhomogeneous pressureless collapse where strong-curvature
naked singularities, the ones that are genuine crushing
singularities, developed from regular initial conditions.
A general treatment for
such a scenario was developed by Joshi and Dwivedi
in 1993 (see Ref.
\refcite{DJ}).
In particular, it became clear that while the homogeneous
pressureless collapse considered by Oppenheimer and Snyder
produced a black hole, a more realistic density profile
with density higher at the center and decreasing as one
moves away can give rise to a naked singularity,
which is an intriguing situation indeed.
While these dust collapse models ignored pressure,
the general techniques developed in the above work to
understand the dynamical evolution of collapse did
find applications later when collapse models with
pressure were considered.

\subsection{Collapse formalism}\label{formalism}
As we know, when nuclear processes end at the core of the star,
the transition from the equilibrium phase to the collapsing phase
happens very rapidly. Therefore, in analytical models describing collapse
it is reasonable to assume that collapse starts from equilibrium with
only gravity acting on the particles of the star and without considering
effects due to other forces.

In this section we summarize and review the key mathematical features
of gravitational collapse in spherical symmetry and review
some of the basic assumptions and equations.
The reader who is already familiar with the formalism or not
interested in the technical details may skip this section.

The most general metric describing a spherically
symmetric collapsing cloud in the comoving coordinates $r$ and $t$
depends upon three functions $\nu(r,t)$, $\psi(r,t)$ and $R(r,t)$,
and is written as,
\begin{equation}\label{metric}
ds^2=-e^{2\nu(t, r)}dt^2+e^{2\psi(t, r)}dr^2+R(t, r)^2d\Omega^2 \; .
\end{equation}
The energy-momentum tensor for a fully general anisotropic
fluid is then written as,
\begin{equation}
T_t^t=-\rho; \; T_r^r=p_r; \; T_\theta^\theta=T_\phi^\phi=p_\theta \; ,
\end{equation}
where $\rho$ is the energy density and $p_r$ and $p_\theta$
are the radial and tangential pressures. We note that the
widely studied dust collapse is obtained for $p_r=p_\theta=0$,
while a perfect fluid collapse
is given by $p_r=p_\theta$. The metric functions
$\nu$, $\psi$ and $R$ are related to the energy-momentum tensor
via the Einstein equations that can be given in the form:
\begin{eqnarray}\label{p}
p_r&=&-\frac{\dot{F}}{R^2\dot{R}} \; ,\\ \label{rho}
\rho&=&\frac{F'}{R^2R'} \; ,\\ \label{nu}
\nu'&=&2\frac{p_\theta-p_r}{\rho+p_r}\frac{R'}{R}-\frac{p_r'}
{\rho+p_r} \; ,\\ \label{G}
2\dot{R}'&=&R'\frac{\dot{G}}{G}+\dot{R}\frac{H'}{H} \; ,\\ \label{F}
F&=&R(1-G+H) \; ,
\end{eqnarray}
where we have defined the functions $H$ and $G$ as,
\begin{equation} \label{HG}
H =e^{-2\nu(r, t)}\dot{R}^2 , \; G=e^{-2\psi(r, t)}R'^2,
\end{equation}
and where $F$ is the Misner-Sharp mass of the system
which
is related to the total amount of matter
enclosed in the comoving
shell labeled by $r$ at the time $t$.

Gravitational collapse is obtained by requiring that
$\dot{R}<0$ and the central shell-focusing singularity is
reached when $R=0$.
At the singularity the energy density and spacetime curvatures blow up.
Divergence of $\rho$ is obtained also whenever $R'=0$, thus
indicating the presence of a `shell-crossing' singularity. Such
singularities are generally found to be gravitationally weak
and do not correspond to a required divergence of the
curvature scalars. Therefore this indicates a breakdown of
the coordinate system being used, rather than a true
and genuine physical singularity.
\cite{cross}
Since we are interested only in the occurrence of
strong curvature shell-focusing singularities,
in the following discussion we will always assume
that $R'> 0$ throughout the collapse.

The system presents an additional degree of freedom
due to scale invariance.
We can then choose the initial time in such a way so
that $R(r, t_i)=r$, and we introduce the scaling function
$v(r,t)$ defined by,
\begin{equation}
    R=rv \;,
\end{equation}
with $v(r,t_i)=1$ (so that the collapse will be described
by $\dot{v}<0$ and the singularity will be obtained at
the value $v=0$).
This is a better definition for the singularity since at
$v\neq 0$ the energy density does not diverge anywhere on the
spacelike surfaces, including at the center $r=0$.
In this manner, for a regular mass function $F$, the
divergence of $\rho$ is reached only at the singularity.
We note that the function $v(t,r)$ is monotonically decreasing
in $t$ close to the formation of the singularity and therefore
can be used as a `time' coordinate replacing $t$ itself.
We can then perform a change of coordinates from $(r,t)$
to $(r,v)$ and thus we have $t=t(r,v)$. In this case the
derivative of $v$
with respect to $r$ in the $(r,t)$ coordinates shall be considered
as a function of the new coordinates, $v'=w(r,v)$.
Then the Misner-Sharp mass can be taken to be a function of the
comoving radius $r$ and the comoving time $t$, expressed via the
`temporal' label $v$ as
\begin{equation}
    F=F(r,v(r,t)) \; .
\end{equation}

To solve the system of Einstein equations we then proceed as follows:
\begin{itemize}

  \item[-] If $p_r$ is related to $\rho$ via an equation of
state, the integration of the same, once we substitute equations
\eqref{p} and \eqref{rho}, will give $F$ as a function of $R$
and its derivatives.  If no equation of state is provided
then $F$ is a fully free function.

  \item[-] $\rho$ and $p_r$ are given from equations
\eqref{p} and \eqref{rho} as functions of $F$ and $R$,
substituting $F$ leaves $\rho$ and $p_r$ as functions
of $R$ and its derivatives.

  \item[-] Integration of the equation \eqref{nu} gives
$\nu$ as a function of $p_\theta$ and $R$ and its derivatives.
If no relation is given for $p_\theta$ (as in the perfect fluid case)
then we can take $p_\theta$ as a free function of $r$ and $v$.

  \item[-] Integration of equation \eqref{G} will generally
give $G(r,v)=b(r)e^{2A(r,v)}$  with $A(r,v)$ given by
$A_{,v}=\nu'\frac{r}{R'}$ and $b(r)$ being a free function coming from
the integration which describes the velocity profile of the particles.

  \item[-] Finally, from equation \eqref{F} we are left with
one differential equation in $R$ and its derivatives that acts as
an equation of motion for the system. Once this is solved the whole
system of Einstein equations is solved.
\end{itemize}

The equation of motion is given by
\begin{equation}\label{Rdot}
    \dot{R}^2=e^{2\nu}\left(\frac{F}{R}+G-1\right) \; ,
\end{equation}
or in terms of the scaling factor $v$ as,
\begin{equation}\label{vdot}
    \dot{v}=-e^\nu\sqrt{\frac{F}{r^3v}+\frac{be^{2A}-1}{r^2}} \; ,
\end{equation}
where the minus sign has been taken since we are dealing with collapse.
In order to have a solution the following `reality condition' must be satisfied:
\begin{equation}
    \left(\frac{F}{r^3v}+\frac{be^{2A}-1}{r^2}\right)>0.
\end{equation}

As is well-known, under completely general conditions
it is not possible to fully solve the system of Einstein
equations globally, the main reason being that these are
a complicated set of non-linear partial differential equations.
Nevertheless, the relevant information about the final fate
of collapse in the general form of matter fields and curvature
invariants can be extracted by an analysis of the behaviour
of the solutions near the singularity and near the
center of the cloud. As we point out below, this can be done
by integrating the Einstein equations by one order, and
by reducing them to a first order system so as to obtain
a time evolution equation for the collapsing cloud
which decides the nature of the final singularity curve
in the case of a continual collapse. This will help us
decide the local visibility or otherwise
of the central singularity.

\subsubsection{Regularity and energy conditions}\label{conditions}
Einstein equations by themselves provide only the relations between
geometry and matter distribution within the collapsing matter cloud,
actually without giving
any statement about what kind of matter we are dealing with,
that is responsible for the given spacetime geometry.
>From a physical perspective, not every type of matter distribution
can be allowed, and some restrictions on possible matter models
must be made based on physical reasonableness.
These restrictions typically come in the form of energy conditions
ensuring the positivity of mass-energy density and the sum of pressure
and energy density. For example the weak energy conditions can be written as:
\begin{equation}
  \rho \geq 0, \;
  \rho+p_r \geq 0, \;
  \rho+p_\theta \geq 0.
\end{equation}
These typically give some constraints on the behaviour of the mass
function $F$. For example, it is easy to see that from the first one,
if we impose the condition of no `shell-crossing' singularities,
it follows that $F'>0$. Also some conditions arise on the allowed
pressures, though positive pressures are obviously always allowed,
imposing the dominant energy condition would further require
that pressures cannot exceed the energy density. At the same time
some negative pressures can be allowed as well.

Further to these, certain regularity conditions must be fulfilled
in order for the matter fields to be well-behaved at
the center of the cloud at the initial epoch from which
collapse evolves.
These are the finiteness of the energy density at
all times anteceding the singularity and regularity of
the Misner-Sharp mass in $r=0$, and they imply that we
must have generically,
\begin{equation}\label{mass1}
    F(r,t)=r^3M(r,v(r,t)) \; ,
\end{equation}
where $M$ is a regular function going to a finite
value $M_0$ at $r=0$.
Also, requiring that the energy density has
no cusps at the center is reflected in the condition,
\begin{equation}
    M'(0,v)=0 \; .
\end{equation}

In Ref.
\refcite{ndim1}
it was shown that vanishing of the pressure gradients near
the center of the cloud implies that the radial and tangential stresses
must assume the same value in the limit of approach to the center.
This requirement comes from the fact that the metric functions
should be at least $\mathcal{C}^2$ at the center of the cloud and
is a consequence of the fact that the Einstein equation \eqref{nu}
for a general fluid contains a term in $p_r-p_\theta$ that
therefore must vanish at $r=0$ thus indicating that in the final
stages of collapse of the core of a star the cloud behaves like a
perfect fluid in proximity of $r=0$.
Further, since the gradient of the pressures must
vanish at $r=0$, we see that $p'\simeq r$ near $r=0$,
which for the metric function $\nu$ implies that
near the center,
\begin{equation}
\nu(r,t)=r^2g(r,v(r,t))+\bar{g}(t) \; ,
\end{equation}
where the function $\bar{g}(t)$ can be absorbed in
a redefinition of the time coordinate $t$. From the above,
via the definition for $A$ we can write $A_{,v}=\frac{2g+rg'}{R'}r^2$.

Regularity at the center implies some requirements for the
velocity profile $b(r)$ as well. Then $b$ can be
written as,
\begin{equation}
    b(r)=1+r^2b_0(r) \; ,
\end{equation}
near $r=0$. We can now interpret the free function
$b(r)$ in relation with the known Lemaitre-Tolman-Bondi (LTB)
dust models.
\cite{LTB1,LTB2,LTB3}
In fact, the cases with $b_0$ constant
are equivalent to the bound ($b_0<0$), unbound ($b_0>0$)
and marginally bound ($b_0=0$) LTB collapse models,
depending on the sign of the quantity $b_0$.

\subsubsection{Initial data and matching}
Specifying the initial conditions from which collapse evolves
consists in prescribing the values of the three metric functions
and of the density and pressure profiles as functions of $r$ on the
initial time surface given by $t=t_i$. Once this is done and the
eventual free functions are provided, the dynamical evolution of
the collapsing cloud is entirely determined.
Specifying initial conditions reduces to defining the following functions:
\begin{eqnarray}\nonumber
    \rho(r, t_i)&=&\rho_i(r), \; p_r(r, t_i)=p_{r_i}(r),  \;
p_\theta(r, t_i)=p_{\theta_i}(r), \\ \nonumber
    R(r, t_i)&=&R_i(r), \; \nu(r, v(r, t_i))=\nu_i(r), \; \psi(r, t_i)=\psi_i(r).
\end{eqnarray}
Since the initial data must obey Einstein equations it
follows that not all of the initial value functions can be
chosen arbitrarily.

As said before the choice of the scale function $v$ is such that $R_i=r$.
Then, for example in the case of perfect fluids,
from equations \eqref{p}, \eqref{rho} and \eqref{mass1}, writing
$M_i(r)=M(r, v(r, t_i))=M(r, 1)$
we get
\begin{equation}
    \rho_i=3M_i(r)+rM_i'(r) \; , \; p_i=-(M_{,v})_i \; .
\end{equation}
>From equation \eqref{nu} we can write,
\begin{equation}
    \nu_i(r)=r^2g(r, v(r, t_i))=r^2g_i(r) \; ,
\end{equation}
which in turn can be related to the function $A$ at the
initial time,
\begin{equation}\label{A}
    A_{,v}(r, v(r, t_i))=(A_{,v})_i=2g_ir^2+g_i'r^3 \; .
\end{equation}
Finally, the initial condition for $\psi$ can be related
to the initial value of the function $A(r,t)$ from equation
\eqref{G} and equation \eqref{A},
\begin{equation}
    A(r, v(r, t_i))=A_i(r)=-\psi_i-\frac{1}{2}\ln b(r) \; .
\end{equation}

Further to this we must require that the initial configuration
is not trapped, since we intend to study the formation of
trapped surfaces during collapse.
We therefore must require
\begin{equation}
    \frac{F_i(r)}{R_i}=r^2M_i(r)< 1 \; ,
\end{equation}
from which we see how the choice of the initial matter
configuration $M_i$ is related to the initial boundary of
the collapsing cloud. Some restrictions on the choices of the
radial boundary must be made in order not to have trapped
surfaces at the initial time.

Finally we mention how the spacetime describing the
collapsing cloud can be matched to a known exterior metric.
In the case of dust (or in the case of collapse with only
tangential pressures) the Misner-Sharp mass is conserved during
collapse and therefore it is possible to have a matching with
a Schwarzschild exterior. On the other hand if we consider
an anisotropic fluid model or a perfect fluid model (thus allowing
for the presence of radial pressures) the Misner-Sharp mass is in
general not conserved during collapse.
In this case matching with an exterior spherically symmetric
solution leads to consider the generalized Vaidya spacetime, where
the mass that is not conserved during collapse is described
as outgoing radiation in the exterior spacetime.
It can be proven that matching to a generalized Vaidya
exterior is always possible when the collapsing cloud is taken
to have a compact support within the boundary taken at $r=r_b$,
and the pressure of the matter is assumed to vanish at the
boundary.
\cite{matching1,matching2,matching3}

\subsubsection{Collapse final states }\label{collapse}
In order to study the final outcome of collapse we turn
our attention to the equation of motion \eqref{vdot}
which can be inverted to give the function $t(r,v)$
that represents the time at which the comoving shell
labeled $r$ reaches the event $v$,
\begin{equation}\label{t}
    t(r,v)=t_i+\int_v^1\frac{e^{-\nu}}{\sqrt{\frac{M}{v}
+\frac{be^{2A}-1}{r^2}}}dv \; .
\end{equation}
Then the time at which the shell labeled by $r$
becomes singular, called the singularity curve, can be written as
\begin{equation}\label{ts}
    t_s(r)=t(r,0)=t_i+\int_0^1\frac{e^{-\nu}}{\sqrt{\frac{M}{v}
+\frac{be^{2A}-1}{r^2}}}dv \; .
\end{equation}
Regularity ensures that, in general, $t(r,v)$
is at least $\mathcal{C}^2$ near the singularity and
therefore can be expanded as,
\begin{equation}\label{expand-t}
    t(r,v)=t(0,v)+\chi_1(v)r+\chi_2(v)r^2+o(r^3) \; .
\end{equation}
The coefficients $\chi_1=\frac{dt}{dr}\rvert_{r=0}$ and
$\chi_2=\frac{1}{2}\frac{d^2t}{dr^2}\rvert_{r=0}$ are determined by the
expansions of the functions $M$, $A$ and $b$ close to the center and it
is their values that determine the visibility of the singularity.
The singularity curve in fact becomes
\begin{equation}\label{expand-ts}
    t_s(r)=t_0+r\chi_1(0)+r^2\chi_2(0)+o(r^3) \; ,
\end{equation}
where $t_0=t(0,0)$ is the time at which the
central shell becomes singular.

For physical purposes, or to study a special case,
we can assume that all the involved quantities
can be expanded in a power series in a close neighborhood of $r=0$
and that energy density and pressures present only quadratic terms in
the expansions. In this case we will obtain $\chi_1=0$ and the final
fate of collapse will be decided by the term $\chi_2$.
Of course, the situations where discontinuities are allowed to
be present can be analyzed also with some small technical
modifications to the above formalism.

\subsubsection{Trapped surfaces and outgoing null geodesics}\label{horizon}
The crucial elements that are necessary in order to
understand if within a certain collapse model there are future
directed null geodesics that are
emanating from the singularity and reaching external observers
in the universe are two: the apparent horizon, which marks the
limit of trapped surfaces, and the
null geodesic curves. If the outgoing light rays coming from
the singularity do not meet
the apparent horizon in their journey within the cloud,
then they can reach the boundary of the cloud and propagate
freely in the outside universe, thus reaching any future observer.

This means that the boundary of the cloud also plays
an important role towards
the local or global visibility of the singularity. Nevertheless
within analytical models
that, for the time being, do not necessarily rely on
observational data, it is
always possible to choose the boundary suitably in order
for the singularity to be globally visible.
The apparent horizon is given by the condition
\begin{equation}\label{ah}
    1-\frac{F}{R}=0 \; ,
\end{equation}
which describes a curve $v_{ah}(r)$ given implicitly by
\begin{equation}
    r^2M(r, v_{ah})-v_{ah}=0 \; .
\end{equation}
Inversely, the apparent horizon curve can be
expressed as the curve $t_{ah}(r)$ which gives the time
at which the shell labeled by $r$ becomes trapped.

>From the above equation it is clear that in general the
apparent horizon curve can have any behaviour, that is, it
could be increasing or decreasing. In order
for the singularity to be visible it is clearly necessary
that the apparent horizon does not form before the singularity.
We shall consider here the most common situation where
the time at which the central shell becomes singular coincides
with the time at which the same becomes trapped. In this case
if the apparent horizon curve is constant or decreasing there
is no possibility for the central singularity to be visible.
On the other hand an increasing apparent horizon curve can allow
for future directed null geodesics to escape the singularity.
All shells with $r>0$,
when they become singular,
are instead necessarily trapped since in
this case $t_s(r)>t_{ah}(r)$.

In order to understand what are the features
that determine the visibility of the
singularity to external observers we can evaluate
the time curve of the apparent horizon in these models as
\begin{equation}\label{t-ah}
    t_{ah}(r)=t_s(r)-\int_0^{v_{ah}(r)}\frac{e^{-\nu}}
{\sqrt{\frac{M}{v}+\frac{be^{2A}-1}{r^2}}}dv \; ,
\end{equation}
where $t_s(r)$ is the singularity time curve, whose
initial point is $t_0=t_s(0)$.
Close to the center the apparent horizon curve behaves
like the singularity curve and therefore a necessary condition
for visibility of the central singularity is that $t_{ah}(r)$
be increasing.
>From the above we see how the presence of pressures affects
the time of the formation of the apparent horizon.
In fact, it is easy to see that all the configurations for which
$\chi_1>0$ (or $\chi_2>0$ in case that $\chi_1=0$) will cause the
trapped surfaces to form at a later stage than the singularity,
thus leaving the door open for the null geodesics to escape
the central singularity.

At this point we turn our attention to radial outgoing null geodesics.
>From the metric \eqref{metric} we can write
\begin{equation}
    \frac{dt_{ng}}{dr}=\pm e^{\psi-\nu} \; ,
\end{equation}
from which we can get the general expression for the null geodesic curve
$t_{ng}(r)$. If we impose the initial condition $t_{ng}(0)=t_0$, that means
that the light ray is originated at the singularity, and the condition
for the geodesic to be future directed (which determines the sign) we can,
in principle, integrate the above equation.
At this point to see if the singularity is visible we just need to check
under which conditions we have $t_{ng}(r)<t_{ah}(r)$ for $r\in(0,r_b)$.

In Ref.
\refcite{ndim1}
it was shown that the condition that $t_{ah}$ is increasing, which is
equivalent to the positivity of the first non-vanishing $\chi$, is
not only necessary but also sufficient for the local visibility of
the central singularity (see also Refs.
\refcite{ndim2}--\refcite{ndim4}).
Furthermore it can be proven that if there
is one outgoing radial null geodesic that escapes the singularity then
there is necessarily a whole family of such geodesics.
\cite{outgoing-geodesics}

On a more general ground, the scenario of collapse of a cloud composed
of a generic fluid offers a wider spectrum of possibilities.
In fact we can see from equation \eqref{ah} that whenever the mass
function $F$ goes to zero as collapse evolves it is possible
to delay the formation of trapped surfaces in such a way that
a portion of the singularity curve $t_s(r)$ becomes timelike.
This leads to the possibility that shells different from the central
one be visible when they become singular.
\cite{timelike}

Nevertheless it must be noted that the pressure of the fluid must
turn negative at some point before the formation of the singularity
in order for the mass function to be entirely radiated away during
collapse. This might seem an unphysical artificial feature but we
note that negative pressures are worth investigating as they could
provide a means to represent classically, within general relativity,
some effects of a quantum theory of gravity that become apparent
close to the formation of the singularity.
\cite{JoshiGoswamiLQG}

\subsubsection{An example: Dust collapse}
If we apply the above formalism to the case of dust collapse
where pressures are set to zero (as for example in the OSD model mentioned
above), the whole structure of Einstein equations simplifies drastically.
We note of course that the density could be inhomogeneous
in space, even when pressures are vanishing.

The spacetime metric in this case is described by
the well-known and widely studied Lemaitre-Tolman-Bondi (LTB)
collapse model.
In this case, $p_r=0$ implies the condition $M=M(r)$ and from
$\nu'=0$ together with the condition $\nu(0)=0$ we get $\nu=0$ identically.
Then we must have $G=b(r)=1+r^2b_0(r)$ and the equation of motion becomes
$\dot{v}=-\sqrt{\frac{M}{v}+b_0}$ which can be explicitly integrated.
The metric in this case takes the form
\begin{equation}
ds^2=-dt^2+\frac{(v+rv')^2}{1+b_0r^2}dr^2+r^2v^2d\Omega^2, \;
\end{equation}
and the singularity and the apparent horizon curves
can be evaluated.
It is easy to see that $t_{ah}(0)=t_s(0)$ and therefore only the central
singularity can eventually be visible.
If we assume that $M(r)$ can be expanded in a series near the
center as $M(r)=M_0+M_1r+M_2r^2+o(r^3)$ (where from regularity we shall
impose $M_1=0$), we can easily evaluate $\chi_1(0)$ and $\chi_2(0)$ which
turn out to be
\begin{eqnarray}
    \chi_1(0)&=&-\frac{1}{2}\int^1_0\frac{M_1+b_{01}v}
{\left(M_0+vb_{00}\right)^{\frac{3}{2}}}\sqrt{v}dv \; , \\
    \chi_2(0)&=&\frac{3}{8}\int^1_0\frac{(M_1+b_{01}v)^2}
{\left(M_0+vb_{00}\right)^{\frac{5}{2}}}\sqrt{v}dv
    -\frac{1}{2}\int^1_0\frac{M_2+b_{02}v}{\left(M_0+vb_{00}\right)^
{\frac{3}{2}}}\sqrt{v}dv \; ,
\end{eqnarray}
where we have expanded $b_0$ in a series as
$b_0=b_{00}+b_{01}r+b_{02}r^2+o(r^3)$.
We then see that it is the
values of the first non-vanishing
terms $M_i$ and $b_{0i}$ (with $i>0$)
which determines the visibility or otherwise of the
central singularity.

Simplifying even further we obtain the homogeneous dust case
that was discussed in section \ref{OSD}. The OSD model is achieved
when $M=M_0$ and $b_0=k$,
then we have $F'=3M_0r^2$, $R'=v(t)$, and the energy density
is homogeneous throughout the collapse and is given by
$\rho=\rho (t)= {3M_0}/{v^3}$.
The spacetime geometry then becomes the Oppenheimer-Snyder-Datt
metric, which is given by,
\begin{equation}
ds^2=-dt^2+\frac{v^2}{1+kr^2}dr^2+r^2v^2d\Omega^2 \; .
\end{equation}
In this case  all the quantities $\chi_i$ identically vanish
thus showing that the singularity curve reduces
to that of a simultaneous singularity, namely $t_s(r)=t_0$.

In the OSD homogeneous collapse case, the trapped surfaces
and the apparent horizon develop much earlier than the formation
of the singularity, thus creating a black hole in the spacetime
as the collapse final state.
But when some inhomogeneities are allowed in the initial
density profile, such as a higher density at the center of the
star, the trapped surface formation is
delayed in a natural manner, thus leaving the final singularity
of collapse to be visible to faraway observers in the universe
(see e.g. Ref.
\refcite{JDM}).

\subsection{Collapse with non-zero pressure}
The next step, in order to study more realistic collapse
scenarios, is that of introducing pressures in the model.
Gravitational collapse with non-zero pressures included
has been discussed and investigated in past years in detail by
many researchers (see for example Ref.
\refcite{JoshiDwivedi99} for a study of the
initial data leading to different outcomes, Ref.
\refcite{JhinganMagli} for a discussion of the Einstein cluster model, Ref.
\refcite{Goncalves} for collapse with only radial pressures, or Refs.
\refcite{Harada-pf,GGMP} for perfect fluid models).
Given the difficulties arising from Einstein equations the
full integration of the system is not possible even in the
simplest cases when pressures are included.
Therefore the general line of inquiry follows a path
which can be broadly outlined as follows: Firstly, the general
structure for Einstein equations to study the spherical collapse
is considered. We describe how the equations can be integrated
up to first order, thus obtaining the equation of motion for
the system. The regularity conditions and energy conditions
that give physically reasonable models are considered and imposed
on the models. The final stages of collapse are then discussed,
evaluating key elements that determine when the outcome will
be a black hole or a naked singularity.

It is seen that there is a specific function which is
related to the tangent of outgoing geodesics at the singularity,
whose sign solely determines the time of formation of trapped
surfaces in relation with the time of formation of the singularity.
We also analyze the occurrence of trapped surfaces during
collapse and the possibility that radial null geodesics do
escape, thus making it visible. We can show how both features
are related to the sign of the above mentioned function,
thus obtaining a necessary and sufficient condition for
the visibility of the singularity.

Collapse models with non-zero pressures were discussed in
detail by various researchers (see Refs.
\refcite{Magli1a}--\refcite{Magli2} for the tangential pressure case, or Refs.
\refcite{Nakao1,Nakao2}).
Considering a form of pressure generated by
rotation of particles within the collapsing cloud, they showed how
the naked singularities develop as collapse end states. Several
models with a realistic equation of state, which specifies how
the density and pressure within the cloud are related, were
also investigated, including a model by Ori and Piran
and by our group (see Refs.
\refcite{Ori1,Ori2}).

Today, we have a general formalism to treat spherical collapse
from initial data. General matter fields can include realistic
features such as pressure and inhomogeneities in density distribution,
and reasonable equations of state of matter are incorporated.
Physicists are even
beginning to consider situations where matter takes on some
other form, such as a fundamental quantum field, or is converted
to radiation in a sudden phase transition in the very late stages
of collapse. What these works show in a generic manner is that
the collapse with non-zero pressures can lead to either the
formation of a black hole or a naked singularity as the final state.
The outcome is decided
by the initial data and the dynamical evolutions as allowed by
the Einstein equations. It turns out that the collapse ends in a
naked singularity in a wide variety of situations.

These conclusions and models are by now fairly widely accepted,
and it is now clear that physically reasonable gravitational
collapse can create naked singularity. However, in its original
statement, Penrose included a condition, namely that `generic'
gravitational collapse would always produce black holes.  Are
the above models generic? In fact, they are in the sense of the
initial data for matter being fully generic as required.
Though the key
difficulty for such a requirement, as such and in general, is that
there is no mathematically well formulated definition of genericity
available in general relativity and gravitation theory, which
one could meaningfully use to answer this question completely
one way or the other.  Stability and genericity are extremely
difficult concepts to formulate precisely in general relativity,
within the framework of a general spacetime with a Lorentzian
metric with an indefinite signature, with no definite and clear
criteria available today. In fact, there are formidable
mathematical difficulties in achieving the same as we shall
discuss later. Therefore, the only way to proceed is by asking
further questions, such as is there any perturbation of spherical
symmetry that would remove these naked singularities, or would
naked singularities form in non-spherical collapse, and so on.

\subsection{Are naked singularities stable and generic?}\label{generic}
Firstly, as we have noted the concepts such as
genericity and stability are at present not well-defined
in general relativity, as they are in the Newtonian gravity.
The major difficulty towards obtaining a unique and well posed
definition comes from the non-uniqueness
of topology, or the concept of `nearness' itself in a given
spacetime manifold.
\cite{HE}

For example, in order to quantify when two spacetime
geometries are nearby, we can define a certain topology on
a space of all spacetime metrics by requiring that the values
of the metric components are `nearby'. On the other hand,
we can also additionally
require that the first $n$ derivatives of the metric
functions are also nearby.
The problem lies in the fact that the resulting topologies will
be different. It is easy to see how this fact is connected
to the basic problem of arriving at a well-formulated definition
of the cosmic censorship itself.
\cite{JoshiSaraykar}

>From this perspective, certain studies have been developed on
the concept of critical collapse (see e.g. Ref.
\refcite{Grund} and references therein)
and certain spacetime geometries with naked
singularities have been studied in the past with the aim to better
understand their genericity and stability. For example in Ref.
\refcite{Christodoulou}
it was shown that if we consider self-similar massless scalar field
collapse (which is a somewhat special model) the initial data leading
to naked singularity has a positive codimension, in a certain space of
initial data, and this led to the conclusion that the occurrence of naked
singularity is not generic in that case. On the other hand in Refs.
\refcite{Sar-Pramana,SG}
it was shown that the naked singularity occurrence is generic for certain
types of matter fields such as inhomogeneous dust and some others.
Therefore, the issue of genericity and stability of
naked singularities in collapse remains wide open for spherically
symmetric models (and even more so if we wish to consider departure from
spherical symmetry) with different forms of matter field (see Ref.
\refcite{Duffy}
for stability of the Cauchy horizon or Ref.
\refcite{Dafermos}
for stability under linear perturbations).

To begin with, one needs to adopt some definition for what
is meant when we talk about stability and genericity of some outcome
of gravitational collapse.
To this purpose the definitions of stability and genericity will
always need to be referred to the specific space of initial data
set that is being considered.
In fact what is stable or generic in a certain space of
initial data (as for example
that of inhomogeneous dust) need not be stable in another, bigger space
(as for example that of perfect fluid collapse).
We say that a certain outcome of collapse in terms of either
black hole or naked singularity from a certain set of
initial data is stable if there exist a whole neighborhood of that initial
data set which leads to the same outcome. The neighborhood,
as said, being defined within the specific space of initial data
that we are considering.

Furthermore, we will say that a certain outcome of collapse
is generic, within a specific space of initial data, if the measure
of the subset of initial data leading to that outcome in the
original initial data space is non-zero.

This definition, though being related to the usual definition of
genericity for dynamical systems, does not coincide with it in the
sense that typically an outcome can be said to be generic in the
sense of dynamical systems if the set of initial data leading
to that outcome is open and dense in the set of all initial data.
With this definition we could prove that both black holes and naked
singularities are `non-generic' already in the space of initial data
for Lemaitre-Tolman-Bondi collapse.

Therefore, with the definitions referred to as above,
we investigated the
initial data sets leading to black holes and naked singularities
in the space of all initial data sets
for different type I matter models (including dust and perfect fluid).
We showed that the initial data set leading the collapse to a naked singularity,
just like the initial data set leading to a black hole,
forms an open subset of a suitable function space comprising of the
initial data, with respect to an appropriate norm which makes the
function space an infinite dimensional Banach space. Furthermore
considering the possible measure theoretic aspects of this open set
it can be shown that a suitable well-defined measure of this set
can be given and it must be strictly positive.
\cite{Sar-Pramana}

\subsubsection{The genericity and stability of collapse outcomes}
We investigate here the genericity and stability aspects
for naked singularities and black holes that arise as the final
states for a complete gravitational collapse of a spherical massive
matter cloud. The form of the matter considered is a general
{\it Type I} matter field, which includes most of the physically
reasonable matter fields such as dust, perfect fluids,
massless scalar fields and
such other physically interesting forms of matter widely used in
gravitation theory
(for a complete definition of different matter fields,
see for example,
\refcite{HE}, section 4.3).
Our aim is to examine and characterize
the above two possible outcomes in terms of stability of the
initial data and genericity. We show that both black holes and
naked singularities are generic outcomes of a complete collapse,
once genericity is defined in a suitable sense in an appropriate
space of functions.



While general relativity may predict the
existence of both black holes and naked singularities
as collapse outcomes, an important question
that needs to be answered
is, what would a realistic continual gravitational collapse
of a massive star in nature would end up with.
Thus the key issue under an active debate now
is the following:
Even if naked singularities did develop as collapse
end states, would they be generic or stable in some
suitably well-defined sense, as permitted by the gravitation
theory? The point here is, if naked singularity formation
in collapse was necessarily `non-generic' in some
appropriately well-defined sense,
then for all practical purposes, a realistic physical
collapse in nature might always end up in a black hole,
whenever a massive star ended its life.

In fact, such a genericity requirement has been
always discussed and desired for any possible mathematical
formulation of CCC right from its inception. However,
the main difficulty here has again been that, there
is no well-defined or precise notion of genericity
available in gravitation theory and the general theory
of relativity as we discussed above. Again, it is
only
various gravitational collapse studies that can provide
us with more insight into this genericity aspect.

\begin{figure}
\centerline{\includegraphics[width=9cm]{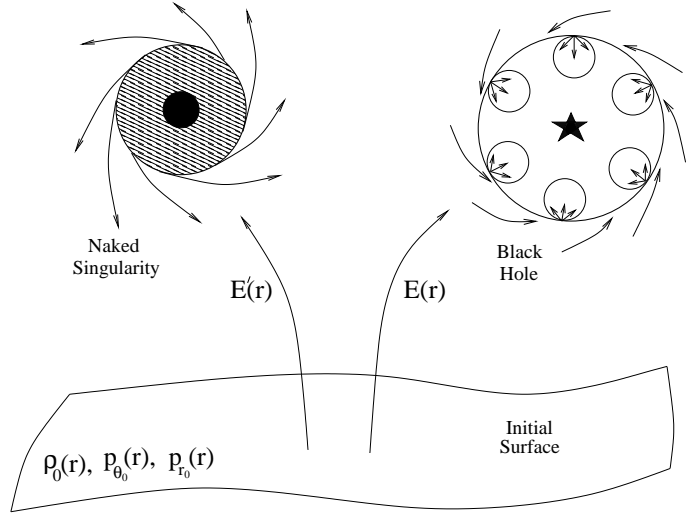}}
\caption{Evolution of spherical collapse for a
general matter field with inhomogeneities and
non-zero pressures included where $\rho_0$, $p_{\theta_0}$ and $p_{r_0}$
specify the initial data set and $E(r)$, $E'(r)$ are possible
velocity profiles. \label{f:three}}
\end{figure}

A result that is relevant here is the following
\cite{JoshiDwivedi99}:
For a spherical gravitational collapse of a
rather general (type I) matter field, satisfying the
energy and regularity conditions, given any regular
density and pressure profiles at the initial epoch,
there always exist classes of velocity profiles for the
collapsing shells and dynamical evolutions as determined
by the Einstein equations, that, depending on the
choice made, take the collapse to either a black hole
or naked singularity final state
(see e.g. Fig. \ref{f:three} for a schematic illustration of
such a scenario).

\begin{figure}
\centerline{\includegraphics[width=9cm]{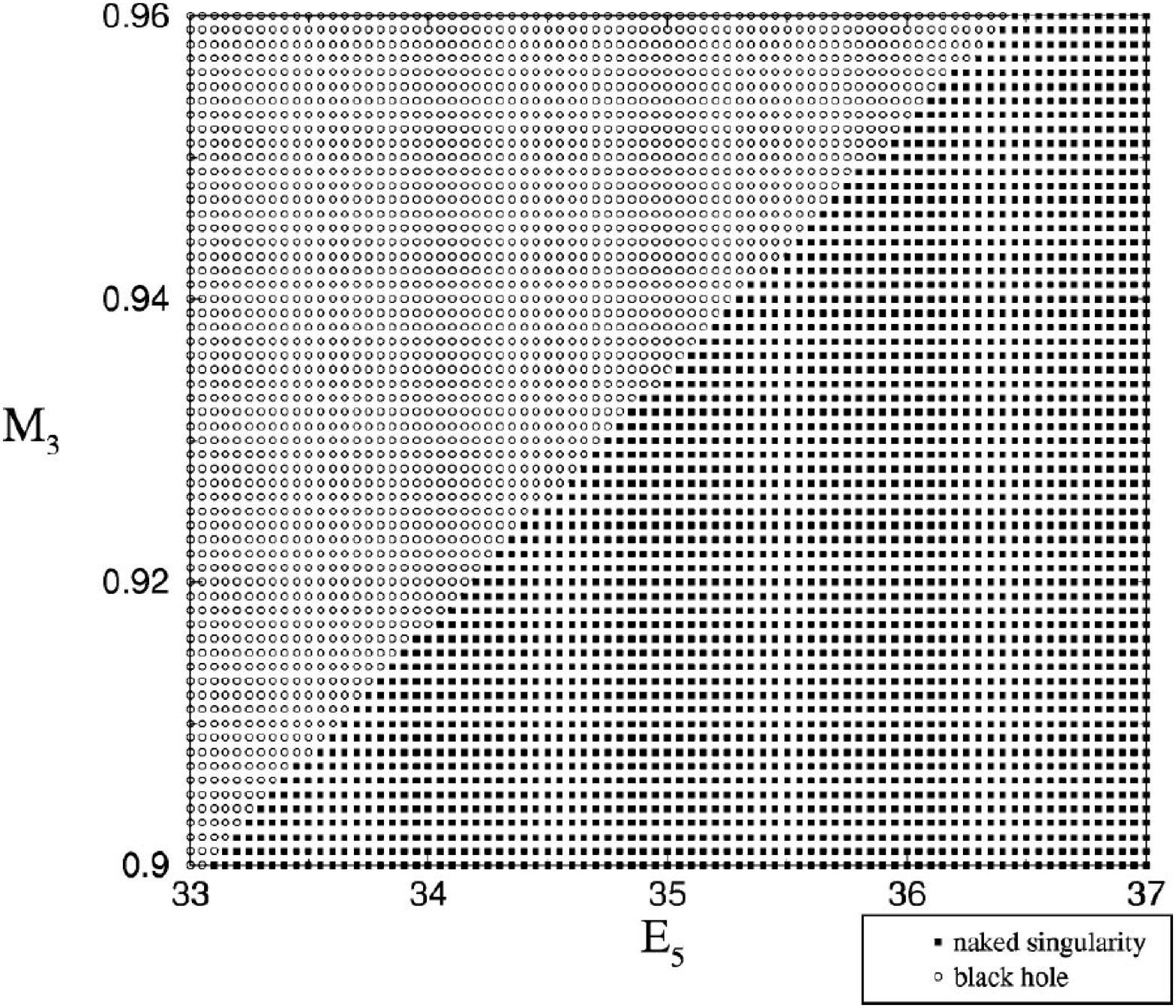}}
\caption{Collapse final states for inhomogeneous
dust in terms of the initial mass and velocity profiles
for the collapsing shells,
where the mass profile is given by $M=\Sigma_{i=3}^{\infty}M_ir^i$
and the velocity profile is given by
$E=-E_2r^2-\Sigma_{j=5}^{\infty}E_jr^j$ with $E_2$ kept fixed.
}
\label{mena}
\end{figure}

Such a distribution of final states of collapse
in terms of the black holes and naked singularities
can be seen much more transparently when we consider
a general inhomogeneous dust collapse,
for example, as discussed in Ref.
\refcite{Tavakol}
(see Fig. \ref{mena}).

What determines fully the final fate of collapse
here are the initial density and velocity profiles
for the collapsing shells of matter.
One can see here clearly how the different choices
of these profiles for the collapsing
cloud distinguish and determine between the two
final states of collapse, and how each of the
black hole and naked singularity states
appear to be `generic' in terms of their being
distributed in the space of final states.
Typically, the result we have here is, given any
regular initial density profile for the collapsing dust
cloud, there are velocity profiles that take the collapse
to a black hole final state, and there are other
velocity profiles that take it to naked singularity
final state. In other words, the overall available
velocity profiles are divided into two distinct
classes, namely the ones which take the given density
profile into black holes, and the other ones that take
the collapse evolution to a naked singularity
final state. The same holds conversely also, namely
if we choose a specific velocity profile, then the
overall density profile space is divided into two
segments, one taking the collapse to black hole final
states and the other taking it to naked singularity
final states. The clarity of results here give us
much understanding on the final fate of a collapsing
matter cloud.

Typically, all stars have a higher density at
the center, which slowly decreases as one moves away.
So it is very useful to incorporate inhomogeneity into
dynamical collapse considerations. However, much more
interesting is the collapse with non-zero pressures
which are very important physical forces within a
collapsing star. We briefly consider below a typical
scenario of collapse with a non-zero pressure component,
and for further details we refer to Ref.
\refcite{JoshiMalafarina11}.

For a possible insight into genericity of naked
singularity formation in collapse
it is interesting to study the final outcomes
of collapse once some kinds of pressures are introduced
in the collapsing cloud.
To this aim, we investigated the effect of introducing small pressure
perturbations, in the form of tangential stresses or perfect fluids,
in the collapse dynamics of the classic
Oppenheimer-Snyder-Datt (OSD) gravitational collapse, which is
an idealized model assuming zero pressure, and which
terminates in a black hole final fate as discussed above.
Thus we study the stability of the OSD black hole under
introduction of small stresses or pressure perturbations.

It is seen explicitly that there exist
classes of stress perturbations such that the
introduction of a smallest tangential pressure within
the collapsing OSD cloud changes the endstate of collapse
to formation of a naked singularity, rather than a black hole.
What follows is that small stress perturbations
within the collapsing cloud change the final fate of
collapse from being a black hole to a naked singularity.
The same model can also be viewed as a first order
perturbation of the known spacetime metric describing the dust cloud.
Thus we can understand here
the role played by pressures in a well-known
gravitational collapse scenario. A specific and physically
reasonable but generic enough class of perturbations is
considered so as to provide a good insight into the
genericity of naked singularity formation in collapse
when the OSD collapse model is perturbed by introduction
of small pressures, either the tangential ones or more
realistic
perfect fluids.
>From this analysis we gain some useful and important
clarification into the structure
of the censorship principle which as yet
remains to be properly understood.

\subsubsection{Instability of the OSD black hole}
With the aim of understanding the stability properties of
known dust collapse models leading to the formation of a black
hole, we studied the effects of introducing small
pressure perturbations in an otherwise pressure-free gravitational
collapse scenario which was to terminate in a black hole final state.

We followed the collapse formalism in spherical symmetry developed
in section \ref{formalism}. The analysis of
pressure perturbations in known collapse models, inhomogeneous
but otherwise pressure-free, shows how collapse final states
in terms of black hole or naked singularity are affected and altered.
This allows us to examine how stable these outcomes are and to understand
within the initial data space the set of conditions
that lead the collapse to a naked singularity or to a black hole in
order to evaluate how abundant are these final states.

Of course many examples of spacetimes with naked singularity can be
found, but their relevance in models describing physically viable
scenarios is still a matter of much debate, as it has been the
case since the first formulation of the CCC.
The reason for this being that
many of these collapse scenarios are, however, restricted by some
simplifying assumptions such as the absence of pressures, as is the
case for dust models. It is well-known that the pressures
cannot be neglected in
realistic models describing stars in equilibrium. It seems natural
therefore that if one wishes to study analytically what happens
during the last stages of the life of a massive star, when its
core collapses under its own gravity thus forming a compact object
as a remnant, pressures must be taken into account.

Therefore, given the fact that naked singularities do arise
as collapse endstates under
physically viable conditions, this further analysis helps us understand
in a clear manner whether these examples are a mere curiosity with no impact
on the general structure of collapse of massive stars or whether they might
be regarded seriously towards possible observations.

Hence, further to early works that showed the occurrence
of naked singularities in dust collapse, much effort has been
devoted to understanding the role played by pressures.
\cite{press}

The presence of pressures is a crucial element towards the description
of realistic sources as we know that stars and compact objects are
generally sustained by matter with strong stresses (either isotropic
or anisotropic). At first it was believed that the naked singularity
scenario could be removed by the introduction of pressures, thus implying
that more realistic matter models would lead only to the formation of a
black hole. We now know that this is not the case. The final outcome
of collapse with pressure is entirely decided by its initial
configuration and allowed dynamical evolutions and it
can be either a black hole or a naked singularity. Furthermore it
is now clear that within each model (be it dust, tangential pressure
or others) the data set leading to naked singularities is not a
subset of `zero measure' of the set of all possible initial data.
Despite all this work we can still say that much more is to be
understood about the role that general pressures play during the final
stages of collapse. Perfect fluid collapse has been studied mostly
under some simplifying assumptions and restrictions in order to
gain an understanding, but a general formalism for perfect fluids
described by a physically valid equation of state is still lacking
due to the intrinsic difficulties arising from Einstein equations.
Considering both radial and tangential pressures is a fundamental
step in order to better understand what happens in the ultra-dense
regions that form at the center of the collapsing cloud prior to
the formation of the singularity. For this reason, perfect fluids
appear as a natural choice since these are the models that are
commonly used to describe gravitating stars in equilibrium and since
it can be shown that near the center of the cloud regularity
implies that matter must behave like a perfect fluid.

At first let us consider the case of tangential pressure perturbations.
Towards understanding the stability or otherwise
of the OSD collapse model under the injection of small
tangential pressure perturbations, we consider the dynamical
development of the collapsing cloud, as governed by the
Einstein equations. The visibility or otherwise of the
final singularity that develops in collapse is determined
by the behaviour of the singularity curve and of the apparent horizon,
as shown in section \ref{formalism}.

In the tangential pressure case we have four equations (since the equation
for the radial pressure \eqref{p} reduces to $p_r=0$ which
implies that $F=F(r)$) in
six unknowns, namely $\rho,\; p_{\theta}, \; R,\; F,\; G,\; H$.
With the definitions of $R$, $H$ and $F$, provided earlier we substitute
these unknowns with $v$, $\nu$ and $M$.

We then have the freedom to specify two free functions, namely the mass profile
$M(r)$ and the tangential pressure $p_\theta(r,v)$.

In this model we can integrate the Einstein equations, at least
up to first order, to reduce them to a first order system,
to obtain the function $v(t,r)$, or equivalently $t(r, v)$,
which in general will depend upon $F(r)$, $b(r)$ and $A(r,v)$.
The function $A(r,v)$ here depends on the
nature of tangential pressure chosen.

Again, since $t(r, v)$ is at least $C^2$ we
can write it as in equation \eqref{expand-t},
and the singularity curve is given by equation \eqref{expand-ts}.
This means that the singularity curve should have a well-defined
tangent at the center.

>From regularity we can write $\nu$ as $\nu(r,v)=r^2g(r,v)$, which,
from Einstein equation \eqref{nu}
with $p_r=0$ gives
\begin{equation}
p_\theta=\left(g+\frac{1}{2}rg'\right)r\rho\frac{R}{R'} \; .
\end{equation}
Therefore we can take $g(r,v)$ as the free function and since it must
be a regular function it can be written near $r=0$ as,
\begin{equation}\label{expand-g}
g(r, v)=g_0(v)+g_1(v)r+g_2(v)r^2+...
\end{equation}

We can now study how the OSD gravitational
collapse scenario gets altered when small pressure
perturbations are introduced in the dynamical evolution of collapse.

If the collapse outcome is not a black hole, the final
singularity of collapse cannot be simultaneous.
We cannot have $v=v(t)$ as was the case for OSD.
Allowing for $v = v(t,r)$ is enough to ensure a pressure
perturbation, therefore, for simplicity, we can maintain unchanged the other
conditions that gave the OSD model (namely $M(r,v)=M_0$ and $b(r)=b_0$).
This way we ensure that we do not depart too much from the OSD scenario,
thus bringing out with more clarity the role played by the pressure
perturbations. Note that this choice also automatically forces the choice of
the pressure perturbation to be a tangential pressure, since having
$M=M_0$ immediately implies $p_r=0$.

We know that the metric function $\nu(t,r)$ must
identically vanish for the dust case. On the other hand,
the above perturbation amounts to allowing for $\nu$, or
equivalently $g$ to be non-zero and small.
Proceeding as above we obtain that the tangential pressure can
be written as $p_\theta= p_1 r^2 + p_2 r^3 +...$, where
$p_1, p_2$ are evaluated in terms of the coefficients of
$M, g$, and $R$ and its derivatives. In fact we get,
\begin{equation}\label{pt}
p_\theta=3\frac{M_0g_0}{vR'^2}r^2+\frac{9}{2}\frac{M_0g_1}{vR'^2}r^3+...
\end{equation}

Considering small pressures in a close neighborhood of the center
we can see how it is the choice of the sign of
the function $g_0$ which ensures positivity or negativity of $p_\theta$.

Then the first order coefficient $\chi_1$ in the expansion
of the singularity curve $t_s(r)$ is obtained as,
\begin{equation}\label{chi}
\chi_1(0)=-\int^1_0\frac{v^{\frac{3}{2}}g_1(v)}
{(M_0+vb_0+2vg_0(v))^{\frac{3}{2}}}dv \; .
\end{equation}
As we have noted above, it is this quantity
$\chi_1(0)$ that governs the nature of the singularity curve,
the apparent horizon curve and the visibility of the singularity.

In general it can be seen that it is the matter
initial data in terms of the density and pressure profiles, and
the velocity of the collapsing shells, that decides the value
of $\chi_1(0)$.
In this case, having fixed $M$ and $b_0$ from the OSD model
the final outcome is entirely decided by the function $g_1$ in
the expansion of the tangential pressure.

Of course, if one wishes to consider more realistic models
where the pressure has only quadratic terms in $r$, the same analysis
can be done following a similar method. In that case one would
have to impose that $g_1=0$ and
at the end we would find that $\chi_1(0)=0$ and it will
be the second term $\chi_2$, which would be depending on $g_2$
to decide the final outcome of collapse.

>From these considerations, it is possible to see
how pressure perturbations can affect the time of formation
of the apparent horizon, and therefore the formation of
a black hole or naked singularity.


As can be seen from above, for all functions $g_1(v)$
for which $\chi_1(0)$ is positive the apparent
horizon appears after the formation of the central
singularity. In this case, the apparent horizon
curve originates at the central singularity at $t=t_0$
and increases with increasing $r$, moving to the future.
Thus we have $t_{ah} > t_0$ for $r > 0$ near the center.
The behaviour of outgoing families of null geodesics
has been analyzed in detail in such a case
when $\chi_1(0)>0$ and we can see that the geodesics
terminate at the singularity in the past. Thus timelike
and null geodesics come out from the singularity,
making it visible to external observers.
\cite{JoshiDwivedi99}

One thus sees that it is the term $g_1$ in the
stresses $p_\theta$ which decides either the black hole or
naked singularity final fate for the collapse. Since we
have the freedom to choose $g$ to be arbitrarily small
it is then easy to see how introducing a generic
small tangential pressure perturbation in the OSD model
can change drastically the final outcome.
In general it is a subspace of the space of all functions
$g_1$ that makes $\chi_1(0)$ positive (which includes all those
that are strictly negative), and that causes the collapse
to end in a naked singularity.
In fact, if we wish to give an example, we see that
for the class of all non-vanishing tangential stresses with
$g_0=0$ and $g_1<0$, even the slightest perturbation of the
Oppenheimer-Snyder-Datt scenario would result in a naked
singularity. Of course while this is an explicit
example, by no means it reduces to be the only class. The important
feature of this class is that it corresponds to a
collapse model for a simple and straightforward perturbation
of the Oppenheimer-Snyder-Datt spacetime metric.

In such a case, the geometry near the center can
be written as,
\begin{equation}\label{pert}
ds^2=-(1-2g_1 r^3)dt^2+\frac{(v+rv')^2}{1+b_0r^2-2g_1 r^3}dr^2+r^2v^2d\Omega^2 \; .
\end{equation}
The metric above satisfies Einstein equations
in the neighborhood of $r=0$ when the function $g_1(v)$
is small and bounded. For example, we can take
$0<|g_1(v)|<\epsilon$, so that the smaller we take the
parameter $\epsilon$ the bigger results the radius
where the approximation is valid.

We can consider now the requirement that a realistic
matter model should satisfy some energy conditions ensuring
the positivity of mass and energy density.
As seen earlier the weak energy condition would imply restrictions
on the density and pressure profiles.
In fact, the energy density as given by the Einstein equation must
be positive. And since $R$ is positive, to ensure positivity
of $\rho$ we require $R'>0$ (no shell crossing singularities) and
$F'>0$ (which near the center is satisfied if $M_0>0$).
The choice of positive $M$ is physically reasonable and
ensures also positivity of the Misner-Sharp mass.
Positivity of $\rho+p_\theta$ is then obtained for small values of $r$
throughout the collapse for any form of $p_\theta$ (including
negative pressures).
In fact, regardless of the values taken by $M$ and $g$,
there will always be a neighborhood
of the center $r=0$ for which $|p_\theta|<\rho$ and therefore
$\rho+p_\theta\geq0$. Therefore we see that greater and greater
negative pressures are allowed by the weak energy condition
as we get closer and closer to the center of the cloud,
or conversely, once we fixed a negative pressure given by $g_1$,
the weak energy condition will provide the small boundary
within which such a pressure is allowed.

What we see here is, in the space of initial data
in terms of the initial matter densities and velocity
profiles, any arbitrarily small neighborhood of the OSD
collapse model, which is going to a black hole final fate,
contains collapse evolutions that go to a naked singularity
final fate. Such an existence of subspaces of collapse solutions,
that go to a naked singularity final state rather than a black
hole, in the arbitrary vicinity of the OSD black hole,
presents an intriguing situation. It gives an idea of the
richness of structure present in the gravitation theory, and
indicates the complex solution space of the Einstein equations
which are a complicated set of highly non-linear partial
differential equations. What we see here is the existence of
classes of stress perturbations such that an arbitrarily
small change from the OSD model is a solution going to naked
singularity.

This then provides an intriguing insight into
the nature of cosmic censorship, namely that the collapse
must be necessarily properly fine-tuned if it is to produce
a simultaneous black hole only, just like the OSD case,
as the collapse final endstate. Traditionally it
was believed that the presence of stresses or pressures
in the collapsing matter cloud would increase the chance of
black hole formation, thereby ruling out dust models that
were found to lead to a naked singularity as collapse endstate.
It now becomes clear that this is actually not the case.
The model described here
shows how the bifurcation line that separates the phase space
of `black hole formation' from that of the `naked singularity
formation' runs directly over the simplest and most studied
of black hole scenarios such as the OSD model.

Of course within all matter models with pressures the case of only tangential
stresses is also somehow artificial and it could be objected that conclusions
drawn from the analysis of only tangential pressure perturbations need not be
definitive since the overall picture might be very different when general
radial pressures are also considered.
For this reason, it is useful to enlarge and extend
the previous analysis to include the case of perfect fluid pressures.
Perfect fluids are widely accepted as rather much more realistic
matter sources, and the inclusion of small perfect fluid pressures
in collapse seems not only reasonable but also necessary, if we wish
to derive some conclusion from this whole analysis.
We therefore asked the question: `How is the OSD collapse
scenario affected by the inclusion of small perfect fluid pressures
in the collapsing cloud?'

If we wish to consider perfect fluid collapse we need to use
the formalism presented in section \ref{formalism} together with the
assumption
\begin{equation}
    p_r=p_\theta=p \; .
\end{equation}
The procedure is similar to the one outlined above for the case
of only non-vanishing tangential pressures but the whole
mathematical structure describing the collapsing cloud is now different.
In fact in the case of perfect fluid pressures we have five equations
for the six unknowns $\rho,\; p, \; v,\; M,\; G,\; \nu$, where now
$M$ is a function of both $r$ and $v$. Therefore by considering collapse
of a perfect fluid we must match the collapsing cloud with an
exterior generalized Vaidya spacetime, which is always
possible.

If we provide an equation of state that links the density to the
pressure the whole system becomes closed and we have no freedom
whatsoever to choose any free function. The evolution is then
uniquely determined by the initial data.

If, on the other hand, we do not provide an equation of state, we
are left with the freedom to choose one free function, typically
the mass profile $M(r,v)$, globally. This means that, in contrast
with the tangential pressure case where the mass profile was conserved
throughout the collapse, here we must specify how the mass of the cloud
evolves during collapse.

The mathematical freedom coming from the absence of an equation of
state has therefore a physical counterpart in that some models might
not be physically viable.
Still, imposing regularity and energy conditions is sufficient (even
though this may be not entirely satisfactory) to study the physical
reasonableness of the models.

>From the investigation of perfect fluid collapse it was found that
not only naked singularities are not ruled out in this case but also
that the separation between the regions in the space of all possible
evolutions that lead to black holes and to naked singularities has
some interesting features.
Here again we see how the introduction of small pressures can
drastically change the final fate of the OSD model.
Still we note that the result is more general, in the sense that adding
a small pressure perturbation to a dust model (not necessarily homogeneous)
leading to a black hole can be enough to change the outcome of collapse to
a naked singularity, and viceversa.

In order to perturb the OSD matter cloud by introducing
the radial pressure perturbations, we have chosen a suitable
mass function $M$, motivated by physical reasons, as the free function.
Such a mass profile is taken to be arbitrarily close to the OSD
scenario, in the sense that we considered an arbitrarily small
pressure, where by `small' we mean that the pressure remains much
smaller than the energy density at all times as the collapse
develops and evolves in time.

The class of perfect fluid pressure perturbations considered is given
by the choice of a mass profile of the form
\begin{equation}
    M=M_0+M_2(v)r^2 \; ,
\end{equation}
where $M_0$ is a constant and we have considered $M_1=0$ in accordance with
most physical models that require only quadratic terms in the
expansions of density and
pressure. Then we consider the class,
\begin{equation}
    M_2(v)=C+\epsilon(v) \; .
\end{equation}
The pressure perturbation is
small if we require that $M_0 \gg \mid M_2 \mid$ at all times.

We note that taking $\epsilon=0$ reduces
the model to inhomogeneous dust, and therefore
further imposing $C=0$ gives the Oppenheimer-Snyder-Datt case.
If we wish to perturb the OSD model we need to take $C=0$ and
$\epsilon\neq 0$.
The initial condition $M_2(1)=0$, corresponding to
$\epsilon(1)=0$, ensures that at the initial epoch we
have an homogeneous dust cloud and that the pressures are
triggered only after collapse commences.

The freedom to chose the free function is here reflected
in the freedom to choose
$\epsilon(v)$ which is related to the pressure perturbation.
In fact, from Einstein equations with the present choice of
the mass profile the
density and pressure are given by
\begin{equation}
p=-\frac{\epsilon_{,v}}{v^2}r^2 \; , \; \rho= \rho_{dust}-p+
\frac{5\epsilon-\epsilon_{,v}v}{v^2(v+rv')}r^2 \; .
\end{equation}
Furthermore, for the sake of clarity, we have considered
a constant velocity
profile given by $b_0(r)=0$, in analogy with the marginally
bound case in LTB models.

Evaluating the singularity curve we obtained $\chi_1(0)=0$,
as expected, and
therefore the final fate of collapse is decided by $\chi_2(0)$
which turns out to be,
\begin{equation}\label{chi2}
    \chi_2(0)=-\frac{1}{2}\int^1_0(C+Y(v))Z(v)dv-\frac{4}{9M_0^2}
\int^1_0 W(v)Z(v)dv \; ,
\end{equation}
where we have defined the functions
\begin{eqnarray}
  Y(v)&=& \left(\epsilon+\frac{2}{3}\epsilon_{,v}v\right) \; , \\
  W(v)&=& \epsilon v (\epsilon+\epsilon_{,v}v) \; , \\
  Z(v)&=& v\left(M_0+\frac{4}{3}\frac{\epsilon v}{M_0}\right)^{-3/2} \; .
\end{eqnarray}

Given this structure there are two possible behaviours
for $\epsilon$:
\begin{itemize}
  \item[-] $\epsilon>0$, which implies $\epsilon_{,v}<0$ and positive pressure,
  \item[-] $\epsilon<0$ giving $\epsilon_{,v}>0$ and negative pressure.
\end{itemize}
It is easy to see that negative pressures
more easily allow for the formation of naked singularities,
nevertheless for positive pressures also it is possible to
construct physically valid evolutions that terminate in a
naked singularity.
Requiring $\chi_2(0)>0$ leads to a set of conditions
to be satisfied by $Y$, $W$ and $Z$ and therefore by
$\epsilon$ and $M_0$.
Imposing these conditions is therefore sufficient to
obtain a naked singularity as the endstate of collapse.

Now coming to the stability of the OSD collapse scenario
let us call $\mathcal{D}$ the
set of physically valid initial data for collapse,
which will in general be split into two subsets given by the
possible final outcomes as
$\mathcal{D}=\mathcal{D}^{BH}\bigcup\mathcal{D}^{NS}$.
Then every point $x\in\mathcal{D}$ is characterized
as $x=\{M_i(r), p_{ri}(r), p_{\theta i}, b(r)\}$, where
for example the OSD initial configuration is given by
$x_{OSD}=\{M_0, 0, 0, b_0\}\in\mathcal{D}^{BH}$.

\begin{figure}
\centerline{\includegraphics[width=9cm]{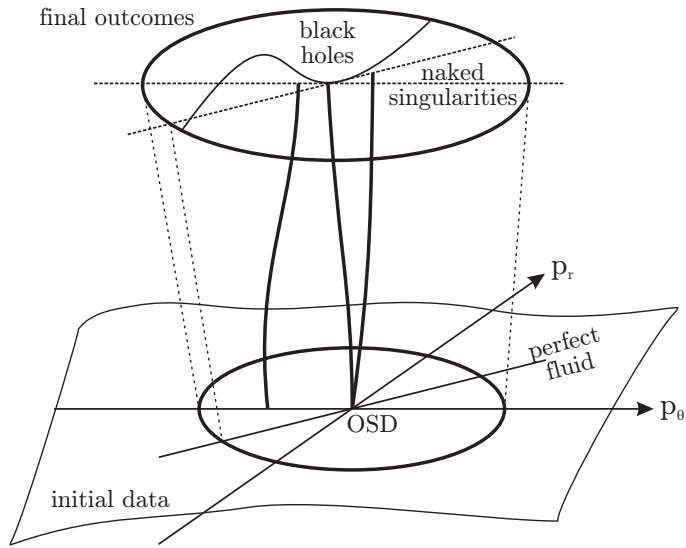}}
\caption{Schematic view of how small pressure perturbations of an
homogeneous dust cloud, both in the form of tangential stresses and
perfect fluids, can lead to the formation of a naked singularity.}
\label{pp}
\end{figure}

What we have shown is that for every neighborhood
$\mathcal{U}(x_{OSD})\subset \mathcal{D}$, however small,
there exist physically valid pressure profiles
with initial data $x\in\mathcal{U}$ such that
$x\in\mathcal{D}^{NS}$ (see Fig. \ref{pp}).
This provides a strong indication that
the OSD dust model is `unstable' in that
the initial configurations for its endstates lie on
the critical surface separating the two possible
outcomes of collapse discussed above.
In this sense we can say that the homogeneous dust collapse
model leading to a black hole is not stable under
the introduction of small pressure perturbations.

It has to be noted of course that the general issue
of stability and genericity of collapse outcomes has been
a deep problem in gravitation theory, and requires
mathematical breakthroughs as well as evolving further
physical understanding of the collapse phenomena.
As noted above, this is again basically connected
with the main difficulty of the cosmic censorship itself,
which is the issue of how to effectively formulate the
same. However, it is also clear from the
discussion above, that a consideration of various collapse
models along the lines as discussed here does yield
considerable insight and inputs in understanding the
gravitational collapse in general relativity
and its final outcomes.

\subsection{The equation of state}
Here we give a brief discussion on possible equations
of state that the collapse may permit. It is known that the
presence of an equation of state introduces a differential
relation for the previously considered mass function which
was as such free, and
that closes the system of Einstein equations. Examples of simple,
astrophysically relevant, linear and polytropic equations of
states have been considered, for which the collapse has been analyzed
to understand its final fate in terms of black hole or a naked
singularity. This treatment outlines the key features of the
above approach and its advantages, and point to possible
future uses of the same for astrophysical and
numerical applications.

An equation of state is a constitutive relation that provides
the link between some state quantities describing the
system. Typically for a collapsing matter cloud an equation of state
is provided once the pressure can be expressed as a function of
the energy density (this could very well mean two different
relations for both the pressures, in the case of anisotropic
collapse). Stars during their equilibrium phase can be described
by such equations of state, both at the level of classical mechanics
\cite{Chandrasekhar},
as well as within the context of general relativity
\cite{Tooper1,Tooper2}.
It is reasonable to suppose then that once the nuclear fuel that
maintains the star in equilibrium is exhausted collapse commences in
a very short time from an initial configuration that is described by
the same physical quantities related by the same equation of state.
Therefore the physical values for $p$ and $\rho$ at the initial
time can be taken from such models at equilibrium. The same holds
true for the other quantities appearing in the equation of state.
All these can be expressed in terms of the thermodynamical variables
of the system such as the temperature and the molecular weight
of the gas.

If we consider collapse of a perfect fluid (thus neglecting
anisotropies that, as we said, must reduce to zero as $r$ goes to
zero), it is easy to see that a barotropic equation of state of
the form $p=Y(\rho)$ introduces a further differential equation
in the system of Einstein equations. This differential equation must
be satisfied by the mass function $M(r,v)$, thus providing the
connection between the equation for $p$ (equation \eqref{p}) and that for
$\rho$ (equation \eqref{rho}) and making them dependent on $R(r,v)$
and its derivatives only.  The dynamics is then entirely fixed by the
initial conditions and therefore we see how solving the equation
of motion is equivalent to solve the whole system of equations.
In the case of anisotropic collapse two equations of state (or one
equation of state and one relation specifying the anisotropy),
one for $p_r$ and one for $p_\theta$, are required in order
to close the system.

The equations of state that are typically considered for perfect
fluids at equilibrium are two. A linear relation of the type
\begin{equation}
    p=k\rho \; ,
\end{equation}
where $k$ is a constant, and a polytropic relation of the type
\begin{equation}
    p=k\rho^\gamma \; ,
\end{equation}
where the exponent $\gamma$ is typically written as
$\gamma=1+1/n$, with $n$ being the polytropic index of the
system. It can be shown that $n\leq 5$ (for $n>5$ the cloud has
no boundary at equilibrium).

As we said the introduction of an equation of state closes
the system of Einstein equations and therefore no freedom to specify
any functions remains. Gravitational collapse of a perfect fluid
with a linear equation of state was studied in Ref.
\refcite{JoshiGoswamiPF}.
The differential equation for $M$ becomes
\begin{equation}\label{eos}
    3\lambda M+\lambda r M_{,r}+[v+(\lambda+1)rv']M_{,v}=0 \; ,
\end{equation}
and it is possible to show that solutions exist and that both
black holes and naked singularities are possible outcomes of collapse
depending on the initial data and the velocity distribution
of the particles.
\cite{GoswamiJoshi04}

The study of collapse of a perfect fluid with polytropic equation
of state is more complicated.
In both cases solving the differential equation for $M$ might prove
to be unattainable. Nevertheless with the assumption that
$M$ can be expanded in a power series near $r=0$, and expanding
$p$ and $\rho$ accordingly, we can obtain a series of differential
equations for each order $M_i$, by equating term by term from
\begin{eqnarray}\nonumber
  p(r,v) &=& p_0(v)+p_2(v)r^2+...= Y_0(v)+Y_2(v)r^2+... \; , \\
  \rho(r,v) &=& \rho_0(v)+\rho_2(v)r^2+... \; ,
\end{eqnarray}
with
\begin{eqnarray}
   Y_0 &=& Y(\rho_0) = -\frac{M_{0,v}}{v^2} \; , \;
Y_2(v)= Y_{,\rho}\rho_2 = -\frac{M_{2,v}}{v^2} \; , \; ...\\
  \rho_0 &=& \frac{3M_0}{v^3} \; , \; \rho_2 = \frac{5M_2}
{v^3}-\frac{3M_0}{v^4}w_{,r}(0,v) \; , \; ...
\end{eqnarray}
This series of differential equations, if they can be satisfied
by all $M_i$, and if the solutions converge to a finite mass function
$M$, solves the problem, thus giving the explicit form of $M$.
One thing that is interesting to note is that if pressures and density
can be expanded in a power series near the center the behaviour
close to $r=0$ approaches that of an homogeneous perfect fluid.

Considering the whole process it is reasonable to assume that
during collapse, which goes from a nearly Newtonian initial state,
to a final state where a strong gravitational field is present, the
equation of state describing the matter will vary, maybe even
abruptly as in some phase transitions (as it happens when the nuclear
saturation limit for the neutron star matter is reached).

Recent developments suggest that close to the formation of the
singularity strong negative pressures might develop and the
gravitational field might act repulsively thus disrupting the
collapsing cloud and dispersing away the infalling matter. This
kind of effect can be inferred already at a classical level
but it is more evident when quantum corrections are considered.
\cite{JoshiGoswamiLQG}
This evidence suggests that the equation of state relating
pressure and density (which must be always positive) evolves in
a non-trivial manner during collapse.

Typically we can expect an adiabatic behaviour with small
adiabatic index at the beginning of collapse when the energy density
is much lower than the nuclear saturation energy. As collapse evolves
it is possible that the equation of state will present a sharp
transition when matter passes from one regime to another, as for
example in the case when the nuclear saturation energy limit is exceeded.
Furthermore towards the end of collapse repulsive forces become
relevant thus giving rise to strong negative pressures and the
speed of sound in the matter cloud approaches the speed of light
(see Ref.
\refcite{Zeld}).


It is clear now that to account for such extreme situations
the usual linear and polytropic equations of state are not enough
and the equations of state describing ultracompact objects are not
well understood.
\cite{Arnett}
Therefore one is left with two options: either dropping the equation
of state altogether, thus allowing for $k$ and $\gamma$ to vary
with $r$ and $t$ (one can always choose the functions in such a way
that they tend to become constant in the limit of weak gravitational
field thus connecting the description close to equilibrium with
the description close to the formation of the singularity), or to
consider more exotic equations of state. Among the latter possibility
gases with exotic properties (such as the Chaplygin gas, see Ref.
\refcite{Chaplygin})
have been also considered in order to account for possible sources
of dark energy and dark matter. Of course dark energy effects are not
likely to be relevant for collapse at stellar scales but they might
become important for bigger objects such as the compact ones
dwelling at the center of galaxies that have masses of the order
of many million solar masses.

\subsection{Non-spherical collapse}
It is true that most of the collapse models analyzed so far
are spherical. However, many
physicists believe that (as Roy Kerr pointed out to PSJ) if cosmic
censorship is to hold as a basic principle of nature, it better holds
in spherical class too, which has wide astrophysical
significance. Thus, analyzing these models could be of great
value from the perspective of censorship,
for example
to isolate the physical features that cause naked singularities.
On the other hand,
there are researchers who believe that the current classes analyzed so
far are already good enough to begin investigating the physics and
astrophysical implications of naked singularities.

While we have a good understanding now of spherical collapse
for a generic matter field, non-spherical collapse remains a major
uncharted territory (regardless of the fact that attempts have been made
since many decades
\cite{Doros,nonspherical}).
Several recent studies have found non-spherical
models that also give rise to naked singularities.
Also, certain analytical solutions in cylindrical symmetry provided
some new examples of exact solutions describing collapse away
from spherical symmetry (see e.g. Ref.
\refcite{Nolan,Prisco}).
Though the physical validity of such models remains doubtful, it
is nevertheless one of the few instances in which we have analytical
dynamical solutions that are not spherical.

The question now is
whether these situations are contrived.  Fast-developing numerical
core collapse models in general relativity could be of help here. The results
so far also show that naked singularities are, in fact, stable to
small perturbations in the initial data of matter fields, to the
introduction of non-zero pressures in the cloud, and so on.
Therefore we have yet to find
and isolate precisely the kind of perturbation that would make a given
naked singularity go away. These situations are what physicists
call `generic' that is, they are not contrived.  A tiny deviation in the
initial data leads to much the same outcome. However, we should
emphasize the general `stability' proof for the naked singularity yet
remains to be achieved.

The key issue when we are dealing with non-spherical collapse
regards what happens to the `deformations' that mark the departure
from spherical symmetry.
This is a major problem analytically since the set of Einstein equations
describing the evolution of the cloud becomes immensely complicated when
deformations and rotation are taken into account.
Though there have been some attempts to study quasi-spherical
and cylindrical collapse in full generality, the physical significance
of such models remains for now somehow obscure
and we can say that the study of non-spherical collapse within
exact solutions of Einstein
field equations is a field where most of the work
still needs to be done.

Since a comprehensive analytical treatment of collapse away from
spherical symmetry is
still missing, the few insights that we have in the final
fate of collapse of a non-spherical cloud come from numerical relativity.
The study of non-spherical collapse is related to cosmic censorship
through the so called `no-hair theorem' which, roughly speaking, says
that any asymptotically flat black hole vacuum solution of Einstein
equations must be identified by only three quantities, namely mass,
charge and angular momentum. This means that other vacuum solutions which are
characterized by some other quantity, such as higher multipole moments
for an axially symmetric spacetime, can have a naked singularity
(as is the case for the class of Weyl metrics).
Therefore one crucial issue with non-spherical collapse regards what happens
to those higher multipole moments during the collapsing phase.

If the collapsing cloud retains its non-spherical shape throughout
collapse, then the final vacuum configuration would not be represented
by the Schwarzschild or Kerr
\cite{Kerr}
metrics.
In this case naked singularities like those present in axially
symmetric vacuum spacetimes (such as the Weyl class or the
Tomimatsu-Sato solution) could be present (see e.g.
\refcite{Herrera07}).
On the other hand, the presence of deformations might
provide a modification in
the density distribution of the collapsing source that
opposes the pull of gravity, thus suggesting that non-spherical
sources might produce some equilibrium configuration
before complete collapse settles to a singularity
(see for example,
\refcite{HerreraPrisco}).

Recently there have been investigations to look for the observational
signatures that such vacuum metrics would possess
(see section \ref{observational}),
and the possibility that the Kerr metric might not be the best
metric to describe the exterior of a rotating star has been suggested.

If, on the other hand, higher order multipole moments are radiated
away during collapse then the final configuration must settle to a
Schwarzschild or Kerr-Newman spacetime. Of course, if this happens to
be the case, there has to be a mechanism, within the evolution
of Einstein equations themselves, which explains why and how the
higher multipole moments are radiated away. Such a mechanism, at
present, is still a matter of speculation.
Furthermore in this case one is still left with the possibility
that the event-like naked singularities, like those that form in
collapse with spherical symmetry, might form. In this sense the
study of small perturbations of spherical symmetry shows that
the final outcomes, whether they are black holes or naked
singularities, are most likely stable.

Some numerical studies of the collapse of a rotating gas cloud
showed that a final configuration with over-spinning angular momentum
is unlikely to form under general
conditions and therefore a Kerr black hole might be a most
probable final outcome than a Kerr naked singularity.
\cite{Rezzolla11}

Nevertheless people have started to study the observational
features of such singular spacetimes (as we discuss in
section \ref{observational})
in the hope to gain a better understanding of what trace they
could leave in the universe if they were allowed to happen. Of course
we still have no proof that such configurations might arise
from collapse of a cloud with regular initial data, but just as
well we have no proof that such configurations must be forbidden.

\subsection{Numerical simulations}\label{numerical}
Since the important work by Shapiro and Teukolsky
\cite{Shapiro1},
who showed that axially symmetric collapse could lead to
the formation of naked singularities, the investigations on
gravitational collapse have been carried out both on the analytical
side as well as on the numerical side. As we have noted, a lot of
questions regarding complete collapse of massive bodies still
remain unanswered from the theoretical point of view (like the
issue of cosmic censorship), as well as from the observational
perspective (like the mechanism behind supernovae explosions),
and it is to answer these questions that computer simulations might
prove very useful.
\cite{Shapiro2,Garfinkle}
An exhaustive treatment of numerical simulations
for gravitational collapse is beyond the scope of the discussion here
and so we will give here a brief summary of results and
open problems, including some references for the reader who may
like to delve deeper into this growing subject. For many
developments and references see e.g. Ref.
\refcite{Alcubierre}.

Due to the difficulties arising from the structure of Einstein
equations when non-trivial situations are considered (such as
non-spherical collapse with rotation, for example), numerical
simulations can provide several insights into problems that are
presently not addressable analytically. Among these the most important
are the final fate of the complete collapse of a massive cloud
to a compact object or the merger of two compact objects, two
phenomena that are related to cosmic censorship as well as to
astrophysics. Furthermore since numerical simulations allow to
consider effects that are generally neglected in the analytical
treatment, but are of much importance in the astrophysical context,
they can be used to study the mechanism behind supernova explosions,
core collapse, and high energy phenomena that occur when a star
dies. All these scenarios are expected to produce gravitational waves,
the holy grail of gravitational physics today, and therefore
numerical modeling of how these gravitational waves can be produced
is of much importance to the success of experiments such as LIGO
and VIRGO. For example, the presence of electromagnetic fields and
accretion disks are crucial elements that will affect the process
by which the core of a star implodes and at present they can be
studied only with numerical simulations, which are therefore the
only mean to provide indications on how some of these high energy
phenomena, such as relativistic jets or gamma-ray bursts, that
are observed in the universe can occur.

As we know the complete collapse of a massive cloud produces
a singularity, that can be naked or visible depending on the initial
conditions. Since diverging quantities cannot be obviously handled
numerically, a crucial issue of numerical simulations is how these
singularities are treated or excised (see Ref.
\refcite{BaiottiRezzolla2006}
and references therein).
A numerical model where the singularity region is removed from
the simulation regardless of the presence of an apparent horizon
will not provide much insight on the issue of cosmic censorship for
those scenarios, while it could still prove to be extremely useful
towards the understanding of the issues of rotation and dissipation
of higher multipole moments that might settle the collapse to a
Kerr black hole or an axially symmetric naked singularity. On the
other hand, numerical models can be used to trace the apparent
horizon in more complicated matter models thus giving some hints
to what happens to trapped surfaces in realistic collapse scenarios
(see Ref.
\refcite{Ortiz}).

There are two main collapse scenarios that have been widely
investigated and that are of crucial relevance from the astrophysical
point of view:
\begin{itemize}
  \item[-] The complete collapse of rotating bodies.
  \item[-] The merger of two compact objects.
\end{itemize}
It is important to notice that when one takes into account enough
elements to make the simulation somehow closer to reality the time of
computation grows enormously thus requiring the use of very powerful
computers. For this reason very often such simulations had to restrict
the space of parameters considered. Typically simulations were carried
out in one or two spatial dimensions only and it was only recently
that full 3D simulations of core collapse supernovae and of binary
mergers allowed to study these phenomena in more detail. Furthermore
for the above reason simulations that consider the microscopic
structure of matter, including neutrino emissions and electroweak
interactions, are generally constrained to be non-relativistic while
fully general relativistic simulations typically neglect the
microscopic details of the matter cloud.

Within the problem of the complete collapse of a rotating body
the main features that have been investigated through numerical
simulations are the influence of rotation and magnetic fields (see Ref.
\refcite{Rezzolla05}
and ref therein).
These affect greatly the last stages of collapse and are responsible
for the waveform of the emission of gravitational waves and the
structure of the accretion disk that surrounds the final compact object.
The way that the outer layers are ejected from the collapsing
spinning body can help us understand some questions that are crucial
to the cosmic censorship such as:
Does the collapse of a rotating object always end in a Kerr spacetime?
Or is it possible that a rapidly rotating object collapses to a
super-spinning Kerr solution?

Further, these simulations provide valuable physical insights
into astrophysical questions such as how the matter ejected during
collapse due to rotation falls back on the compact object in the
form of an accretion disk, or what kind of features these accretion
disks will present in terms of thickness, angular momentum and light
emission. Or also, how the presence of a magnetic field will
affect the infall of particles and the formation of high energy jets.

Recently the attention has been posed towards the issue of
binary mergers. These models describe the merger of two inspiraling
compact objects that can be taken to be two black holes, two neutron
stars or one black hole and a neutron stars. The effect of unequal
masses for the initial objects on the final configurations have
been studied and a variety of scenario have been proposed in which
the system settles to a final Kerr black hole with an accretion disk.
These mergers are believed to occur frequently in the universe and
are a major source of gravitational waves. Therefore the simulation
of the waveforms emitted during the merger is one of the crucial
results of these simulations, though it is not the only one. In fact
accretion disks and the mechanism by which the matter from the disk
accretes onto the final compact object is thought to be one possible
mechanism for the production of gamma-ray bursts (see Ref.
\refcite{Rezzolla10}).
Furthermore, crucial to the cosmic censorship conjecture, the possible
formation of superspinning Kerr spacetimes have been studied.
In principle such a spacetime could result from the merger of two
rapidly rotating compact objects (see Ref.
\refcite{Motoyuki}),
from the increase in angular momentum due to the inflow of
particles from the accretion disk, or from the complete collapse
of a star with high angular momentum (see Ref.
\refcite{Rezzolla11}).
In practice there are no definitive results on the possibility
of the formation of superspinning Kerr spacetimes, though there are
arguments that suggest that overspinning a Kerr black hole might
not be possible. The discussion on these issues is still very much
open. What has become clear is that the mechanism by which such
a final configuration could in principle be produced would be
very different from that of producing a Kerr black hole.

Finally, we mention here some work that has been developed towards
the goal of a viable description of the processes that lead to type
II supernovae explosions. Many different settings of numerical
simulations have been carried out considering the microscopic effects,
in order to provide a viable model for the production of the
shockwaves that create these supernovae. These simulations describe
the core collapse of a star from the microscopic point of view, taking
into account nuclear interactions and neutrino productions. Within
this field a lot of progress has been made in the last few years
(with the computational power that finally enabled for full 3D
simulations) and a lot of attention has been put on the influence
of neutrino production during the explosion (see Ref.
\refcite{Sekiguchi}
and ref therein).
Nevertheless these simulations still face one crucial problem in
the fact that the efficiency of the explosion, even taking `neutrino
heating' into account, is not enough to produce a supernova. What
typically happens in the simulations is that the explosion dies out
after a few hundred kilometers, and thus no known mechanism so far has
been able to revive the shockwave that would produce the supernova.
\cite{Ott11}

Almost all models for type II supernovae explosions consider a
shockwave that is generated when the infalling matter coming from
the outer shells reaches the inner compact core. The compact object
at the center constitutes a barrier onto which the matter from the
outer layers bounces, thus creating the shockwave. The energy of the
shockwave is related to the size of the core that creates it.
Nevertheless, during the process of complete collapse, it is still
possible that another wall, a quantum-gravity limit, exists at a
shorter scale. This, as we have seen in previous sections, is where
general relativity breaks down predicting a singularity. It is
possible that another shockwave be created once this limit is
reached. Therefore if the structure of collapse is such that no
horizon exists at that time it is also possible that such a shockwave
propagates outwards, providing the missing energy for the explosion to occur.
As far as we are aware no simulations has been carried out taking these
possible effects into account and therefore at present we do not know if
these constitute a viable solution to the problem of the missing energy
for type II supernovae explosions.

\section{Structure of naked singularities}
As such the collapse models studied so far and the
resulting singularities exhibit a wide variety
of interesting structure.
The thing that emerges clearly is that not all
singularities are the same in their properties and
therefore their
physical implications as well as their connection
to the CCC will also be different from case to case.
In some models, only a part of
the singularity is visible, rest covered in the horizon eventually,
but elsewhere they can remain visible forever. This depends
on the form of matter collapsing, the equation of state used,
and such other properties. Basically they seem to come in all
varieties, depending on different models of collapse
considered. Typically, the naked singularity
develops in the geometric center of collapse to begin with, but
later it can spread to other regions, or get covered,
as the collapse progresses.

\subsection{Are they always massless or with negative mass?}
A question is asked sometimes on the mass of naked singularities,
namely if they are always `massless', because typically in
spherical collapse the naked singularities forming at the center
of the cloud has the function $F$ tending to a vanishing value
in the limit of approach to the singularity.

We note, however, that this need not always be the case.
For example, in higher dimensional gravity, timelike naked
singularities do develop which are massive both within Einstein's
theory (see e.g. Refs.
\refcite{Debnath,Ghosh})
as well as in alternatives theory of gravity (see Ref.
\cite{maeda}).
Massive singularities are present in solutions of Einstein
field equations in classical relativity as well, the most famous
of them probably being the one appearing in Kerr's solution, though
at present we do not know if such a naked singularity can form
via some physically viable dynamical process.

In any case, in our opinion what really matters when a
singularity is visible is that super-ultra-dense regions where
densities and curvatures really blow up are visible to external
observers in the universe. That is of actual physical consequence,
rather than the `mass' of the singularity itself, which is
actually not an object or part of the spacetime, and in fact
it may even be resolved when quantum gravity effects are
taken into account.

Given the examples of massive timelike singularities
and others such as above, and also other examples of collapse
evolutions available so far, it is clear that naked
singularities, when they
form in collapse come in many types of varieties and properties
depending on the physical situations as well as the form of
the matter considered and such other factors, and it is not
possible to actually rule out any of these by some kind
of a theorem. Therefore, in
our view, imposing such restrictions will not help
preserve the cosmic censorship conjecture.

Also, there is some discussion in the literature on the
negative mass Schwarzschild singularities. It is asked sometimes
if naked singularities, whenever they form,
must always have a negative mass similarly to the specific
Schwarzschild case mentioned.
As we have clarified above, that is not the case.
Generically, if we take the case of a spherically symmetric
collapse, the mass function has a vanishing or positive
value in the limit of approach to the singularity which
is visible.

In fact, one would like to impose all possible physical
reasonability conditions, such as the positivity of the mass-energy
density, regularity of the initial data from which the collapse
evolves, and such other regularity conditions.
We note that constructing models with naked singularities and negative
mass is of course possible, but the physical validity of the same
would be quite unclear.
Then the idea is to see whether naked singularities still develop
in gravitational collapse that is developing from regular
initial data and under physically viable conditions.
The answer that follows is in affirmative
as shown by many studies we discussed above. These singularities
have no negative mass. In fact, the negative mass Schwarzschild
singularity is not obtained as a result of any dynamical collapse
evolution, and to that extent it is not physical.

\subsection{An object or an event?}
Are naked singularities always pointlike in time,
or also extended? This is related to the question of whether
they are like an object or an event.
In typical spherically symmetric collapse models when
considered in comoving coordinates, the first point of the
singularity curve is visible, from which families of
non-spacelike curves come out. The later points in comoving
time, of the singularity curve, get hidden under the horizon.

Thus one can ask whether naked singularities, whenever
they form, are always momentary or they could also be extended in time.
We point out here that depending on the collapse scenario and the
form of matter and the equation of state considered, the naked singularity
could be pointlike, or it could be also extended in the comoving
time. In particular, the timelike singularities, when they form in
collapse are extended typically (see e.g. Ref.
\refcite{timelike}).

Even when they are pointlike or null singularities, the structure
of the families of geodesics coming out from the singularity have been analyzed
in detail (see for example Refs.
\refcite{JD-Vaidya1}--\refcite{MenaNolan2}
It is found that there are non-zero measure of non-spacelike
curves coming out even from a pointlike
naked singularity, and as far as the faraway external observer is
concerned, once he gets to see the first ray from the singularity,
for all times to come the null or timelike paths from singularity
will keep reaching the observer.

This issue is also related to whether the naked singularities are
always null, when they form in collapse. As we pointed out, they
could be very much timelike, as well as extended in the space, rather
than being just null and pointlike.

We note that as opposed to the naked singularities developing
in collapse which are sometimes like an event, those occurring in
the super-spinning Kerr geometry or many other vacuum models are
ever-lasting, and in that sense they are like an `object'. While
such solutions do occur in general relativity, till recently
what was not clear is whether such object-like naked singularities
do arise from dynamical physical processes in gravity physics.
In this connection, it is relevant to note that in Section \ref{observational}
we shall discuss a class of gravitational collapse models
which give rise to the final configuration which contains an
ever-lasting naked singularity which develops in collapse
from a regular initial data.

\subsection{Do naked singularities violate causality?}
One of the issues that is often considered in connection
to the occurrence of naked singularities is that of the possible
violation of causality that can arise in certain spacetimes.

Roughly speaking, causality violations occur in those
spacetimes which contain closed timelike curves. An observer moving
on such a curve would eventually find himself at the initial event
in space and time even though locally his clock never went backwards.
The most well known examples of such spacetimes are the Godel universe
\cite{Godel}
and the solution found by Tipler for a spacetime outside
a rotating cylinder.
\cite{Tipler}
Also, closed timelike curves are present in familiar solutions
such as the Kerr metric. The singularity theorems by Penrose and
Hawking show that singularities form generically from Einstein
equations even when causality is protected, but these theorems
are not very helpful when exploring the
connection between causality violations and singularities.

While closed timelike curves are present in the Kerr spacetime,
they are confined inside the horizon for the Kerr black
hole. Therefore an observer entering the horizon would be able to
travel back in time but would not be able to return to the outside
universe, thus preserving the causal structure of the same.
In some cases, closed timelike curves can be accessed by any observer
in the Kerr metric, allowing therefore for the theoretical possibility
of time travel.
\cite{HE}
>From such an example, it could be tempting to think that naked
singularities would typically allow for causality violations to occur
in the universe, thus making them undesirable.
It is possible that the presence of the horizon would hide the closed
timelike curves thus disconnecting them from the rest of the
universe. Although this is true for some manifolds, there
are also dynamical spacetimes such as the LTB models and others,
which admit naked singularities but which have no closed timelike
curves. It follows that in general there is no direct connection
between these two phenomena of causality violation in a spacetime
and the occurrence of naked singularities.

We note that even when singularity theorems assume some causality
condition, the spacetime singularities and causality violation are
independent phenomena.
Therefore, in general, the connection if any between the
spacetime singularities, and in particular naked singularities
and causality violations is far from clear and could be much
more subtle. One of the main reasons could reside in the fact that
different kinds of naked singularities need to be treated
differently. For example, what holds true for the Kerr solution need
not be true for the Lemaitre-Tolman-Bondi metric. A spacetime without
rotation, such as the LTB collapse model, could not have closed
timelike curves and would therefore respect causality regardless of
the presence of a naked singularity.

Furthermore it is clear that causality violations need to be better
defined and studied in order to understand what is really undesirable
about them (see for example Ref.
\refcite{Monroe}).
The possibility of closed timelike curves by itself could not be
enough to make a spacetime `physically unrealistic'. In fact if we want
to summarize we can in principle divide causality violations into
three main categories:
\begin{itemize}
  \item[-] Microscopic causality violations: in the case that closed
timelike curves can occur at a microscopic level, thus being resolved
or included by an eventual theory of quantum gravity.
  \item[-] Local scale causality violations: in the case that closed
timelike curves can occur at planetary or galactic level, thus giving
rise to the possibility of time travel with all the connected paradoxes.
  \item[-] Cosmological causality violations: in the case that these can
occur only on time scales comparable with the life of the universe and
thus having no bearing whatsoever on our local picture of the cosmos.
These closed timelike curves cannot be ruled out in principle.
\end{itemize}

As noted by many authors, global causality requirements for the whole
universe might be too restrictive since we are able to experience only
a limited, local, portion of the universe. From the fact that
causality holds here and now it might be far fetched to conclude that
causality holds globally for the entire universe. Furthermore
non-local correlations have been studied within the framework of quantum
mechanics since many years and it seems not impossible that at
a microscopic level a full theory of quantum mechanics coupled to
gravity might allow for some kind of causality violations to occur.
Therefore we are left only with the second class
of causality violations, namely the local scale violations, to be
considered as undesirable in principle. Within this class (which for
example excludes the Godel solution), can we say that naked
singularities and closed timelike curves are linked in some way?
Examples of spacetimes without singularities but with closed timelike
curves are known since the early days of general relativity (see Ref.
\refcite{Bonnor}
and references therein), but such solutions cannot be obtained from
the evolution of a regular matter cloud. In fact it was shown that the
occurrence of singularities is a necessary condition for closed timelike
curves to evolve from regular initial data
\cite{Tipler2,Tipler3},
therefore suggesting that in dynamical configurations the formation
of singularities is connected to the appearance of closed timelike
curves. Nevertheless the possible relation that might link the
occurrence of closed timelike curves with the behaviour of the
horizon that could eventually cover the singularity has not been
investigated so far, leaving at present the issue of causality
violation separated from the fate of cosmic censorship.

\subsection{Local versus global visibility}
A singularity is said to be locally naked if there exist
outgoing null or timelike trajectories that reach some observer
in the spacetime. In this sense, for example, the singularity
in the Reissner-Nordstrom spacetime is locally naked since
observers located within the radius of the event horizon,
but not outside it, can be reached by nonspacelike geodesics coming
from the singularity. On the other hand a singularity is said to
be globally visible if there exist outgoing nonspacelike
geodesics that reach observers at future spatial or null infinity.
In this case no horizon is present before the singularity
and the light rays coming from the singularity can be
seen by any future observer.

Global and local visibility are of course related to
the cosmic censorship conjecture itself. As said, cosmic censorship
comes in various forms and formulations and over the years some forms
have been proposed that allow for the existence of locally
visible singularities. The main distinction that is usually
made is between two formulations of the conjecture that are
called the Strong Cosmic Censorship and Weak Cosmic Censorship.
The form of the conjecture that one assumes in the analysis has
implications for the local and global properties of the spacetime
containing the singularity that one is going to study.

In fact, the weak form of the CCC postulates that a singularity
cannot be seen by any observers at null infinity, thus
allowing for locally naked singularities to occur. In this case
when we study the formation of the singularity and of the apparent
horizon, we need not worry about the global structure of the horizon
itself. Actually, proving that there are future directed outgoing
null geodesics emanating from the singularity is enough to
ensure local visibility.

\begin{figure}[hh]
\includegraphics[scale=0.70]{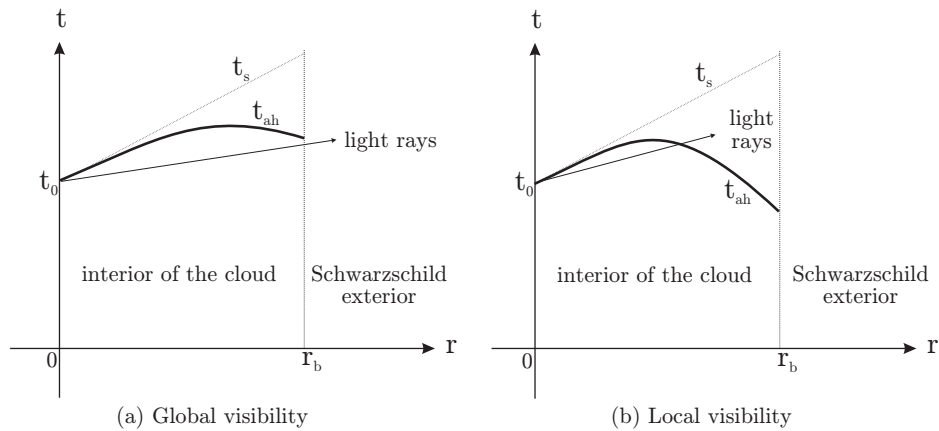}
\caption{(a) Globally visible singularity: light rays emitted
from the central singularity can propagate inside the cloud
until they reach the boundary, from which they can travel freely
to reach observers at future infinity.
(b) Locally visible singularity: light rays emitted from the
central singularity can propagate inside the cloud but must fall
into the apparent horizon before they reach the boundary.}
\label{visibility}
\end{figure}

As an example, one could consider the LTB collapse scenario.
As we have shown earlier it is the first term in the expansion of
the mass profile that determines the local visibility of the
singularity. This holds regardless of where the boundary of the
cloud is taken. On the other hand, a careful analysis of the behaviour
of the apparent horizon shows that for some of these matter models
the horizon curve, while being increasing in a neighborhood of
the center, can become decreasing afterwards, thus covering the
singularity eventually from observers at future infinity
(see Fig. \ref{visibility}).

It was shown (see Ref.
\refcite{DJ,Giambo,Deshingkar})
that for a fixed value of the boundary there is a range
of values of the first non-vanishing term in the expansion of
the mass profile for which the singularity is only locally naked
and a range of values for which it is globally naked.
>From a mathematical point of view, this is not relevant since in
dust models we always have the freedom to choose the boundary as we
like. Therefore for any given matter profile we can always
ensure global visibility by taking $r_b$ such that $t_{ah}(r)'>0$
for all $r\in[0,r_b]$ (see Fig. \ref{visibility}).

The issue is somewhat different when we apply the formalism to
the description of a collapsing star. First of all, the presence of
pressures (that must vanish at the boundary) might affect the way
the boundary and the mass profile are related. Furthermore, in this
case the total mass and the radius of the boundary are fixed by
realistic physical values for a star, depending on the stellar model
chosen. It can then very well happen that even though the singularity
that forms at the end of collapse is locally visible it might not
be globally visible.
Recently, Jinghan et al.
\cite{jinghan}
pointed out that if we idealize a star as composed only
by dust such a scenario could happen.

On the other hand, as we have seen, the LTB model is very
restrictive as far as the matter content is concerned, since dust is
not a very realistic form of matter for a dense object as the
core of a star. Furthermore, as indicated previously, assuming that the
same equation of state (in this case dust) holds throughout the
whole star is also unrealistic, since outer layers will be less dense
and made of different matter constituents as compared to the inner
layers. So if we apply the same reasoning as above, adopting
the LTB model not to describe the whole star but only for its core
we are once again left with some degree of freedom as to where to
take the boundary of the core.

Therefore it is easy to see that for more elaborate and realistic
matter models, which include pressures and different layers, there is
still no concluding evidence in one direction or the other, and the
issue of global visibility for a given collapse model remains
very much open.

\subsection{Can energy come out of a naked singularity?}
We note here that even if a naked singularity forms in
collapse, the singularity by itself does not necessarily have to radiate
matter or energy thus bearing a signature that is detectable from
far away in the universe.
Such singularities forming at the end of collapse, unlike naked
singularities such as the ones in the superspinning Kerr metric, are
more like events rather than an object, being a moment when
collapsing matter reaches its final doom, something like the
big bang in reverse. Questions such as what will
come out of a naked singularity are then not really meaningful;
`things' do not have to come out of it. What we really see is not
the singularity itself but the signature of processes that occur
in the extreme conditions of matter near this epoch, such as
the shockwaves due to inhomogeneities in this ultra-dense
medium, or quantum gravity effects in its vicinity.

Generally, when considering a complete gravitational collapse,
it is often assumed that the boundaries set by the
electromagnetic and nuclear forces are surpassed and nothing can
halt the collapse. Therefore, in the ultra high density regions that
surround the singularity, the pull of gravity is thought to be
so strong that nothing can escape it. One would be inclined to think
then that even if such naked singularities exists they would bear
no impact on the rest of the universe since no signal can escape
their gravitational field. Nevertheless, there are other ways by which
naked singularities could in principle leave a trace in the outside
universe, and from the study of exact solutions of the field equations
we can see that there are scenarios in which the process of
formation of the singularity can be accompanied by the
emission of energy.

Firstly, we note that if we regard infinities as the mathematical
signature of the breakdown of a theory then it is reasonable to
suppose that some other physical effects will account for those divergencies
thus resolving them. Therefore general relativity might not be the
best tool to describe the final instants of collapse when the
singularity forms. Generally it is believed that it will take some
quantum theory of gravity to resolve the singularity, but there is
no reason at present to rule out also some other classical
gravitational effects coming from some corrections to general
relativity (see e.g. the Eddington gravity alternative in Ref.
\refcite{Eddington}
for an example of an alternative theory without matter singularities).
Hence, once a suitable theory of gravity is able to resolve the
divergence of the energy density that originates the singularity, it
is very well possible that some new kind of barrier arises at short
scales (such as the Planck scale for an eventual theory of quantum
gravity), thus disrupting the singularity formation and creating
a shockwave through which the matter-energy that was collapsing is
ejected away. Such a shockwave would be immensely energetic and
it would bear a clear signature of the level at which it has arisen.

This point raises the issue of the possible observational features
that a naked singularity could have. Did we observe similar phenomena
in the universe already? The answer is we do not know as yet. Since
the stellar structure in realistic cases is much more complicated
than in the idealized analytical models and it involves many layers
of different materials with different properties, it is obvious that
whatever might come from such a Planck-scale event would be scattered,
absorbed and emitted many times before an actual signal comes out of
the surface of the star. Therefore, of the many highly energetic
events that are known to happen when a star collapses under its own
gravity, we do not know if any of them bears the signature of the
Planck scale physics that is happening very close to the center.
For example, we
still do not have as of now a comprehensive model that explains how
supernovae explode. Computer simulations describing what happens at
the core of the star when it explodes have found difficulties in
producing the amount of energy necessary for the shockwaves to
propagate through all the layers of the star, thus generating
the explosion
(see section \ref{numerical} for more details).
It seems not unreasonable to suppose that a barrier at a scale
smaller than the Schwarzschild radius might provide this missing
energy, nevertheless there are still no studies in this direction.
A similar reasoning might apply also to gamma-ray bursts, a very
energetic phenomenon created from the core collapse of a star that
is still far away from being well-understood.
\cite{JoshiDadhich}
It could be that the
gamma-ray bursts are created in the exploding outer layers of the star,
and therefore it is not a direct effect of some possible quantum
barrier. Nevertheless the mechanism by which the explosion is
related to the collapse of the inner core is still not well-understood
and it is indeed possible that certain kind of phenomena are linked
to certain types of collapse.

On the other hand, even without calling for some modified theory
of gravity or quantum gravity, it is still possible that some energy
comes out from the singularity by classical effects only. In fact, if
we allow for negative pressures to occur during collapse at a purely
classical level, we find immediately that the mass profile of the
collapsing star must be radiated away during the process.

\begin{figure}[hh]
\includegraphics[scale=1]{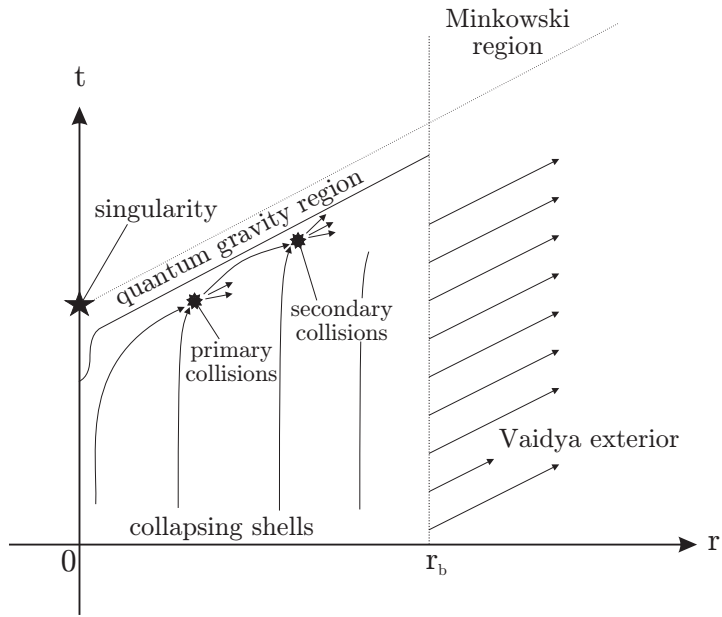}
\caption{The collapsing shells are repelled by the quantum gravity region.
 When the highly energetic particles traveling outwards collide with the particles
 from outer collapsing shells, fountains of collisions are created at arbitrarily
high center of mass energies.}
\label{collision}
\end{figure}

Negative pressures close to the formation of the singularity could
expel the inner shells whose particles would then collide with the
infalling outer shells with very high energies. Particle collisions
would then occur close to the Cauchy horizon with arbitrarily high
center of mass energy thus turning the collapsing cloud into an
immense particle accelerator (see Fig. \ref{collision}).

As an example we discuss here a model that was presented in Ref.
\refcite{PJM}
where a perfect fluid collapse was considered in spherical symmetry
with a matter function given by a perturbation of an homogeneous perfect
fluid close to the center as:
\begin{equation}
    M(r,v)=M_0(v)+M_2(v)r^2 ;\
\end{equation}
This corresponds to a variable relation between energy and pressure:
\begin{equation}\label{k}
    \frac{p(r,v)}{\rho(r,v)}=k(r,v) \; .
\end{equation}
Recall that in perfect fluid collapse if the equation of state is not
specified we have the freedom to choose one function globally to solve
the system of Einstein equations.
In this case, considering equation \eqref{k} as the equivalent of
an equation of state
we can take $k(r,v)$ as the free function that once chosen will specify
$M(r,v)$ via the differential equation \eqref{eos}.
Writing $k$ as
\begin{equation}
    k(r,v)=k_0(v)+k_2(v)r^2 \; ,
\end{equation}
and taking $k_0(v)$ and $k_2(v)$ as the free functions for the system
from the expansion of equation \eqref{eos} we obtain the two differential
equations
\begin{eqnarray}
  0 &=& 3k_{0}(v)M_{0}(v)+M_{0,v} v  \; , \\
  0 &=& 3k_2M_0 +5k_0M_2+M_{2,v}v+(1+k_0)M_{0,v}w_{,r}(0,v) \; ,
\end{eqnarray}
where $w(r,v)=v'(r,v(r,t))$.
The first one can be solved for $M_{0}(v)$ once
$k_{0}(v)$ is specified and this gives
\begin{equation}
    M_{0}(v)= C_{1}e^{-3\int_v^1{\frac{k_{0}}{v}dv}} \; .
\end{equation}


In the second equation we use the freedom to specify $k_{2}(v)$ and choose a
class that allows us to integrate the differential equation. Thus we define
\begin{equation}
    k_2=(1+k_0)k_0\frac{w_{,r}(0,v)}{v} \; ,
\end{equation}
and obtain
\begin{equation}
    M_{2}(v)=\tilde{C}_{2}e^{-5 \int_v^1{\frac{k_{0}}{v}dv}} \; .
\end{equation}

This solves the system up to second order in $r$ and $M(r,v)$ results
as follows,
\begin{equation}\label{mass}
    M(r,v)= C_{1}e^{-3\Phi(v)}\left[1+ r^2 C_{2}e^{-2\Phi(v)}\right] \; ,
\end{equation}
where we have set $\Phi(v)=\int_v^1{\frac{k_0(v)}{v}dv}$
and $C_2=\frac{\tilde{C}_2}{C_1}$.

We see that positivity of $C_1$ is enough to ensure positivity
of $M$ near the center, while $C_2<0$ would
imply that the density is decreasing radially outwards thus
imposing some conditions on the boundary of the
cloud once $C_2$ is set.

We can choose $k$ in such a way that at the beginning of
collapse it is positive and almost constant,
thus resembling a perfect fluid with linear equation of state.
Then negative pressures are triggered
close to the singularity by $k$ turning negative. This can easily
be done by suitably choosing the
terms in the expansion of $k$.

As an example consider $k_0(v)$ to be expanded also near the singularity as
\begin{equation}
    k_{0}(v)=k_{00}+k_{01}v+k_{02}v^2+... \; ,
\end{equation}
and for the sake of clarity all higher order terms are assumed to vanish.

Since the formation of trapped surfaces is governed by equation \eqref{ah}
and with the present choice of $k_{0}$ we get
\begin{equation}
  \frac{F}{R}=r^2C_{1} v^{-3k_{00}-1}
  e^{-3k_{01}v-\frac{3}{2}k_{02}v^2}
\left[ 1+r^2C_{2} v^{-2k_{00}} e^{-2k_{01}v-\frac{2}{2}k_{02}v^2}\right] \; ,
\end{equation}
it is easy to see that the requirement that $\frac{F}{R}<1$
near the singularity imposes that
$k_{00} \leq -\frac{1}{3}$, which is a condition for avoidance
of trapped surfaces. In this case not only the central shell
but also all other shells are not trapped when they become singular.
In fact the special case where $k_{00}=-\frac{1}{3}$ in the
limit of approach of singularity gives
$\frac{F}{R} \longrightarrow r^2 C_{1}$
which implies a finite radius for the trapped surfaces at all
times. Choosing the boundary so that
$r_{b}<\frac{1}{\sqrt{C_{1}}}$ is enough to avoid trapped
surfaces during the whole collapse.
On the other hand when $k_{00}<-\frac{1}{3}$ the radius of the
apparent horizon diverges at the singularity
thus indicating that to avoid trapped surfaces no restrictions
are imposed on the boundary.
Finally, as seen in section \ref{formalism}, for values of
$k_{00}>-\frac{1}{3}$ only the central
shell could be visible when becoming singular, depending
on the sign of $\chi_2(0)$, which in turn
would depend upon $C_2$.

The coefficients $k_{01}$ and $k_{02}$ can then be chosen
so that the pressure is positive at the initial time
(and for $v$ close to $1$), while it decreases during collapse
and becomes negative at the latest stages of collapse
eventually, thus leading to the formation of a naked singularity.

The presence of negative pressure requires that the function
$M$ must vanish at the singularity
which suggests that the whole mass of the core gets radiated
away during the last stages of collapse.
Therefore a finite amount of energy escapes from the
collapsing process. Furthermore the repulsive effects
near the singularity would cause the infalling shells to be
expelled and thus creating the conditions for particles
escaping the central region to collide with particles
of the infalling outer shells.

Particles escaping the vicinity of the singularity by
this mechanism would have to be intrinsically very energetic
since they originate in the
region where quantum gravity is dominant. These particles,
when traveling close to the Cauchy horizon would retain their
energy even at larger distances from the singularity as they
travel in the region of spacetime which can be described by a
metric which is Minkowskian with small perturbations.

\section{Current Status of Cosmic Censorship}
In view of the developments such as those discussed above,
it is natural to ask here what is the present status of the cosmic
censorship hypothesis, as implied by the current work on
gravitational collapse that has taken place in recent years.

\subsection{Does Cosmic Censorship hold?}
Firstly, we are compelled to ask at this stage the
following question: Is the cosmic censorship conjecture
correct or not as a basic principle of nature?
Clearly, the answer either way has profound implications for
fundamental physics and cosmology.

What comes out unambiguously
from the work so far is, censorship is certainly not correct in
an unqualified form, as it is sometimes taken to be.
It is quite clear by now from the work so far that singularities
appear in a wide variety
of forms and from a wide array of scenarios in general
relativity and to rule out all naked singularities at once
with a single theorem seems
at present not feasible. Therefore if cosmic censorship
holds, that will be in a highly refined and fine-tuned
form only, with suitable conditions and a formulation
that is yet to be achieved. The work on gravitational collapse
so far will play an important role here to achieve such a
formulation. On the other hand, investigating the quantum and
astrophysical processes in the vicinity of such visible
ultra-dense regions predicted by dynamical collapse models
can give rise to intriguing physical consequences
as we point out and discuss in the next section.

The implications of these developments and the
gravitational collapse studies for censorship are certainly
important. We can now say with confidence that one cannot
formulate censorship in a rather general way such as,
`Collapse of any massive star makes a black hole only', or,
`Any physically realistic gravitational collapse must end in
a black hole only', as there are now many counter-examples
to such statements. It follows that any formulation for
cosmic censorship must carefully specify when a black hole
will develop in collapse.
Specifically, one must examine the collapse scenarios carefully
and isolate the features that cause a naked singularity
to arise. Physicists believed for a long time that
spherical collapse must yield a black hole only, and naked
singularities arise if at all in non-spherical collapse
only. We now know this is clearly not the case. This explains
why earlier efforts to formulate a general theorem
for CCC failed.

The point, in other words is the following.
If one is considering the
gravitational collapse of a massive matter cloud, there will
be specific and fine-tuned conditions which one must set on
the initial data such as the matter densities and pressure profiles,
and the allowed dynamical evolutions of the Einstein equations,
so that the collapse endstate will be a black hole only.
The key aim such conditions or fine-tuning will achieve is that
these will ensure that in a continual general relativistic
collapse, when a spacetime singularity develops as collapse
endstate, the event horizon will develop necessarily prior
to the epoch of the occurrence of the singularity.

As an example of such a fine-tuning, we noted the example
of dust collapse. If
we require the density to be not increasing from the
center of the cloud outwards and we must get the black
hole as the collapse
endstate, then the density profile must be so fine-tuned
so that it is fully homogeneous at the initial epoch from where
the collapse develops, and then the velocity profile for the
collapsing shells is to be so tuned, so that the density
remains necessarily homogeneous at all later epochs. Only then
one gets a black hole final state. In all other cases,
a naked singularity develops as the final outcome of
collapse. Similar fine-tunings will be necessary for more general
collapse of different forms of matter, depending on the nature
of the stress-energy tensor and the equation of state it
follows.

Under such a scenario, a general statement for cosmic
censorship conjecture seems very difficult or nearly impossible
to formulate. The best one can do, probably is to specify a set
of conditions for a given collapse scenario so that it
terminates into a black hole. It would be then natural and
appropriate to accept that at the level of theory, both black holes
and naked singularities do occur as the final fate of a
continual gravitational collapse of a massive matter cloud
within the framework of general relativity.

\subsection{Why naked singularities form?}
It is natural then to ask here, what is really the physics
that causes a naked singularity to develop in collapse, rather
than a black hole. We need to know how particles and energy are
allowed to escape from extremely strong gravity fields.
We have examined this issue in some detail to bring out the
role of inhomogeneities and non-zero pressures.

In Newtonian gravity, it is only the matter density that
determines the gravitational field. In Einstein theory, however,
density is just one attribute of the overall gravitational field,
and the various quantities such as the space-time curvatures play
an equally important role in dictating what the overall nature
of the field is.

What our results show is that an inhomogeneous density profile
and the resulting shearing effects in the matter cloud, as implied
by the evolution of the cloud governed by the Einstein equations,
could delay the trapping of light and matter which can then
escape away even from very close to the singularity
(For a definition of shearing effects in matter fields,
see for example
\refcite{JDM}).

This is a general relativistic effect wherein even if the densities are
very high, inhomogeneity creates paths for light or matter to escape,
leading to a naked singularity rather than a black hole. If the
amount of inhomogeneity is very small, below a critical limit,
a black hole will form,
but with sufficient inhomogeneity, trapping of material is delayed
and a naked singularity arises.

While such a criticality can be clearly seen in the dust
models, it actually comes out very transparently in the models
where matter converts fully to radiation, originally constructed
in 1940s by Vaidya
\cite{Vaidya1,Vaidya2}
to model a radiating star. In these models, it is the rate
of collapse, that is how fast or slow the cloud is collapsing,
that decides between the formation of black hole and naked
singularity. For further details on these models
we refer to Ref.
\refcite{Global}.

Let us consider the physics that lead to naked singularity
formation in gravitational collapse. For transparency and clarity,
we will discuss a specific model depicting collapse of a dust
cloud in the absence of pressure, based on our general treatment
of this case (see Refs.
\refcite{DJ,JDM}).
The cloud begins collapse from a position of rest,
corresponding to the phase in a massive star's life when it
has exhausted its internal fuel and the gravity takes over.
This is a classical system governed by general relativity
with all physical regularity conditions being satisfied, such as
positivity of energy density and regularity of density and curvatures
at the initial epoch when the collapse begins. If the density
were taken to be completely homogeneous at the initial time,
this would be exactly the Oppenheimer-Snyder model, with the
collapse giving rise to a black hole. Let us now consider the
physically realistic situation, where the density of the star is
higher at the center and decreases as one moves away.

To determine the collapse end state, the trapping of light
and matter is to be understood as gravity fields become more and more
powerful, and the equation of the trapped region in spacetime is
determined, which decides the formation of the event horizon developing
dynamically as the collapse evolves. The key factor is the timing of
the horizon during collapse. If the horizon forms well before the
final singularity, the outcome is a black hole, but if it is delayed
as the collapse evolves, we have visible ultra-dense regions forming in
the universe. It turns out that sufficient inhomogeneity in the
density distribution at the initial time delays the horizon. If
the density decreases fast enough away from center of the star,
the final outcome is a naked singularity, but in a slow decrease
or nearly homogeneous case, a black hole results.

To understand the collapse evolution, see Figs. \ref{f:one} and \ref{f:two},
which correspond to the homogeneous and inhomogeneous cases respectively.
The arms of the cones denote paths of the ingoing and outgoing light
rays. In the homogeneous case, the collapse begins with regular
initial conditions where densities and curvatures are finite.
As the collapse progresses, these increase and the focusing effect
on light and matter grows. Then, there comes a time when a
region in the spacetime starts developing such that the light
from the same is simply unable to escape to any faraway
observer in the universe, but stays confined and trapped only in a
finite extent. Once light is trapped, the same always
happens for matter, namely for the timelike geodesics as well,
with its trajectory only within the cones. This corresponds
to the light rays from surface of the star at a phase when the radius
of the star has contracted to a distance proportional to its mass.
As the collapse progresses, since the density is only time dependent,
the entire cloud finally is crushed simultaneously to a singularity
in future. The trapping of light and matter, however, occurs well
before the final singularity developed, and therefore the singularity
is well-hidden inside the black hole that thus formed, with no light
signals or matter escaping from the ultra-dense regions near the
spacetime singularity.

The collapse, however, develops quite differently once density
is no longer homogeneous, as shown in Fig. \ref{f:two}.
The light paths are
shown as collapse progresses and as the star gets denser and denser.
The entire cloud now no longer collapses simultaneously to the
singularity as density is not homogeneous. The different matter
shells arrive at different times at the singularity, one after the
other with shells at larger radius coming later. As a result, at
no stage before the epoch of formation of singularity are the light
rays ever trapped, and rays and particles can come out in principle
from the super ultra-dense regions which are arbitrarily near to the
spacetime singularity. This lack of trapping happens due to a deficit
in the focusing effect of gravity on light and matter. Due to
inhomogeneity and decreasing density of the star away from center,
there is never enough total mass at any given epoch to cause full
light trapping, prior to the epoch of singularity formation.
We analyzed this general relativistic
effect of delay of horizon formation and related it to the spacetime
shear. Physically, it may correspond to creation of general
relativistic shocks due to inhomogeneity in such ultra-dense
regions which may allow for escape and ejection of light and
matter despite very high matter densities.

The general result that we have for this system is that two
parameters fully determine the evolution and the final fate of
collapse. These are the mass and velocity functions of the star,
specifying the total mass within a given area radius and the
velocities of the collapsing shells respectively at the initial epoch.
Generically, given any initial density distribution, there is a
non-zero-measure collection of velocity functions that take the star
to either a naked singularity or a black hole final state depending
on the choice made, which is freely available with all regularity
conditions being satisfied. This shows the genericity and stability
of the naked singularity within the given framework. Similar
features arise also when pressures and other physical equations
of state are incorporated for general collapsing matter fields,
as we noted earlier.

When a naked singularity develops, it is characterized by
three important attributes.
\begin{itemize}
\item[-]{First, there are families of infinitely
many light and particle trajectories coming out from the
singularity region.}
\item[-]{Second, it is a genuine curvature singularity in
that the densities and curvatures become infinite and grow
very powerfully in the limit of approach to the singularity
along these paths.}
\item[-]{Finally, the overall spacetime satisfies all physical
regularity conditions, so this becomes an interesting framework
to study actual physical processes in the ultra-strong
gravity regions.}
\end{itemize}

\subsection{Reformulate Cosmic Censorship?}
It is obvious that naked singularities are a general feature
arising in general relativistic gravitational collapse if we
do not impose any restrictions on the structure of the spacetime
or the energy-momentum.
Nevertheless it has become clear now that even standard physical
requirements such as regularity of the initial data or energy
conditions are not enough to guarantee the absence of formation
of strong curvature naked singularities as endstates of collapse.

Therefore, as a result of the gravitational collapse studies
carried out in the past years,
there have been some efforts to reformulate the
cosmic censorship conjecture.
As we noted earlier, the CCC does not hold in an unqualified
form within the framework of the Einstein gravity.

One would wish then to investigate the possibility that there
exist an alternative formulation of CCC that, while restricting
to a narrower array of scenarios (thus allowing for certain types
of naked singularities to occur), still can be proved mathematically.

One way here would be to specify a set of conditions
so that the evolving collapse from regular initial data for
a given matter field has the horizon developing necessarily
prior to the development of the singularity. While some general
indications are available here for spherical collapse
\cite{GJ}
a general formulation in this direction is not clear as yet.
It is clear that from an astrophysical perspective, such an effort
will be crucial to ensure the validity and applicability of
the black hole physics.
Another idea would be to consider cosmic censorship only for
certain specific matter models.
\cite{Rendall}

A still more radical proposal is to consider the cosmic
censorship in vacuum general relativity only where the
spacetime contains no matter fields at all. Alternatively,
one may allow for only selective `suitable' matter
fields, such a Maxwell field or massless scalar fields
only, and then ask whether the censorship will hold
\cite{Wald2}.
One would like to ask whether in a pure gravity framework,
without any matter fields, whether the cosmic censorship holds.
If this can be achieved, then the idea would be to disregard
any naked singularities arising in the gravitational collapse
of matter clouds as `singularities arising from matter' rather
than the pure gravity itself, which would obey the CCC.

There is, however, no essential progress in this direction
also so far as to how to formulate properly such a statement
in a mathematically rigorous manner, and then to proceed
to any possible proof for the same.

A similar idea, also suggested by Wald
\cite{Wald3}
is placed somehow in the middle.
This proposal claims that fluid models such as perfect
fluids and dust be
disregarded because they fail to give an account of the
microscopic properties of matter. The naked singularities
occurring in these collapse models would then be due to the lack
of a description of fundamental interactions that would resolve
them either at a classical or at a quantum level.

This idea is much in line with what we mentioned before when
we suggested that either classical effects or quantum
correction would intervene to resolve the singularity.
Nevertheless, considering only classical general relativity,
the only fundamental fields that can be studied in this respect
are the electromagnetic field and maybe scalar fields (and
naked singularities have been shown to occur in massless
scalar field collapse, see Ref.
\refcite{Chris2,Chop}, however, it is claimed that
these are `non-generic').

On the other hand, a full quantum treatment of the last stages
of collapse is missing at present, leaving the possible
reformulation of CCC only at the stage of a proposal.
The fundamental point here is that this formulation lacks
any connection with the issue of horizon formation.

In fact, if we agree to the idea that the singularities must be
resolved when we take into account the microscopic structure
of matter, then there is no need to conjecture that they can
arise in some cases, being hidden within an horizon, and not
in other cases, and the whole issue of cosmic censorship becomes
rather futile.

Another important point, from an astrophysical perspective,
is that, disregarding all the matter forms such as dust and
perfect fluids with different reasonable and well-motivated
equations of state, which have been widely used and studied
in astrophysical contexts will also be far from being widely
acceptable. In fact, collapse models with matter fields such
as these and others have been studied and applied in astrophysics
for decades, and to rule them all out just for the sake of
possible formulation of a hypothesis would not seem
to be reasonable.

The key point here is, in the end the physical problem
we need to study and investigate is that of the final fate
of a massive collapsing star when it shrinks gravitationally
at the end of its life cycle. The massless scalar fields
that are claimed to be `suitable' for the purpose of a
cosmic censorship statement are not observed in nature,
and it will be far from reasonable to assume or claim that
very massive stars are composed of or are dominated by
massless scalar fields only, even in the later stages of
their gravitational collapse.
In fact, matter fields such as perfect fluids, radiation
collapse models, and even dust collapse would be considered
quite useful and important to study this physical problem
of collapse of a star, rather than ruling them all out.
This is one of the main reasons why so many studies
have been already conducted for past many years for collapse
of matter clouds with forms of matter such as above in
the gravitation theory.

\section{Astrophysical and observational perspectives}\label{observational}
We have pointed out in the considerations here that the
final fate of gravitational collapse of a massive star continues
to be an exciting research frontier in the black hole
physics and gravitation theory today. The outcomes here
will be fundamentally important to the basic theory and
astrophysical applications of black hole physics, and for
modern gravitation physics. We highlighted certain key
challenges in the field, and also several recent interesting
developments were reviewed. Of course, by no means the issues
and the list given here are complete or exhaustive in any manner,
and there are several other interesting problems in
the field as well.

We like to mention here a few points which we think require
most immediate attention, and which will have possibly a maximum
impact on the future development in the field:

1. The genericity of the collapse outcomes, in terms of
black holes and naked singularities needs to be understood very
carefully and in further detail. It is by and large well-accepted
now, that the general theory of relativity does allow and
gives rise to both black holes and naked singularities as final
fate of a continual gravitational collapse, evolving from a
regular initial data, and under reasonable physical conditions.
What is not fully clear as yet is the distribution
of these outcomes in the space of all allowed outcomes
of collapse. The collapse models discussed above and considerations
we gave here would be of some help in this direction, and may
throw some light on the distribution of black holes and
naked singularity solutions in the initial data space.

2. Many of the models of gravitational collapse analyzed so
far are mainly of spherical symmetric collapse. Therefore, the
non-spherical collapse needs to be understood in a much better
manner. While there are some models which illustrate what
the departures from spherical symmetry could do (see e.g. Ref.
\refcite{JoshiKrolak96}),
some other analytical models of collapse for matter
clouds with cylindrical symmetry have also been studied (see Ref.
\refcite{Nolan}).
Though we note that on the whole, not very much is known
for non-spherical collapse. Probably numerical relativity could
be of help in this direction
See for example Ref.
\refcite{BaiottiRezzolla2006}
for a discussion on the evolving developments as related
to applications of numerical methods to gravitational collapse
issues. Also, other alternative would be to use global
methods to deal with the spacetime geometry involved, as used
in the case of singularity theorems in general relativity.

3. At the very least, the collapse models studied so
far do help us gain much insight into the structure of the
cosmic censorship, whatever final form it may have.

But on the other hand, there have also been attempts
where researchers have explored physical applications
and implications of the naked singularities investigated
so far (see e.g. Refs.
\refcite{Nakao2,Harada00,Harada01}
and also references in there).
If we could find astrophysical applications of the models
that predict naked singularities as collapse final fate, and
possibly try to test the same through observational methods
and the signatures predicted, that could offer a very
interesting avenue to get further insight into this
problem as a whole.

4. An attractive recent possibility in that connection
is to explore the naked singularities as possible particle
accelerators (see Refs.
\refcite{PJM,PJ10,PJ11a,PJ11b}).
	
Also, the accretion discs around a naked singularity,
wherein the matter particles are attracted towards or repulsed
away from the singularities with great velocities could provide
an excellent venue to test such effects and may lead to
predictions of important observational signatures to
distinguish the black holes and naked singularities in
astrophysical phenomena (see, e.g. Refs.
\refcite{Kovacs,Pugliese}).

5. Finally, further considerations on quantum gravity effects
in the vicinity of naked singularities, which are super-ultra-strong
gravity regions, could yield intriguing theoretical insights
into the phenomena of gravitational collapse (see Ref.
\refcite{JoshiGoswamiLQG}),
as we shall discuss below in Section \ref{QG} .

\subsection{Observable signatures of naked singularities}
As we have seen naked singularities that appear in exact solutions
of Einstein equations can take various different forms. It is then
reasonable to suppose that their observational features, if present,
might be various as well. If such objects do exist in the universe
it is therefore crucial to any astrophysical endeavour to study
how they might interact with the surrounding environment in order
to understand whether they can be observed and how.
\cite{Krolak-obs}
The key questions when considering observational features of
naked singularities therefore are:
\begin{itemize}
  \item[-]{Is there any observational signature coming out
from the vicinity of a singularity?}
  \item[-]{If there is, can we distinguish it from that of
some other astrophysical objects?}
\end{itemize}

As said, the singularity itself is not part of the spacetime,
therefore here we are considering the ultra-dense region surrounding
the singularity where relativity breaks down and quantum gravitational
effects (or effects coming from modified gravity) become relevant.
It would seem then that a theory of quantum gravity is needed in
order to resolve the quantities that diverge and to make predictions
on the observational signature of such objects. Nevertheless it is
possible that a naked singularity presents some observable features
already at a classical level, as it is also possible to study some
eventual expected effects in a classical relativistic framework.
Furthermore it is also possible that some modifications to general
relativity, though not including quantum effects, could provide
a framework where to understand better these extreme regimes and
make predictions on their observational features.

Roughly speaking then the observability of a naked singularity
might at present be studied by the use of three separate strategies:
\begin{itemize}
  \item[-]{Considering relativistic effects at extremely high densities.}
  \item[-]{Studying effects due to modified theories of gravity for
interiors or very strong gravitational fields.}
  \item[-]{Modeling possible quantum-gravity effects (either in
the semiclassical approximation or by adapting some theory of quantum
gravity, such as loop quantum gravity or string gravity,
to some toy models).}
\end{itemize}

So where could the observational signatures of naked singularities
lie? If we look for the sign of singularities such as the ones that
appear at the end of collapse, we have to consider explosive and high
energy events. In fact such models expose the ultra-high density
region at the time of formation of the singularity while the outer
shells are still falling towards the center. In such a case,
shockwaves emanating from the superdense region at scales smaller than
the Schwarzschild radius (that could be due to quantum effects or
repulsive classical effects) and collisions of particles near the
Cauchy horizon could have effects on the outer layers. These would be
considerably
different from those appearing during the formation of a black hole,
where the most dense regions are confined within the horizon and
thus unable to communicate with the exterior.

If, on the other hand, we consider singularities such as the
superspinning Kerr solution we can look for different kinds of
observational signatures. Among these the most prominent features deal
with the way the singularity could affect incoming particles, either
in the form of light bending (such as in gravitational lensing),
particle collisions close to the singularity, or properties of
accretion disks (as it will be shown in next section).

Within this class of `long lived' singularities there is another
kind which presents intriguing possibilities, namely interior
solutions that describe a regular source with a singular center.
This is an entirely different kind of final state, as opposed to
the vacuum solutions such as the Kerr metric, in the fact that it
involves a finite matter cloud with a boundary larger than the
horizon and which presents a singularity at its center.
Interior solutions of Einstein field equations have been considered
for decades as sources of the gravitational field. One key requirement
in building such interiors used to be that the matter density should
be regular all the way to the center of the source. Nevertheless
there is no reason to impose that a stable configuration cannot evolve
which presents a singularity at the center. These singularities, once
again, will model the region of arbitrarily high density that develops
at the center of the object and where gravity exhibits extreme
properties as outlined before. It is then possible that such a source
could be observationally different from a corresponding source with
regular interior or from a vacuum solution such as a black hole.
Furthermore if the overall density of the cloud is low enough, as
it is the case for supermassive objects like the ones residing at
the center of galaxies, then processes happening close to the center,
such as particle collisions and lensing, could be visible to
faraway observers.

In fact one key point when dealing with these models is understanding
what kind of phenomena we are trying to describe. Namely, what kind of
singularity are we looking at? Within gravitational collapse many
different phenomena can be described. Of course there are the smaller,
explosive, short lived phenomena like the collapse of the core of a
star, which has been the main topic of this review. These will
typically involve very high energies and very short time scales. On
the other hand, the fairly recent discovery of active galactic nuclei
has led astronomers to suppose that objects like supermassive black
holes exist at the center of galaxies. These objects are still not
very well understood today and it is indeed very plausible that
processes such as gravitational collapse, on a much larger scale and
on much longer time scales, are crucial to their development.

\subsection{Can we test censorship using Astronomical Observations?}
With so many high technology power missions to observe the
cosmos, can we not just observe the skies carefully to determine
the validity or otherwise of the cosmic censorship?

To answer this question, first of all we need once again to
distinguish the two main candidates for naked singularities, which
are very different in nature and which would bear very different
observational signatures. The first candidate, that has been widely
discussed above, is the naked singularity that results at the end
of spherical collapse. When we are dealing with collapse of a star
such a naked singularity will happen at the center of the collapsing
cloud and it will most likely eventually be covered by the event
horizon when collapse ends. Therefore the observational signature of
such an event must be in the form of a short-lived explosive event.
Similar phenomena are observed in the universe, for example
gamma-ray bursts that are believed to originate from core collapse
of massive stars. But a link between such events and the possible
existence of a naked singularity at the core of the collapse has
never been thoroughly investigated.

On the other side, we have scenarios such as the super-spinning
Kerr solution, or some axially symmetric vacuum metrics, that are
derived from exact solutions of Einstein's field equations and
which generally present naked singularities. We still do not know if
such configurations could arise from a dynamical process such as
collapse. Therefore it is legitimate to ask whether these metrics
could represent some real object existing in the universe. The
question of how these kind of objects could form from the collapse
of an initially regular star is still unanswered and very little
is known on the behaviour of a realistic source during the final
stages of collapse. Therefore the fact that such exact solutions could
or could not arise from the evolution of a regular matter source
remains in the domain of speculation. Nevertheless recently scientists
have turned their attention to these vacuum solutions in order to
understand what observational properties they would have and in
order to see if they could be detected by current observations.

The Kerr metric is the monopole solution for a source with
angular momentum. In this case the key quantity that decides the
nature of a compact object or the super massive central object
at the center in a galaxy,
is the ratio of its mass and the spin angular momentum per unit
mass for the same, $a=J/M^2$. For a rotating object described by the
Kerr metric, if the angular momentum to mass ratio is smaller
than one, final state of collapse is a black hole, but for
a larger ratio it is a naked singularity.
In principle such a Kerr naked singularity could form from
three different processes. The first one is the complete collapse
of a rotating star with
mass exceeding the neutron star limit. The possibility that collapse of
such an object forms a superspinning Kerr has been suggested in Ref.
\refcite{Horava}
and investigated in some numerical simulations (see Ref.
\refcite{Rezzolla05}).
The other possibilities by which a Kerr naked singularity can form
are given by the inflow of angular momentum due to accreting particles
and the merger of rapidly spinning compact objects
(see section \ref{numerical} for some references on numerical simulations
of these processes).

A number of proposals to measure the mass and spin ratio
for compact objects and for the galactic center have been made by
different authors. Krolak (see Ref. 123),
suggested using pulsar observations, gravitational waves and
the spectra of X-rays binaries to test $a$ for the center of
our galaxy, Maeda
\cite{Hioki},
proposed using the shadow cast by the compact object to test
$a$ in stellar mass objects, Takahashi and others
\cite{Takahashi},
have suggested using the X-ray energy spectrum emitted by the
accretion disk, and finally Werner and Petters
\cite{Werner}
suggested using certain observable properties of gravitational
lensing that depend upon $a$.

So could such naked singularities form in some realistic
physical processes?
>From the measurement of X-ray binary systems there are indications
that black holes with a mass to spin ratio very close to unity exist
in the universe (see for example Ref.
\refcite{McClintock}).
Furthermore compact objects such as neutron stars with $a$ greater
or smaller than one can exist.
Therefore, given the observational information that we have on
the mass to angular momentum ratio for these compact objects,
it is interesting to study how the parameter $a$ which determines
the separation of the black hole Kerr spacetime from the Kerr naked
singularity, is affected when the object undergoes an inflow of mass
and angular momentum due to an accretion disk.

The inflow of angular momentum due to accretion disks
has been studied in order to understand if the particles falling onto
a black hole can contribute to speed up the angular
momentum, thus reaching the critical limit and removing the horizon.
The idea of overspinning a black hole traces back to a thought experiment
proposed by Wald
\cite{Wald1},
and it has important implications for cosmic censorship (for details see
section \ref{QG}).

>From the astrophysical perspective these examples represent
one possible way by which a Kerr naked singularity could be obtained.
If this process proved to be physically viable then the investigation of
observational properties of the Kerr naked singularities would become
crucial for astrophysics.
On the same line, the inverse procedure has also been investigated.
This is the suggestion that the process of slowing down a Kerr naked
singularity to a black hole, by means of infalling counter-rotating
particles has a greater efficiency than the opposite process.
\cite{Stuchlik}

Furthermore some studies have been carried out in order to
understand how the
process of angular momentum transfer works during the merger of two compact
objects (such as neutron stars) with high spin. If the angular momentum is
not dissipated away during the merger, these events could give rise to a final
configuration in the form of a super-spinning Kerr naked singularity.
Numerical simulations have shown that the `sub-Kerr' models are more likely
to be unstable and merge into a Kerr black hole than the corresponding
super-spinning ones.
\cite{Rezzolla11}

Of course the Kerr metric is not the only vacuum exact solution with angular
momentum. The study of exact solutions with higher multipole moments is
connected with the issue of the no hair theorems and the formation of similar
final configurations is speculative just as much as that of the Kerr solution
(see  Refs.
\refcite{Herrera1,Herrera2}).
There have been some attempts to study observational properties of metrics with
higher multipole moments (see e.g. Ref.
\refcite{Bambi10}),
but at present the fact that rotating compact
objects in the universe are represented by the Kerr metric or by some more
complicated axially symmetric spacetime remains unclear (see Ref.
\refcite{Bambi11a} for galactic nuclei and Ref.
\refcite{Bambi11b} for stellar mass objects).
Nevertheless such metrics have received some attention in recent
times and some attempt at answering the question whether they might
bear an observational signature
different from that of a Kerr metric has been made (see Ref.
\refcite{Bambi11c}).

The basic issue here is that of sensitivity, namely how
accurately and precisely can we measure and determine these
parameters. A number of present and future astronomical
missions could be of help. One of these is the Square-Kilometer
Array (SKA) radio telescope, which will offer a possibility
here, with a collecting area exceeding a factor of hundred
compared to existing ones. The SKA astronomers point out they
will have the sensitivity desired to measure the required quantities
very precisely to determine the vital fundamental issues in
gravitation physics such as the cosmic censorship, and
to decide on its validity or otherwise.

Other missions that could in principle provide a huge
amount of observational data are those that are currently hunting
for the gravitational waves. Gravitational wave astronomy
has yet to claim its first detection of waves, nevertheless
in the coming years it is very likely that the first
observations will be made by the experiments
such as LIGO and VIRGO that are currently still below
the threshold for observation. Then gravitational wave
astronomy will become an active field with possibly large
amounts of data to be checked against theoretical
predictions and it appears almost certain that this
will have a strong impact on open theoretical issues
such as the Cosmic Censorship problem
\cite{LIGO}.

If we see objects with
sufficiently high angular momentum compared to their mass,
than the motivation for expecting the existence of an event
horizon in collapse will reduce drastically, and we may
be able to answer if naked singularities exist in the
astrophysical reality. The prime survey targets here will
include the innermost regions of our Galaxy, and globular
clusters which are vast collections of oldest stars.

\subsection{Distinguishing black holes and naked singularities}
While gravitational collapse has been investigated
extensively over past decades within the framework of
gravitation theory, not much is known today about the observable
signatures that would distinguish the black holes from
the naked singularities, which are hypothetical astrophysical
objects in nature predicted by the gravitation theory. We discussed
and reviewed above first the existence, and then the genericity
and stability of occurrence of naked singularities in a
gravitational collapse.

The question of what observational signatures would then
emerge and distinguish the black holes from naked singularities
is then necessary to be investigated, and we must explore
what special astrophysical consequences the latter may have.

With respect to the singularities that result at the end
of a gravitational collapse, and which might have some observational
signatures in the form of an explosive event, at present, we do
not have the tools to say, even in principle, if they would look
different from an explosive event originated from a collapse
scenario where the central ultra-high dense region is
trapped within an horizon.

One can argue that since gravity is attractive and nothing
opposes the gravity at the very last stages of collapse, therefore
nothing could in principle come out of the singular central region
that develops at the end of collapse.
Furthermore the core of the collapsing star would be so immensely dense
that the mean free path of a particle trying to escape would be
extremely short, thus making it unlikely that anything can
escape from that region.
On the other hand, we have seen that particle collisions
close to the Cauchy horizon can happen under suitable conditions
and that these can bear the signature of the singularity.

Furthermore at present we are not able to consider quantum effects
that will appear at the core when the density reaches critical values.
There are indications that these corrections will give rise to
strong negative pressures that can disrupt collapse and create a
shockwave that might propagate to the outer shells, thus dissipating
away the mass of the star.

Finally, we have noted that simulations of core collapse supernovae
so far have not been able to duplicate what happens at the last stages
of collapse, and that in the computer simulation models some energy
is missing in such a way that the explosion cannot in fact take
place. Therefore it is possible, in principle, that this energy
could be provided by some shockwave emanating from the central
non-trapped ultra dense region, and that naked singularities would
be in fact the fuel of type II supernovae.

At present this is a matter of speculation and if we wish to find a
way to distinguish the black holes from naked singularities, it
might help to consider axially symmetric naked singular solutions
such as the Kerr spacetime. However, we do not know at present if they
can actually be obtained from physical dynamical situations such
as collapse, accretion, or merger.

Therefore, assuming that a Kerr naked singularity or some other
object with a visible singularity exists in the universe,
the question is: How can we distinguish it from a black hole or
another compact object of the same mass?

In theory there are three different kinds of observations that one could
devise in order to distinguish a naked singularity from a black hole.
The first one relies on the study of accretion disks. It has been
shown that the accretion
properties of particles falling onto a naked singularity would be very
different from those of black hole of the same mass (see for example Refs.
\refcite{Pugliese,JMN}),
and the resulting accretion disks would also be observationally different.
In fact the properties of accretion disks have been studied
in terms of the radiant energy, flux and luminosity, in a Kerr-like
geometry with a naked singularity (see Ref.
\refcite{Kovacs}),
and the differences from a black hole accretion
disk have been investigated.
Also, the presence of a naked singularity gives rise to powerful
repulsive forces that create an outflow of particles from the
accretion disk on the equatorial plane.
This outflow that is otherwise not present in the black hole case,
could be in principle distinguished from the jets of particles
that are thought to be ejected from black hole's polar region and
which are due to strong electromagnetic
fields (see Ref.
\refcite{Bambi09,Bambi10b}).
Furthermore, when charged test particles are considered
the accretion disk's properties
for the naked singularity present in the Reissner-Nordstrom spacetime
have been shown to be observationally different from those
of black holes (see Refs.
\refcite{PuglieseRN,PJ10}).

The second way of distinguishing black holes from naked singularities relies
on gravitational lensing. It has been argued that when the spacetime does not
possess a photon sphere, then the lensing features of light passing close to
the singularity will be observationally different from those of a black hole
(see Ref.
\refcite{Virbhadra}).
This method, however, does not appear to be very effective when a
photon sphere is present in the spacetime.
Assuming that a Kerr-like solution of Einstein equations with
massless scalar field exists at the center of galaxies,
its lensing properties were studied
and it was found that there are effects due to the presence of
both the rotation and scalar field that would affect the behavior of
the bending angle of the light ray, thus making those objects
observationally different from black holes (see Ref.
\refcite{Gyulchev}).

Finally, a third way of distinguishing black holes from naked singularities
comes from particle collisions and particle acceleration in the vicinity of
the singularity. In fact, it is possible that the repulsive effects due to
the singularity can deviate a class of infalling particles, making
these outgoing eventually. These could then collide with some
ingoing particle, and the energy of collision
could be arbitrarily high, depending on the impact parameter of the
outgoing particle with respect to the singularity. The net effect is
thus creating a very high energy collision that resembles that of
an immense particle accelerator and that would be impossible in
the vicinity of a Kerr black hole.
\cite{PJ11a,PJ11b}

Black hole candidates from observations belong to two classes. There are
stellar mass black holes that are thought to form either from collapse of
very massive stars or from merger and accretion processes.
In the cases when these candidates have an orbiting companion, we can
measure their mass from the orbital periods of the companion.
Similar measurement can be made when the black hole candidate is
surrounded by a gas disk.
Then there are supermassive black holes that are thought to exist
at the center of galaxies and whose total mass, that can be estimated
from the orbital period of nearby stars, is measured in
millions of solar masses.

For both kinds of black hole candidates the angular momentum
can be inferred from the analysis of the observed X-ray emission,
though this measurement relies on
some assumptions and it is not clear whether Kerr naked singularities
could present a similar spectrum (see Ref.
\refcite{Takahashi}).

The theoretical models described above could then provide some framework where
to test observations coming from stellar and supermassive  black hole
candidates.
In fact, some observations in the millimeter wavelength of the supermassive
object at the galactic center already suggests that the Kerr limit might be
broken (see for example Ref.
\refcite{Doeleman}).
>From investigations such as these, we see that naked singular
spacetimes cannot be ruled out with the present knowledge that we
have of these sources.

Of course, the Kerr naked singularity is not the only possible
alternative to a black hole.
As we mentioned before, naked singularities with non-vanishing higher
multipole moments can be considered, but also other more exotic objects
can provide some insight on the nature of these compact objects.
In this connection, compact exotic objects such as boson stars
or gravastars have been also considered (see Ref.
\refcite{Visser}
and references therein), though they could be unlikely
candidates due either to
the absence of evidence that the matter required to create
such configurations can exist,
or to the possibility that some dynamical instability
against perturbations can arise for certain matter models,
though these could be stable to perturbations for certain
physical choices of parameters (see Ref.
\refcite{Visser04} and references therein for
a further discussion). In general, the
issues of such stabilities are complex and
need much more careful attention.

Nevertheless, equilibrium configurations of regular
matter sustained by pressures are predicted by general
relativity, and these could also provide an alternative
to model black hole candidates such as the
galactic center, as we discuss next.

\subsection{Equilibrium configuration from collapse}
Here we will review a class of solutions of Einstein equations
with naked singularities that arise asymptotically for suitable matter
models when the pressures opposing collapse manage to halt the process,
thus originating an equilibrium configuration
(for details see Ref.
\refcite{JMN}).

This class of models describes how an initially regular matter
cloud can evolve to form an equilibrium configuration with a finite
radius slightly larger than the Schwarzschild radius and with a naked
singularity at the center. This process can require arbitrarily long
times for the equilibrium configuration to be finally achieved, and
the average density of the cloud towards the final equilibrium can
remain small, therefore suggesting that such a model could describe
the formation of supermassive objects other than a black hole that
could delve at the center of galaxies. Such an object, if it actually
existed, would bear some different properties for the surrounding
accretion disk as opposed to those of the accretion disk around a
black hole, thus making it in principle observationally
distinguishable.

\begin{figure}[hh]
\includegraphics[scale=0.75]{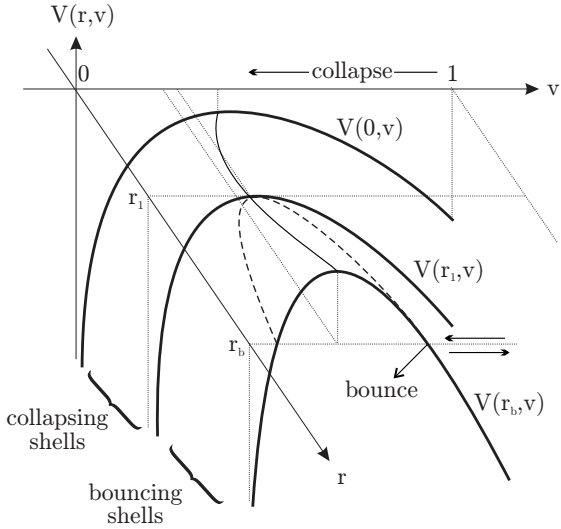}
\caption{The effective potential for a general matter cloud.
The inner shells collapse to the final singularity which may be covered
by an horizon or not. The outer shells halt and bounce back at a finite time.
There is one limiting shell ($r_1$) for which collapse
halts without bouncing back.}
\label{potential}
\end{figure}

As seen in section \ref{formalism}, the system of Einstein equations for
a collapsing cloud can be reduced to solution of one first order
partial differential equation acting as an equation of motion. We can
then rewrite the equation in the form of an effective potential
for any fixed shell $r$ as,
\begin{equation}
     V(r,v)=-\dot{v}^2=-e^{2\nu}\left(\frac{M}{v}+\frac{G-1}{r^2}\right) \; .
\end{equation}
As mentioned earlier, complete gravitational collapse is obtained
when we require $\dot{v}<0$ for all shells at all times.
However, starting with a collapsing configuration, every shell
will have in general three possible evolutions:
\begin{enumerate}
  \item $\dot{v}<0$ at all times. Then the shell collapses
to the singularity as described earlier.
  \item $\dot{v}=0$ at a certain $\bar{t}$, while $\ddot{v}\neq 0$.
In this case the shell labeled by $r$ halts its collapse and bounces
back re-expanding.
  \item $\dot{v}=\ddot{v}=0$ at a certain $\bar{t}$. In this case
collapse slows down until the shell reaches an equilibrium configuration.
\end{enumerate}
The most general scenario for a collapsing object therefore has
the inner core subject to continuous collapse while the outer shell
can halt and bounce back (see Fig. \ref{potential}).

\begin{figure}[hh]
\includegraphics[scale=0.75]{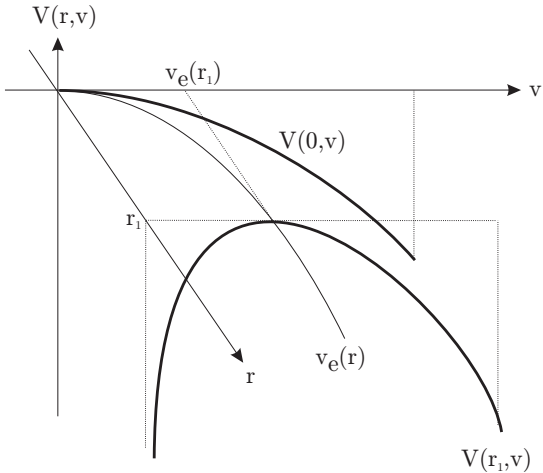}
\caption{The effective potential for a matter cloud originating
an equilibrium configuration. All shells reach a stop where collapse halts
(like for the limiting shell in the general case) and the cloud settles
to a static configuration.}\label{potential2}
\end{figure}

It is easy to see that the parameters describing the matter
content of the cloud can be suitably chosen in such a way that every
shell stops collapsing, reaching an equilibrium state given by
$\dot{v}=\ddot{v}=0$ as in the third situation mentioned above. This
is in fact the case when the effective potential $V(r,v)$ has an
extremum at zero for every value of $r$. What is found is that in this
case the cloud slows down and collapse halts giving rise to an
equilibrium configuration that is described by an interior metric for
the Schwarzschild spacetime. The final equilibrium configuration
is given by $v_e(r)$ such that $V(r,v_e(r))=0$ (see Fig. \ref{potential2}).

>From the usual analysis of dynamical systems close to equilibrium
we can see that the time needed to stop collapse turns out to be
infinite and therefore the equilibrium configuration is achieved only
asymptotically. This fact, together with the fine tuning necessary to
halt the collapse for every shell, means that such a configuration
must not be taken as a serious candidate for a realistic object.
Nevertheless we note that any collapsing configuration arbitrarily
close to the equilibrium solution will have a continuous collapse that
is arbitrarily slow. In this sense the asymptotic limit of the
equilibrium configuration represents an approximation of this
physically valid, slowly evolving scenario. It is then interesting
to study the properties of such static interiors that approximate a
dynamical scenario. Since we know that collapse of the core of a
massive star happens within a very short time, these models probably
cannot be used to describe a similar situation. Nevertheless we know
that immensely massive compact objects exist at the center of
galaxies. These objects are long lived and compact enough to be
possibly represented by such models.

Further, we note that as the equilibrium configuration is achieved
from a matter distribution with regular initial data, we need not
impose that the central shell be regular (as it is often done when
looking for interior solutions for Schwarzschild). A singular
central shell, where the density diverges, can also arise from such
a collapse. In this case the central shell would represent a region
of the compact object where quantum gravity takes hold and the
observational features of such an object might be considerably
different from those or a regular compact source or a black hole.

Models of these kind, where a collapsing configuration approaches
a static limit, are possible for any kind of matter cloud with
pressures. It is easy to see that within the framework of dust
collapse there is no way for the particles to counteract the pull of
gravity thus halting collapse, and therefore a static configuration
is not possible. On the other hand if we include pressures, for
example in the form of tangential pressures or perfect fluids, it
is possible for such pressures to balance the attraction of gravity,
thus generating a static configuration.

As an illustrative example we shall consider here the case of a
cloud sustained only by tangential pressures. This choice will
simplify considerably the structure of Einstein equations and it
has its own interesting physical features. In fact a cloud composed
of counter-rotating particles that move on circular orbits around
the center is well approximated by the models with only tangential
pressures. The choice of such a matter model simplifies considerably
the field equations, since the amount of matter contained within
any shell labeled by $r$ is conserved, thus implying that $M=M(r)$.
Furthermore from Einstein equations we know that we have the freedom
to choose two free functions, namely the mass profile and the
pressure profile (see section \ref{formalism}).

Therefore we can choose $M(r)$ and $p_\theta$ at equilibrium
by imposing the equilibrium conditions,
\begin{equation}\label{static}
    V=V_{,v}=0,
\end{equation}
with
\begin{equation}
    V_{,v}=e^{2\nu}\left(\frac{M}{v^2}-\frac{G_{,v}}{r^2}\right)
    -2\nu_{,v}e^{2\nu}\left(\frac{M}{v}+\frac{G-1}{r^2}\right) \; .
\end{equation}
In fact from these two equations, via Einstein equation \eqref{G}
we get the two equations that
fix the $G$ and $G_{,v}$ at equilibrium in terms of the equilibrium
solution $v_e(r)$ as,
\begin{eqnarray}\label{ve}
  G_{e}(r)&=&G(r, v_e(r))=1-\frac{r^2M(r)}{v_e(r)} \; , \\ \label{M}
  (G_{,v})_e &=&G_{,v}(r, v_e(r)) = \frac{M(r)r^2}{v_e^2} \; ,
\end{eqnarray}
Then from equations \eqref{rho} and \eqref{nu} evaluated at equilibrium,
we obtain the energy density and tangential pressure at equilibrium as
\begin{eqnarray}
\rho_e(r)&=&\frac{3M+rM'}{v_e^2(v_e+rv_e')} \; , \\
p_{\theta e}(r)&=& \frac{1}{4}\rho_e v_e \frac{(G_{,v})_e}{G_e}
=\frac{1}{4}\frac{r^2M(3M+rM')}{v_e^2(v_e+rv_e')(v_e-r^2M)} \; .
\end{eqnarray}
On the other hand, once we have chosen $M(r)$ and $v_e(r)$,
these two functions entirely determine the other quantities at
equilibrium, namely $\rho_e(r),
G_e(r), (G,_v)_e(r)$ and $p_{\theta e}(r)$.  Though we still have
the freedom of two free functions during the evolution, we can see
that one, namely the mass function $M(r)$, we have already chosen.
Therefore, in order to achieve
the desired final configuration, the class of allowed pressures
$p_\theta(r,v)$ needs to be chosen in such a way that
$p_\theta(r,v) \to p_{\theta e}(r)$ as $t\to\infty$, where the final
pressure $p_{\theta e}(r)$ is determined by the choice of $v_e(r)$
as explained above. For this entire class of pressures the dynamical
gravitational collapse will necessarily tend
asymptotically to the static geometry specified by $M(r)$
and $v_e(r)$. As said the above equilibrium configuration need
not be necessarily regular at the center.
In fact since $M$ is finite at $r=0$ we see that whenever we
have $v_e(0)=0$, the energy density
at equilibrium diverges at $r=0$ and the final state presents
a central singularity which has been
obtained as the result of collapse from regular initial data.

\begin{figure}[hh]
\includegraphics[scale=0.75]{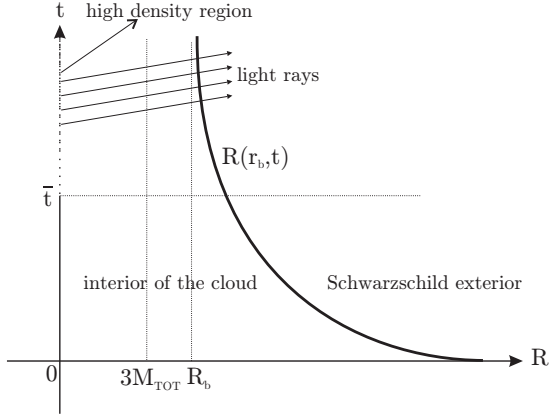}
\caption{The static configuration is achieved asymptotically
and settles to a final radius greater than the photon sphere
for the Schwarzschild black hole. The central density becomes
arbitrarily large and approaches a singularity.}
\label{equilibrium}
\end{figure}

Static interior solutions of the Schwarzschild spacetime sustained
only by tangential pressures were studied in past
\cite{florides},
and the most general metric in this case is easily written.
Therefore the class of equilibrium configurations described above must belong
to this family of static metrics.
The physical importance of these equilibrium configurations comes
from the fact that for sufficiently
large values of $t$ the static models present an arbitrarily
close approximation of the slowly collapsing cloud
(see Fig. \ref{equilibrium}).
Furthermore the ultra-high density region that develops at the
center of the cloud is always visible. It is in this region that
classical general relativity may eventually break down when densities
high enough are reached. These densities are always obtained in
a very large but finite time, before the actual mathematical
singularity occurs, and it is the visibility of such regions
during collapse which is the main reason for the study of the
physical properties of such objects.

With the aim of studying the physical properties of these objects,
we studied
a specific equilibrium configuration given by $F(r)=M_0r^3$ and
$v_e(r) = cr^\alpha$, where, for the sake of clarity, we have imposed
the scaling of the boundary in such a way that $c=1$ and we have
taken $\alpha>0$.  The case $\alpha=0$ corresponds to a regular
solution with positive constant density.
Density and pressure at equilibrium become
\begin{eqnarray}
  \rho_e &=& \frac{3M_0}{(\alpha+1)}\frac{1}{R^{\frac{3\alpha}{\alpha+1}}} \; ,\\
  p_{\theta e} &=& \frac{3M_0^2}{4(\alpha
    +1)}\frac{R^{\frac{2-4\alpha}
{\alpha+1}}}{\left(1-M_0R^{\frac{2-\alpha}{\alpha+1}}\right)} \; .
\end{eqnarray}
>From the above we see that different values of $\alpha$ lead to different
profiles for the pressure as $r\rightarrow 0$. In fact for $\alpha<1/2$
we have $p_{\theta e}\rightarrow 0$, while for $\alpha>1/2$ we have
$p_{\theta e}\rightarrow +\infty$.

As an example we then considered the case $\alpha=2$ and studied
the motion of test particles in circular
orbits in the equatorial plane and the properties of accretion disks.
The energy density in this case is given by $\rho_e = M_0/R^2$ while
the pressure satisfies a linear equation of state, $p_{\theta e}=k\rho_e$, with
$4k=M_0/(1-M_0)$.
>From the Misner-Sharp mass at the boundary we get the condition
$2\mathbf{M_{TOT}}/R_b = M_0$ and the condition to
avoid an event horizon is given by $M_0<1$.  Furthermore,
the weak energy condition holds when $k\geq -1$,
(corresponding to $M_0\leq 4/3$) and the effective sound
speed $c_\theta$, given by $c_\theta^2 = p_{\theta e}/\rho_e
= k$, is less than unity if $M_0 < 4/5$.

Outgoing radial null geodesics in this spacetime are given by,
\begin{equation}
\frac{dR}{dt}=(1-M_0)\left(\frac{R}{R_b}\right)^{\frac{M_0}{2(1-M_0)}} \; .
\end{equation}
and there are light rays escaping the singularity for all values of $M_0<2/3$
since the comoving time required by a photon
to reach the boundary can be evaluated as $t_b=2R_b/(2-3M_0)<+\infty$.
Furthermore the Kretschmann scalar for this naked
singularity model is given by
\begin{equation}
    K = \frac{1}{4} \frac{M_0^2(28-60M_0+33M_0^2)}{(M_0-1)^2 R^4} \; .
\end{equation}
thus indicating the presence of a strong central singularity.

Evaluating the two conserved quantities, we got energy
per unit mass, $E= u_t = e^{2\nu} (dt/d\tau)$ and angular momentum
per unit mass, $\ell = u_\phi = R^2(d\phi/d\tau)$.
For circular orbits, we set $dR/d\tau=0$, so we require the remaining
terms in the above equation to add up to zero. In addition, their sum
should achieve an extremum at radius $R$.  This gives the two
conditions
\begin{eqnarray}\label{E2}
  E^2 &=& \frac{2(1-M_0)^2}{(2-3M_0)}\left(\frac{R}{R_b}
\right)^{M_0/(1-M_0)} \; ,\\ \label{L2}
  \frac{\ell^2}{R_b^2} &=& \frac{M_0}{(2-3M_0)}\left(\frac{R}{R_b}\right)^2 \; .
\end{eqnarray}
The normalization condition $u^\alpha u_\alpha=-1$ then gives
\begin{equation}
\frac{1}{G}\left(\frac{dR}{d\tau}\right)^2-E^2e^{-2\nu}
+\left(1+\frac{\ell^2}{R^2}\right)=0 \; . \label{circ}
\end{equation}
where $R_b$ is the physical radius corresponding to the boundary of
the matter cloud in the final equilibrium state.
Note that both the energy per unit mass $E$
and the angular momentum per unit mass $\ell$ vanish in
the limit of $R=0$, meaning that no mass or rotation is added
to the central singularity by the accretion disk.
The process of accretion does not affect the naked singularity,
which can be considered stable in this respect.
If we want the circular orbit to be stable, we must
require that the term involving $E^2$ be less divergent as $R\to0$
as compared to the term involving $\ell^2$.
This then gives the following results for $R<R_b$:
\begin{eqnarray}
{\rm Stable~circular~orbits:}&~&\qquad M_0\leq 2/3, \\
{\rm Unstable~circular~orbits:}&~&\qquad M_0> 2/3.
\end{eqnarray}
Depending on the value of $M_0$, either all circular
orbits in the interior of this object are stable,
or all are unstable. Note that, apart from having unstable
circular orbits, models with $M_0>2/3$ also give
negative values of $E^2$ and $\ell^2$.

It is known that a standard thin accretion disk can exist only
at those radii where stable circular orbits are available.
\cite{NovikovThorne,Page}
Therefore for a Schwarzschild black hole of mass
$\mathbf{M_{TOT}}=M_0R_b/2$, an accretion disk will have its inner
edge at $R=6\mathbf{M_{TOT}}$, and inside this radius
the gas plunges or free-falls.

Therefore depending on where we perform the matching with the
exterior Schwarzschild solution we have two possible scenarios:
\begin{itemize}
  \item[1-] $M_0\leq1/3$: In this case, the external Schwarzschild
metric has stable circular orbits all the way to the boundary $R=R_b$
where it meets the interior solution which again allows for stable
circular orbits down to $R=0$. The accretion disk will continue
into the interior extending to $R=0$, without an inner edge.
Assuming the matter cloud is transparent to radiation (as it is
reasonable since in order to have an accretion disk the matter in the
cloud must be weakly interacting), faraway observers will receive
radiation coming from all radii until the center. The observed
spectrum will obviously be very different from that of a disk
surrounding a black hole of the same mass.
  \item[2-] $1/3<M_0\leq2/3$: In this case, an accretion disk
will follow the Novikov-Thorne solution until $R=6\mathbf{M_{TOT}}$
where the gas will plunge towards smaller radii. Once the gas reaches
the boundary of the interior solution at $R=R_b$, circular orbits
are once again allowed and the gas will shock and circularize to continue
accreting on circular orbits all the way to $R=0$. Again, since the
accretion disk in this model consists of two distinct segments with a
radial gap in between, we expect it to be observationally
distinguishable from the previous case and from a black hole.
\end{itemize}

The interesting point is that these naked singularity models are easily
distinguishable from a black hole of the same mass.
Therefore it could be useful and interesting to use observational data on
astrophysical black hole candidates to test if the presence
of a similar naked singularity is possible.

\section{Cosmic puzzles and the new perspective}
We thus find from the detailed collapse calculations of recent
years that the final fate of a collapsing star could be a naked
singularity when the initial data is appropriate. Apart from
its physical relevance, this violation of censorship also
has profound philosophical aspects, such as the issue of
predictability in the universe, the questions related to the
nature and structure of singularities, and on possible validity
or otherwise of classical gravity description in the vicinity of
a naked singularity.

It is sometimes argued that breakdown of censorship means
violation of predictability in spacetime,
since the presence of naked singularities, together with geodesic
incompleteness does not allow to uniquely predict the
future evolution of the spacetime.
Therefore
we have no direct handle to know what a
naked singularity may radiate and emit unless we study the physics
in such ultra-dense regions. One would be able then to predict
only partly but not fully the universe in the future of a given
epoch of time. We consider some of these issues below.

\begin{figure}
\centerline{\includegraphics[width=9cm]{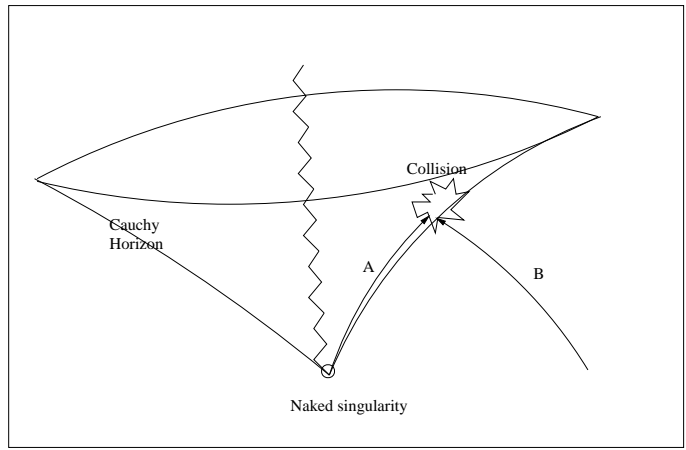}}
\caption{The existence of a naked singularity is
typically characterized by existence of a Cauchy horizon
in the spacetime. Very high energy particle collisions
can occur close to such a Cauchy horizon.} \label{f:four}
\end{figure}

A concern usually expressed is that if naked
singularities occurred as the final fate of gravitational
collapse, that would break the predictability in the
spacetime, because the naked singularity is characterized by the
existence of light rays and particles that emerge from the
same. Typically, in all the collapse models discussed
above, there is a family of future directed non-spacelike
curves that reach external observers, and when extended
in the past these meet the singularity.
The first light ray that comes out from the singularity
marks the boundary of the region that can be predicted
from a regular initial Cauchy surface in the spacetime,
and that is called the Cauchy horizon for the spacetime.
The causal structure of the spacetime would differ
significantly in the two cases, when there is a Cauchy
horizon and when there is none. A typical
gravitational collapse to a naked singularity, with
the Cauchy horizon forming is shown in Fig \ref{f:four}.

Here we would like to mention certain recent intriguing
results in connection to the existence of a Cauchy horizon
in a spacetime when there is a naked singularity resulting as
final fate of collapse. Let us suppose the collapse resulted
in a naked singularity. In that case, there are classes
of models where there will be an outflow of energy and
radiations of high velocity particles close to the Cauchy
horizon, which is a null hypersurface in the spacetime.
Such particles, when they collide with incoming particles,
would give rise to a very high center of mass energy
of collisions. The closer we are to the Cauchy horizon,
the higher is the center of mass energy of collisions.
In the limit of approach to the Cauchy horizon, these
energies approach arbitrarily high values and could reach
the Planck scale energies (see for example Refs.
\refcite{PJM,PJ10,PJ11a,PJ11b}).

The point here is, given a regular initial data on a
spacelike hypersurface, one would like to predict the future
and past evolutions in the spacetime for all times
(see for example Ref.
\refcite{HE}
for a discussion).
Such a requirement is termed as the global hyperbolicity
of the spacetime. A globally hyperbolic spacetime is a fully
predictable universe, it admits a Cauchy surface, the data
on which can be evolved for all times in the past as well
as in future. Simple enough spacetimes such as the Minkowski
or Schwarzschild are globally hyperbolic, but the Reissner-Nordstrom
or Kerr geometries are not globally hyperbolic. For further
details on these issues, we refer to Refs.
\refcite{HE,Joshi2008}.

The key role that the event horizon of a black hole plays
is that it hides the super-ultra-dense region formed in collapse
from us. So the fact that we do not understand such regions
has no effect on our ability to predict what happens
in the universe at large. But if no such horizon exists, then
the ultra-dense region might, in fact, play an important and
even decisive role in the rest of the universe, and our ignorance
of such regions would become of more than merely academic
interest.

Yet such an unpredictability is common in general relativity,
and not always directly related to censorship violation. Even black
holes themselves need not fully respect predictability when
they rotate or have some charge. For example, if we drop an
electric charge into an uncharged black hole, the spacetime geometry
radically changes and is no longer predictable from a regular
initial epoch of time. A charged black hole admits a naked
singularity which is visible to an observer within the horizon,
and similar situation holds when the black hole is rotating.
There is an important debate going on now
for many years, if one could over-charge or over-rotate a black
hole so that the singularity
visible to observers within the horizon becomes visible to external
far away observers too. We discuss this in some detail below.
Another point is, if such a black hole was big enough on a
cosmological scale, the observer within the horizon could
survive in principle for millions of years happily without
actually falling into the singularity, and would thus be able
to observe the naked singularity for a long time. Thus,
only purest of pure black holes with no charge or rotation at
all respect the full predictability, and all other physically
realistic ones with charge or rotation actually do not.
As such, there are very many models of the universe in cosmology
and relativity that are not totally predictable from a given
spacelike hypersurface in the past. In these universes, the
spacetime cannot be neatly separated into space and time foliation
so as to allow initial data at a given moment of time to
fully determine the future.

In our view, the real breakdown of predictability is the
occurrence of spacetime singularity itself, which indicates
the true limitation of the classical gravity theory. It does
not matter really whether it is hidden within an event horizon
or not. The real solution of the problem would then be the
resolution of singularity itself, through either a quantum
theory of gravity or in some way at the classical level
itself.
\cite{Harada04}

Actually, the cosmic censorship way to predictability,
that of `hiding the singularity within a black hole', and
then thinking that we restored the spacetime predictability
may not be the real solution, or at best it may be only a partial
solution to the key issue of predictability in spacetime universes.
In fact, it may be just shifting the problem elsewhere, and some
of the current major paradoxes faced by the black hole physics
such as the information paradox, the various puzzles regarding
the nature of the Hawking radiation, and other issues could
as well be a manifestation of the same.

Another issue is that censorship has been the foundation
for the laws of black holes such as the area theorem and others,
and their astrophysical applications. But these are not free
of major paradoxes. First, all the matter entering a black hole
must of necessity collapse into a spacetime singularity of
infinite density and curvatures, where all known laws of physics
break down. This was a reason why many gravitation theorists
of 1940s and 1950s objected to black hole formation, and Einstein
himself repeatedly argued against such a final fate of a collapsing
star, writing a paper in 1939 to this effect. Second, as is
well-known and has been widely discussed in the past few years,
a black hole, by potentially destroying information, appears
to contradict the basic principles of quantum theory. In that
sense, the very formation of a black hole itself with a
singularity within it appears to come laden with inherent
problems. It is far from clear how one would resolve these
basic troubles even if censorship were correct.

In view of such problems with the black hole paradigm,
a possibility worth considering is the delay or avoidance of
horizon formation as the star collapses under gravity. This
happens when collapse to a naked singularity takes place, namely,
where the horizon does not form early enough or is avoided.
In such a case, if the star could radiate away most of its mass
in the late stages of collapse, this may offer a way out of
the black hole conundrum, while also resolving the singularity
issue, because now there is no mass left to form the singularity.
\cite{PJM}

It has been observed recently that in the vicinity
of the event horizon for an extreme Kerr black hole,
if the test particles arrive with fine tuned velocities,
they could undergo very high energy collisions with other
incoming particles. In that case,
as in the similar case with a naked singularity
mentioned above,
the possibility arises that one could see the Planck scale physics
or ultra-high energy physics effects near the event horizon,
given suitable circumstances (see for example Refs.
\refcite{Banados}--\refcite{Zaslavskii}).

What we mentioned above related to the particle collisions
near Cauchy horizon is a similar scenario where the background
geometry is that of a naked singularity. These results could
mean that in strong
gravity regimes, such as those of black holes or naked
singularities developing in gravitational collapse,
there may be a possibility of observation for ultra-high energy
physics effects, which would be very difficult to see in
near future in terrestrial laboratories.

While such a phenomena gives rise to the prospect of
observing the Planck scale physics near the Cauchy horizon
in the gravitational collapse framework, it also raises the
following intriguing question.
If extreme high energy collisions do take place very close
to the null surface which is the Cauchy horizon,
and if the idea of a singularity being the limit of the
classical theory where quantum effects become relevant is valid
then in a certain sense
these collision models are
essentially equivalent to creating
a singularity at the Cauchy horizon itself. In that case, all or
at least some of the Cauchy horizon would be converted into
a spacetime singularity, and would effectively mark the end
of the spacetime itself. In such a case, the spacetime
manifold terminates at the Cauchy horizon, whenever a naked
singularity is created in gravitational collapse. Since
the Cauchy horizon marks in this case the boundary of the
spacetime itself, the predictability is then restored for
the spacetime, because the rest of the spacetime below and
in the past of such a horizon is any way predictable
before the Cauchy horizon formed.

There is another important aspect to the cosmic
censorship problem that we discuss below now briefly.
Let us suppose a black hole did form as the final end state of
the gravitational collapse of a massive star. There is a
constraint in this case for the horizon to survive, namely
that the black hole must not contain too much of charge or
should not spin too fast. In the case otherwise, the horizon
cannot be sustained, it will breakdown and the singularity
within will be visible.

Even if the black hole formed with a small enough charge and
angular momentum to begin with, there is this key astrophysical
process in its surroundings, namely that of accretion of
matter around it which is the lot of debris and outer layers
of the collapsing star. This matter around the black hole
will as such fall into the same with great velocity, which
could be classical or quantized, and with
charge and angular momentum. Such in-falling particles
could `charge-up' or `over-spin' the black hole, thus eliminating
the event horizon. Thus, the very fundamental characteristic
of a black hole, namely its trait of gobbling up the matter
all around it and keep growing could become its own
nemesis and a cause of its destruction.

Thus, even if a massive star collapsed into a black hole rather
than a naked singularity, important issues remain such as the
stability of the same to throwing in particles with charge or
large angular momentum, and whether that can convert the
black hole into a naked singularity by eliminating its event
horizon. Many researchers have claimed this is possible, and
gave models to create naked singularities this way. But
there are others who claim there are physical effects
which would save the black hole from over-spinning this way
to destroy itself, and the issue is very much out to the jury.
The point is, in general, stability of the event horizon and
black holes continues to be an important issue for
black holes formed in gravitational collapse. For a recent
discussion on some of these issues, we refer to Refs.
\refcite{Matsas07}--\refcite{Barausse}
and references therein.

The primary concern of the censorship hypothesis is of course
formation of black holes only as collapse endstates, and their
stability as above is a secondary issue. Therefore, what this
means for cosmic censorship is, the collapsing massive star should
not retain or carry too much of charge or spin, otherwise it
will necessarily end up as a naked singularity, rather than
a black hole final state.

It is clear that the black hole and naked singularity
outcomes of a complete gravitational collapse for a massive
star are very different from each other physically, and
would have quite different observational signatures.
In the naked singularity case, if it occurs in nature,
we have the possibility to observe the physical effects
happening in the vicinity of the ultra dense regions that form
in the very final stages of collapse. However, in a black
hole scenario, such regions are necessarily hidden
within the event horizon of gravity. The fact that a
slightest stress perturbation of the OSD collapse could
change the collapse final outcome drastically, as we
noted in section \ref{generic}, changing it from black
hole formation to a naked singularity, means that the
naked singularity final state for a collapsing star
must be studied very carefully to deduce its physical
consequences, which are not well understood so far.

It is, however, widely believed that when we have
a reasonable and complete quantum theory of gravity
available, all spacetime singularities, whether naked
or those hidden inside black holes, will be resolved away.
As of now, it remains an open question if the quantum
gravity will remove naked singularities.
After all, the occurrence of spacetime singularities
could be a purely classical phenomenon, and
whether they are naked or covered should not be relevant,
because quantum gravity will possibly remove them
all any way. But one may argue that looking at the
problem this way is missing the real issue.
It is possible that in a suitable quantum gravity theory
the singularities will be smeared out, though this has been
not realized so far. Also there are indications that
in quantum gravity also the singularities may not
after all go away.

\begin{figure}
\centerline{\includegraphics[scale=0.7]{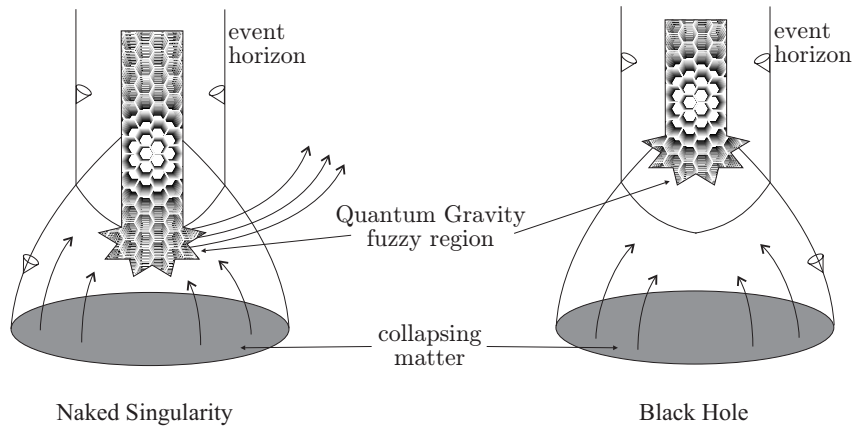}}
\caption{Even if the naked singularity is resolved by
the quantum gravity effects, the ultra-strong
gravity region that developed in gravitational collapse
will still be visible to external observers
in the universe. \label{f:five}}
\end{figure}

In any case, the important and real issue is,
whether the extreme strong gravity regions formed due
to gravitational collapse are visible to faraway observers
or not. It is quite clear that the gravitational collapse
would certainly proceed classically, at least till the
quantum gravity starts governing and dominating the
dynamical evolution at the scales of the order
of the Planck length, {\it i.e.} till the extreme gravity
configurations have been already developed due to
collapse. The key point is, it is the visibility or
otherwise of such ultra-dense regions that is under
discussion, whether they are classical or quantum
(see Fig \ref{f:five}).

What is important is, classical gravity implies
necessarily the existence of ultra-strong gravity
regions, where both classical and quantum gravity come into
their own. In fact, if naked singularities do develop in
gravitational collapse, then in a literal sense we come
face-to-face with the laws of quantum gravity, whenever
such an event occurs in the universe.
\cite{Wald2}

In this way, the gravitational collapse phenomenon
has the potential to provide us with a possibility of
actually testing the laws of quantum gravity.
In the case of a black hole developing in the
collapse of a finite sized object such as a massive star,
such strong gravity regions are necessarily hidden
behind an event horizon of gravity, and this would be
well before the physical conditions became extreme
near the spacetime singularity.
In that case, the quantum effects, even if they caused
qualitative changes closer to singularity, will be
of no physical consequences. This is because no causal
communications are then allowed from such regions. On
the other hand, if the causal structure were that
of a naked singularity, then the communications from
such a quantum gravity dominated extreme curvature
ball would be visible in principle. This will be so
either through direct physical processes near a
strong curvature naked singularity, or via the
secondary effects, such as the shocks produced in
the surrounding medium.

\section{Star collapse: A Lab for quantum gravity?}\label{QG}
At present, we have no mechanism or a complete theory to
deal with both quantum effects and the intense force of gravity
together. However, a quantum gravity theory should take
over from the purely classical theory that is general relativity,
in the very advanced stages of a gravitational collapse,
when densities and spacetime curvatures assume extreme values.
It is possible that a spacetime singularity basically
represents the incompleteness of the classical theory and
when quantum effects are combined with the gravitational
force, the classical singularity may be resolved.

Therefore, more than the existence of a naked singularity,
the important physical issue then is whether the extreme
gravity regions formed in the gravitational collapse of a
massive star are visible to external observers in the universe.
An affirmative answer here would mean that such a collapse
provides a good laboratory to study quantum gravity effects in
the cosmos, which may possibly generate clues for an as yet
unknown theory of quantum gravity. Quantum gravity theories
in the making, such as the string theory or loop quantum
gravity in fact are badly in need of some kind of an observational
input, without which it is nearly impossible to constrain
the plethora of possibilities.

We could say quite realistically that a laboratory similar
to that provided by the early universe is created in the collapse
of a massive star. However, the big bang, which is also
a naked singularity in that it is in principle visible to all
observers, happened only once in the life of the universe
and is therefore a unique event. But a naked singularity of
gravitational collapse could offer an opportunity to explore
and observe the quantum gravity effects every time a massive
star in the universe ends its life.

The important questions one could ask are: If in realistic
astrophysical situations the star terminates as a naked singularity,
would there be any observable consequences which reflect the
quantum gravity signatures in the ultra-strong gravity region?
Do naked singularities have physical properties different
from those of a black hole? Such questions underlie our
study of gravitational collapse.

>From this perspective, the recent results on collapse indicate
certain exciting implications for astrophysics, fundamental physics
and quantum gravity. Consider the scenario when a collapsing star
terminates into a naked singularity, making the ultra-strong super
gravity region visible to external observers. In this context, we
considered a cloud that collapsed to a naked singularity
final state, and introduced loop quantum gravity effects.
\cite{JoshiGoswamiLQG}
It turned out that the quantum effects generated an extremely
powerful repulsive force within the cloud. Classically the cloud would
have terminated into a naked singularity, but quantum effects
caused a burstlike emission of matter in the very last phases of
collapse, thus dispersing the star and dissolving the naked
singularity. The density remained finite and the spacetime
singularity was eventually avoided.  One could expect this to
be a fundamental feature of other quantum gravity theories
as well, but more work would be required to confirm such a
conjecture.

For a realistic star, its final catastrophic collapse takes
place in matter of seconds. A star that lived millions of
years thus collapses in only tens of seconds. In the very last
fraction of a microsecond, almost a quarter of its total mass
must be emitted due to quantum effects, and therefore this
would appear like a massive, abrupt burst to an external observer
far away. Typically, such a burst will also carry with it specific
signatures of quantum effects taking place in such ultra-dense
regions. In our case, these included a sudden dip in the
intensity of emission just before the final burstlike evaporation
due to quantum gravity.

The question is, whether such unique astrophysical signatures
can be detected by modern experiments, and if so, what they tell
on quantum gravity, and if there are any new insights into
other aspects of cosmology and fundamental theories such as
string theory.

The stage at which the burst occurs in the gravitational
collapse and the energy which would be emitted depend on a
quantization parameter in the theory. For all theoretically favored
choices of this parameter, the energy emitted in the radiation
would be extreme. The emission mechanism of the burst
has to be further investigated. The constituents of the burst
may include extreme energy gamma-rays, cosmic rays, and neutrinos
which raise an interesting possibility to use upcoming experiments
such as Extreme Universe Space Observatory (EUSO) which may have
needed sensitivity, and which is expected to be operational soon,
to provide us test of this prediction. If tested these experiments
would provide a proof of quantum gravity. Future astronomical
experiments could then constrain the parameters in quantum gravity
in the same way as particle accelerators at CERN and Fermilab
constrain the parameters for the Standard Model.

Interestingly, these experiments which may constrain the
value of the quantization parameter in the theory would also
have consequences for cosmology, because loop quantum cosmology
on which our work is based also changes the picture of
cosmological dynamics in the very early Universe. For example,
it can change the way inflation occurs and it has been shown
that such a change has observable signatures in the Cosmic
Microwave Background Radiation (CMBR), which is the suppression
of power at large scales in CMBR as observed by the Wilkinson
Microwave Anisotropy Probe (WMAP). Observable effects of quantum
gravity in CMBR are also controlled by the same quantization
parameter which determines the details of energy emission in
the burst we considered, and the way a dying star would dim
before a burst occurs. Thus any constraints on the quantization
parameters would have direct consequence for cosmology
as well as astrophysics.

The key point is, because the very final ultra-dense regions
of the star are no longer hidden within a horizon as in the black
hole case, the exciting possibility of observing these quantum
effects arises now, independently of the quantum gravity theory
used. An astrophysical connection to extreme high energy
phenomena in the universe, such as the gamma-rays bursts that defy
any explanations so far, may not be ruled out. Some researchers
have also examined the possible generation of gravity waves
from such ultra-strong gravity regions.
\cite{INH}

Such a resolution of naked singularity through quantum gravity
would be a philosophically satisfying possibility. Then, whenever
a massive star undergoes a gravitational collapse, this might
create a laboratory for quantum gravity in the form of a
{\it Quantum Star}
\cite{SciAm},
that we may be able to possibly access.
This would also suggest intriguing connections to high
energy astrophysical phenomena. The present situation poses
one of the most interesting
challenges which have emerged through the recent work on
gravitational collapse.

\section{Concluding remarks}
We hope the considerations here have shown that gravitational
collapse, which essentially is the investigation of dynamical
evolutions of matter fields
under the force of gravity in the spacetime,
provides one of the most exciting research frontiers in gravitation
physics and high energy astrophysics.

There are issues here which have deep relevance both for
theory as well as observational aspects in astrophysics and
cosmology. Also these problems are of relevance
for basics of gravitation theory and quantum gravity, and these
inspire a philosophical interest and inquiry into the
nature and structure of spacetime, causality, and predictability
in the universe.

Active research is already happening in many of these
areas as the discussion here has pointed out. Some of the most
interesting areas from our own personal perspective are: Genericity
and stability of collapse outcomes, examining the quantum gravity
effects near singularities, observational and astrophysical
signatures of the collapse outcomes, and several other related
issues.

In particular, one of the interesting and important questions
would be, if naked singularities which are hypothetical astrophysical
objects, did actually form in nature, what distinct observational
signatures would these present. In other words, how one would
distinguish the black holes from naked singularities would be an
important issue. There have already been some developments and efforts
on this issue in recent years as we have indicated above.
The point here is, there are already very high energy astrophysical
phenomena being observed today, with several observational missions
working both from ground and space. The black holes and naked
singularities, which are logical consequences of the general
theory of relativity as we consider the gravitational collapse
of a massive star, would appear to be the leading candidates to
explain these phenomena. Thus the observational signatures
that each of these would present, and their astrophysical
consequences would be naturally of much interest for the future
theoretical as well as computational research, and for their
applications.

In our view, there is a scope therefore for both theoretical
as well as numerical investigations in these frontier areas, which
may have much to tell for our quest on basic issues in quantum
gravity, fundamental physics and gravity theories, and towards
the expanding frontiers of modern high energy astrophysical
observations.

\noindent \textbf{Acknowledgments:}\\
We would like to thank many colleagues and
collaborators for very many discussions which have contributed
greatly to shape our understanding of the questions discussed
here. In particular, we thank Peter Biermann, Abhay Ashtekar,
Ramesh Narayan, Jayant V. Narlikar, and also Gary Horowitz
and Kip Thorne for detailed discussions and comments during
recent ICGC meeting at Goa. Figure acknowledgements are,
Fig. \ref{f:three} is from Ref. \refcite{Joshi2008},
Fig. \ref{mena} is from Ref. \refcite{Tavakol},
Figures \ref{potential2} and \ref{equilibrium} are
from Ref. \refcite{JMN} and Fig. \ref{f:five}
is from Ref. \refcite{PJM}.

\end{document}